\documentclass[atoms,review,accept,moreauthors,pdftex,12pt,a4paper]{mdpi}

\usepackage{amsmath,amssymb}
\usepackage[normalem]{ulem}		

\usepackage{booktabs}
\usepackage{multirow}
\usepackage{soul}
\usepackage{subfigure}
\makeatletter
\renewcommand{\@thesubfigure}{\normalsize(\textbf{\alph{subfigure}})}
\makeatother

\lastpage{x}
\doinum{10.3390/------}
\pubvolume{xx}
\pubyear{2015}
\externaleditor{{Academic Editor}: Jonathan Goldwin and Duncan O'Dell}
\history{Received: 31 July 2015  / Accepted: 15 December 2015 / Published: xx}

\def \beq{\begin{equation}}
\def \eeq{\end{equation}}
\def \bse{\begin{subequations}}
\def \ese{\end{subequations}}
\def \bea{\begin{eqnarray}}
\def \eea{\end{eqnarray}}
\def \bem{\begin{displaymath}}
\def \eem{\end{displaymath}}
\def \bem{\begin{pmatrix}}
\def \eem{\end{pmatrix}}
\def \bb{\bibitem}
\def \bs{\boldsymbol}
\def \nn{\nonumber}
\def \mf{\tilde{J}_1}
\def \mj{\tilde{J}_0}

\def \ha{\hat{a}}

\def \hh{\hat{H}}
\def \mH{\hat{\mathcal{H}}}
\def \hp{\hat{\Psi}}
\def \hb{\hat{b}}
\def \hB{\hat{\mathcal{T}}}
\def \hN{\hat{\mathcal{N}}}

\newcommand{\upa}{\uparrow}
\newcommand{\dna}{\downarrow}
\newcommand{\pdag}{\phantom\dagger}

\newcommand{\ket}[1]{|#1\rangle}

\newcommand{\ketbra}[2]{| #1 \rangle \langle #2 |}
\newcommand{\expect}[1]{\langle#1\rangle}

\newcommand{\si}[1]{\hp^{\pdag}_{#1} (\bs x)}
\newcommand{\dsi}[1]{\hp^\dag_{#1} (\bs x)}

\def \Ps{\hat{\Psi}(\boldsymbol{r})}
\def \Pds{\hat{\Psi}^{\dagger}(\boldsymbol{r})}
\def \i{{\int}d^2{\bf r}}

\def \c{\hat{c}_{n,m}}

\def \cd{\hat{c}_{n,m}^{\dagger}}
\def \cdp{\hat{c}_{n',m'}^{\dagger}}

\Title{Cavity Optomechanics with Ultra Cold Atoms in Synthetic Abelian and Non-Abelian Gauge Field}

\Author{Bikash Padhi $^{1}$ and
 Sankalpa Ghosh $^{2,}$*}

\address{%
$^{1}$ Physics Department, University of Illinois at Urbana-Champaign, West Green Street, Urbana, \linebreak IL 61801-3080, USA; {E-Mail: bpadhi2@illinois.edu} \\
$^{2}$ Physics Department, Indian Institute of Technology, Hauz Khas, New Delhi 110016, India}

\corres{E-Mail: sankalpa@physics.iitd.ac.in.}

\abstract{ In this article we present a pedagogical discussion of some of the optomechanical properties of a high finesse cavity loaded with ultracold atoms in laser induced synthetic gauge fields of different types.
Essentially, the subject matter of this article is an~amalgam of two sub-fields of atomic molecular and optical (AMO) physics namely, the cavity optomechanics with ultracold atoms and ultracold atoms in synthetic gauge field. After~ providing a brief introduction to either of these fields we shall show how and what properties of these trapped ultracold atoms can be studied by looking at the cavity (optomechanical or transmission) spectrum. In presence of abelian synthetic gauge field we discuss the cold-atom analogue of Shubnikov de Haas oscillation and its detection through cavity spectrum.
Then, in the presence of a non-abelian synthetic gauge field (spin-orbit coupling), we see when the electromagnetic field inside the cavity is quantized, it provides a quantum optical lattice for the atoms, leading to the formation of different quantum magnetic phases.  We also discuss how these phases can be explored by studying the cavity transmission~ spectrum.
}


\begin{document}

\section{Introduction}

The fact that electromagnetic radiation can apply forces on mechanical objects through radiation pressure follows directly from Maxwell's equations and was experimentally verified more than a century
ago \cite{radpres1, radpres2}. Cavity optomechanics \cite{MG, Aspelmeyer} concerns itself with the
general method of controlling the mechanical degrees of various objects through the effect of light by coupling them with the resonant modes of Fabry-P\'erot
cavity. The field is referred as cavity quantum optomechanics \cite{COMrev1, COMrev2} when such mechanical degrees of freedom are quantized.  As a result, cavity optomechanics offers a route to determine and
control the quantum states of microscopic as well as macroscopic object. This is why using cavity optomechanical technique it is possible to do hypersensitive measurement down to the size limited only by quantum mechanics.

It is also possible to conduct fundamental tests of quantum mechanics on massive objects consisting of macroscopic number of atoms \cite{COMrev2, COMrev3, DSKrev}. The type of systems where these techniques can be applied are therefore very diverse such as movable mirrors on a cantilever or nanobeam, membranes or nanowires inside a cavity,
nanobeams or vibrating plate capacitors in superconducting microwave resonators, atom or atomic cloud in a cavity \textit{etc}. and the typical scale of coupling frequency can vary from few Hz to GHz.  Therefore, it is no wonder that cavity quantum optomechanics has not only emerged as an extremely fascinating branch in experimental or theoretical physics but also it comes with a number of promising practical implementation.

\begin{figure}[H]
\centering
\includegraphics[width=0.85\columnwidth , height=
0.55\columnwidth]{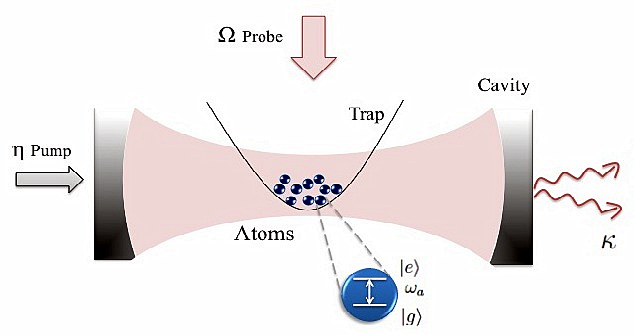}
\caption{A typical configuration for trapped ultracold atoms in cavity. The purpose of using a probe laser and a pump laser are explained later. The $\kappa$ measured the photon decaying (``leaking'') out of the cavity. The inset shows a two level atom trapped in the cavity field.}
\label{atomcavity}
\end{figure}

An interesting example of such a cavity quantum optomechanical system is the one consisting of  a~two level atom placed inside a Fabry-P\'erot cavity interacting with a single resonating mode
of the electromagnetic wave.
Such a system is described by well known Jaynes-Cummings model,
a simple but a highly illustrative model in quantum optics to demonstrate light-atom interaction \cite{JCM}.
This model and it's generalization for an $N$-atom system can lead to the realization two most well known applications of cavity optomechanics. Since the atom-photon coupling inside a cavity is a function of the spatial co-ordinate inside the cavity, cavity optomechanical technique can sense the atomic position in a standing wave of light and thereby enabling to perform very sensitive measurement associated with a microscopic object \cite{atpos}. The extension of the above model to a macroscopic number ($N \gg 1$) of such atom  (or an atomic cloud)
placed inside a cavity and a coherent state of atoms can be created by placing all the atoms in a single quantum mechanical state that leads to the well known
Dicke or Tavis-Cummings Model~\cite{Dicke, TC}. The corresponding cavity optomechanical system can now make quantum measurements on macroscopic object. In such $N$-atom system cavity-atom coupling parameter scales with the number of atoms $N$ as compared to single atom-photon coupling and becomes much enhanced. Thus
cavity optomechanics with such macroscopic objects realizes strongly-coupled cavity optomechanical system. Schematic of a typical set up is drawn in Figure \ref{atomcavity}.

It is natural that with the discovery of ultracold atomic Bose-Einstein condensate (BEC), \cite{BEC} which is a coherent matter wave made out of macroscopic number
of atoms, there would be efforts to couple such ultracold atomic condensate with macroscopic number of atoms to a high-finesse cavity and to do resulting cavity optomechanics. In this direction, the experiments done
in Berkeley group \cite{DSK1, DSK2},  the atomic ensemble is split into several harmonic traps each trap confining a macroscopic number of atoms. In such cases
the collective atomic degrees of freedom that optomechanically couples with the cavity mode is the sum of the center of mass degrees of freedom of various sub ensembles centered at various harmonic traps. The variation of the optomechanical coupling strength with the equilibrium position of the atomic ensemble was demonstrated in subsequent experiments \cite{DSK3}.

In another experiment  the Z\"urich group
loaded a uniform, stationary and weakly interacting BEC in a high finesse cavity \cite{Eslng} couple the cavity mode with selected low lying Bogoliubov modes of this condensate. The resulting cavity transmission spectrum was analyzed to study the oscillation between the ground state and excited state of such BEC and from there the scaling of the cavity-atom coupling was derived.
All these experiments took place in the dispersive regime of the cavity-atom interaction where the photons are scattered by the atomic ensemble inside the cavity and therefore a study
of the cavity transmission spectrum can be used to identify the quantum many-body state of such ultracold atoms inside the cavity \cite{Mekhov1, Mekhov2}. Particularly it has been pointed out that
detection of quantum many body states of ultracold atoms in this forms a class of quantum non-demolition measurement ( for details see~\cite{RitschRMP} and another article in this special issue \cite{Atom-Mekhov} also addresses this aspect).
Subsequently more detailed analysis of the angle-resolved cavity transmission spectrum and the related quantum diffraction of single mode electromagnetic wave by an atomic ensemble in optical lattice was carried on \cite{Adhip}.

It has also been pointed out that bi-stability in the cavity transmission spectrum alters the modulation of standing wave profile of the electromagnetic field inside a cavity. Due
to strong atom-photon coupling  in the system this can induce/alter quantum phase transition such as Superfluid to Mott insulator in ultracold atomic ensemble loaded into it \cite{Larson, Chen1}.
Theoretical analysis of cavity optomechanical effects~that can result from putting ultracold fermionic atoms inside such cavity was also carried \linebreak on \cite{Meystre} subsequently.

With this progress in cavity optomechanics with ultracold atomic condensate, it is very natural to investigate as to what extent the cavity optomechanical
method can enrich our learning about one of the most fascinating sub-fields of the ultracold atomic physics, namely ultracold atoms in synthetic/artificial gauge field (for review see \cite{dalibardrev, SGRS, Spielman}). Study of the behavior of ultracold atoms in such synthetic gauge field
not only helps one to understand the behavior of such systems in various types of laser induced artificial scalar and vector potential, but it also allows one to quantum simulate a number of exotic phenomena in condensed matter and high energy physics, but now in the ultracold atomic systems with higher control on experimental parameters \cite{QS}. Given this significance it is therefore natural to \linebreak ask if cavity optomechanical techniques can provide new ways to generate such synthetic gauge \linebreak field \cite{Jonas, ring, dong}.
One may also try to investigate what modification cavity does to quantum phases of ultracold atomic system in synthetic gauge field or what type of quantum phenomena can be simulated by placing synthetically gauged ultracold atoms in a cavity \cite{ourPRL, ourPRA}.

To facilitate this direction of investigation, in the first part of this review article we provide a general introduction on
the two related sub-fields of ultracold atoms  (a) cavity optomechanics with ultracold atoms and (b) ultracold atoms in synthetic gauge field. In the later sections we discuss a number of interesting theoretical proposals that came up very recently by gluing these subtopics and pointing out possible future directions.

The rest of the review article is organized as follows. In Section \ref{cavopto} we shall discuss the cavity optomechanics with ultracold atoms after introducing the field of cavity optomechanics in general and how it can be achieved by placing a single atom inside a cavity. In the next section, Section \ref{AbnonAb} we shall provide brief but pedagogical discussion of ultracold atoms in synthetic abelian and non-abelian gauge field. As a particular case of non-abelian gauge field we discuss here briefly the synthetically generated spin-orbit coupling and NIST method of generating
such synthetically spin-orbit coupled ultracold Bose Einstein condensate.
After introducing these background materials, in the subsequent section, Section  \ref{COUCF} we discuss in detail some recent works where it's shown how ultracold fermionic atoms in a synthetic abelian gauge field placed in an optical cavity can lead to the cold atom analogue of well known electronic phenomena like Shubnokov de-Hass oscillation. Then in Section \ref{SOC} we discuss in detail the properties of spin-orbit coupled ultracold bosons inside an optical cavity in the non-interacting limit as well as with interaction. We finally summarize our discussion and point out future directions.

\section{Cavity Optomechanics with Cold Atoms}\label{cavopto}
\vspace{-12pt}

\subsection{Introduction to Cavity Optomechanics}
\vspace{-12pt}
\subsubsection{Classical Treatment}

The prototype cavity optomechanical system consists of a laser driven single mode Fabry-P\'erot cavity with single mode resonant frequency $\omega_{c}$ and length $L$
whose one end-mirror is moving because of the radiation pressure or for any other external reason and therefore represents a mechanical motion. That~is why the system is called optomechanics. It is conventional to consider the mechanical motion to be simple harmonic under the assumption that
the typical damping rate of this mechanical motion is much slower than the damping rate $\kappa$ of the inter-cavity field. This model is very generic in nature and can be realized in a large number of systems \cite{COMrev2}
with optomechanical coupling frequency $\Omega_{m}$ ranges from kHz-GHz. Using this harmonic approximation we can model the motion of the end-mirror {as}
\beq x(t)  \approx x_{0} \sin (\Omega_{m}t). \nn \eeq

For a classical monochromatic pump laser with frequency $\omega_{L}$ and amplitude $E_{in}$, the inter-cavity field obeys the equation
of motion
\beq  \frac{ d E(t)}{dt} = [i ( \Delta + G x(t)) - \frac{\kappa}{2} ] E(t) + \sqrt{\kappa} E_{in} \eeq

The steady state solution ( $\frac{d E(t)}{dt}=0$) is given by
\beq E = \frac {\sqrt{\kappa} E_{in}}{-i ( \Delta + Gx) + \frac{\kappa}{2}} \nn \eeq
where $\Delta = \omega_{L} - \omega_{c}$ is the pump cavity detuning.
Thus the normalized transmitted intensity from the Fabry-P\'erot cavity is given by

\beq I_{out} = |E|^{2} = \frac{\kappa I_{in}} { ( \Delta + Gx)^{2} + (\frac{\kappa}{2})^{2} }  \label{TI1} \eeq
where $I_{in}= |E_{in}|^{2}= \frac{P}{\hbar \omega_{L}}$ with $P$ as the input laser power driving the cavity mode. The above result can also be easily generalized
for the case of quantized fields, in which case $E$ will be interpreted as square root of the inter-cavity photon number, $E = \sqrt{\langle \hat{a}^{\dagger} \hat{a} \rangle }$ with
$\hat{a}$ as the usual annihilation operator for the photon. In the next subsection we shall treat the problem quantum mechanically.

\subsubsection{Quantum Treatment}

When the above system is treated quantum mechanically, the simplest quantum mechanical Hamiltonian ( upto the lowest order) for such an optomechanical system is \cite{Law-Eff}
\beq H_{qm}= \hbar \omega( \hat{x}) \hat{a}^{\dagger} \hat{a} + \hbar \Omega_{m} \hat{b}^{\dagger} \hat{b} \nonumber \eeq

The first term represents the quantized single-mode electromagnetic field ( monochromatic photon) in the cavity. The second term represents the
quantized simple harmonic oscillator that represents the mechanical motion of the moving mirror at the one end of the cavity with the oscillation frequency $\Omega_{m}$.
The functional dependence of the
cavity frequency $\omega$ on the displacement operator $\hat{x}$ of the mechanical oscillator is due to the fact that because of this displacement the cavity length and hence the
cavity frequency will change. This also explains why there will be an optomechanical coupling. One can write $\hat{x}$ in terms of the annihilation and creation operator for the mechanical
oscillator as
\beq \hat{x} = \ell_{m} ( \hat{b} + \hat{b}^{\dagger} ) \nn \eeq
where $\ell_{m}$ is the typical size of the mechanical zero point fluctuation.
Expanding the frequency $\omega$ in terms of $\hat{x}$  and only retaining the linear term one gets
\beq \omega (\hat{x}) = \omega_{c} ( 1- \frac{\hat{x}}{L}) \nn \eeq

Here $L$ is the length of the cavity. Once this expression is inserted in the cavity, the Hamiltonian contains a term like
\beq H_{rp} = - \hat{F} \hat{x},~~~ \hat{F} = \frac{\hbar \omega_{c}}{L} \hat{a}^{\dagger}\hat{a} \nn  \eeq

The standard optomechanical Hamiltonian to the lowest order becomes
\beq H_{om} = \hbar \omega_{c} \hat{a}^{\dagger} \hat{a} + \hbar \Omega \hat{b}^{\dagger} \hat{b} - \hbar g_{0} (\hat{a}^{\dagger} \hat{a})( \hat{b} + \hat{b}^{\dagger} ) \label{hom} \eeq
where the last term represents optomechanical coupling and $g_{0} = \omega_{c} \frac{\ell_{m}}{L}$ represents the strength of the optomechanical coupling. To summarize, the above situation
represents the case of a driven system where the mechanical coupling determines the resonant frequency. A large number of physical systems therefore can be found to implement such scenario.  However, since our aim is to study the properties of ultracold atoms in such a cavity in the next Section \ref{atcav} we shall discuss how such cavity optomechanical system can be realized with single two level atom inside a single mode cavity, a~prototype system in Quantum Optics.

\subsubsection{Two Level Atom in a Single Mode Cavity}\label{atcav}
The system of a two level atom inside a single mode Fabry-P\'erot cavity is described by the well known Jayens-Cummings model.
This is particularly a good approximation when the atomic vapor is dilute and and the electromagnetic interaction is very weak.
The ground and the excited state of the unperturbed atom is $| g \rangle$ and $ |e \rangle$ with energy $E_{g}$ and $E_{e}$ respectively.
The atomic Hamiltonian reads \linebreak (setting the ground state energy to $0$).
\beq \hat{H}_{at} = \hbar \omega_{a} | e \rangle \langle e|.  \nn \eeq

  Here $\omega_{a} = \frac{E_{e} - E_{g}}{\hbar}$ is the atomic transition frequency.
One can introduce  Pauli spin operators to describe the atomic excitation and de-excitation process in such two-level system as
\bea \sigma_{x} & = &  |e \rangle \langle g | +  |g \rangle \langle e | \nonumber \\
         \sigma_{y} & = & -i (|e \rangle \langle g | -  |g \rangle \langle e |) \nonumber \\
         \sigma_{z} & = & |(e \rangle \langle e | -  |g \rangle \langle g | ) \nonumber \eea
with $ \sigma_{+} = \sigma_{x} + i \sigma_{y}, \sigma_{-}= \sigma_{x}-i\sigma_{y},  [\sigma_{+}, \sigma_{-}]=\sigma_{z}, [\sigma_{z}, \sigma_{+}]=2\sigma_{+}$.
The atomic dipole operator $\hat{\bs{d}}= e\hat{\bs{r}}$, where $\bs{r}_{e}$ is the electron position operator relative to the center of mass of the atom, can be written
in this two dimensional Hilbert space as
\beq \hat{\bs{d}} = \bs{d}( \hat{\sigma}_{+} + \hat{\sigma_{-}}) \nn \eeq
with $\bs{d} = \langle e | \hat{\bs{d}} | g \rangle$. Now consider such atom in cavity which consists
the electromagnetic field to be confined in one dimension by two reflecting mirrors.
The cavity Hamiltonian is given as

Assuming the cavity axis as $x$-axis and the electromagnetic wave having a sinusoidal mode profile along that axis
the electric field can be written as
\beq \hat{E}(x) = \sqrt{\frac{\hbar \omega_{c}}{2 \epsilon_{0} V_{c}}} ( \hat{a} + \hat{a}^{\dagger})\cos (kx) \nn \eeq
where $\omega_{c} =c k$. Here $c$ is the velocity of light, $V_{c}$ is the cavity volume and $\epsilon_{0}$ is the free space permeability.
 With this the cavity atom coupling term will be
\beq \hat{H}_{at-cav} = \hbar g(x) ( \hat{a}^{\dagger} \sigma_{-} + a \sigma_{+}) \label{cavityatom1} \eeq
with
\beq g(x) = -\sqrt{\frac{\hbar \omega_{c}}{2 \epsilon_{0}V_{c}}} \bs{d} \frac{\cos (kx)}{\hbar}  \eeq

The above atom-photon coupling strength
can be defined as $g(x) = g_{0} \cos (kx)$ where $g_{0}$ is called the vacuum Rabi frequency.
The full Hamiltonian, namely
\beq \hat{H}_{JC} = \hat{H}_{at} + \hat{H}_{c} + \hat{H}_{at-cav} \label{JCham1} \eeq
is called the Jaynes-Cummings Hamiltonian \cite{JCM, JCH}.

To model an ultracold atomic condensate inside a cavity we shall first consider the single particle Hamiltonian in such system \cite{thesis1}.
The corresponding single particle Hamiltonian in the frame of the pump laser oscillating with frequency $\omega_{p}$ takes the form
\beq H = H_{mec} +  H'_{JC} + H_{L} + H_{P} \label{coldJC} \eeq

Here $H_{mech}= \frac{p^{2}}{2m}$ describes the center of mass motion dynamics in absence of coupling with electromagnetic field. And
\bea H'_{JC} & = & -\hbar \Delta_{a} \sigma_{+} \sigma_{-}  - \hbar \Delta_{c} \hat{a}^{\dagger} \hat{a} + \hbar g(x) ( \hat{a}^{\dagger} \sigma_{-} + a \sigma_{+})
 \\
H_{L} &= & \hbar \Omega ( \sigma_{+} + \sigma_{-})  \\
H_{P} & = & \hbar \eta (a^{\dagger} + a)  \eea

Here $\Delta_{a}=\omega_{p} - \omega_{0}$ is the atom-pump detuning and $\Delta_{c}= \omega_{p} - \omega_{c}$ is the cavity-pump detuning. \linebreak $H_{L}$ describes the dipole interaction with the Laser with a coupling strength $\Omega$.

The Heisenberg equation of motion for the dipole and the photon operators are, respectively \linebreak (for ease of discussion we have omitted the terms that give dissipation and decoherence due to cavity-environment coupling)
\bea \dot{\sigma}_{+} & = & -\frac{i}{\hbar}[\sigma_{+}, H]  \label{dipole} \\
\dot{\hat{a}} & = & -\frac{i}{\hbar} [a, H]  \label{photon} \eea

For large detunings $( |\Delta_{a} | \gg g_{0}, \Delta_{c}$, the atomic transition to the excited state is
suppressed. One can then set the LHS of Equation \eqref{dipole} to zero and take $\sigma_{z}(t)  \approx 1$.
This is known as adiabatic elimination of the excited level \cite{Mekhov1, Mekhov2, ourPRA}.
Under this situation one gets
\beq  \sigma_{+} = \frac{g(x) \hat{a}^{\dagger} + \Omega}{\Delta_{a}} \nn \eeq

Substituting this result in the Hamiltonian in Equation \eqref{coldJC} one gets
\bea H'_{atom-cav} & = & \frac{p^{2}}{2m} + \hbar[U_{0} \cos^{2} (kx) - \Delta_{c}]\hat{a}^{+} \hat{a} + \hbar S_{0} \cos (kx) ( \hat{a}^{\dagger} + \hat{a}) \nn \\
             & + & \hbar \eta (\hat{a}^{\dagger} + \hat{a}) + \hbar \frac{\Omega^{2}}{\Delta_{a}} \eea

 Now it is manifest from this form of the Hamiltonian (specifically, from the detuning terms) that the presence of the atoms has effectively shifted the cavity resonance.
Let us concentrate on the first two terms of these Hamiltonian. It clearly represents an optomechanical system if we compare it  with the Hamiltonian in Equation \eqref{hom}.
Thus cavity optomechanical properties can be realized by placing a single atom in a cavity where the mechanical degrees of freedom are now represented by the atom.
In the next subsection we shall show how it can be generalized for the case of ultracold atomic condensed and moreover strong coupling regime of the cavity optomechanics
can be realized with such condensate.

\subsubsection{Ultracold Atoms in a Cavity}

A relatively new development in the field of cavity quantum optomechanics is the ultracold atomic BEC in a cavity and the resulting optomechanics.
To understand we first generalize the discussion in the previous case for a system consists of $N$ two level atoms in a cavity. The collection of such $N$-atom can be thought as a cloud of super-atom with mass $N \, m_{a}$ sitting at the atomic co-ordinate $x_{a}$ where we are describing the dynamics of the center of mass of this
atomic cloud. After carrying out the adiabatic elimination of the excited states of such $N$ two-level atomic system cavity-super-atom coupling term through the dipole interaction takes the form
\beq \hat{V}_{dip} = \hbar N \frac{g_{0}^{2}}{\Delta_{ca}} \cos^{2} ( k\hat{x}_{a} - k \hat{x}_{M} - kL) \hat{a}^{+}\hat{a} \eeq
where $\hat{x}_{M}$ is the position operator corresponding to mechanical object coupled with one end of the \linebreak cavity \cite{ML} and $L$ is the cavity length.


Let us consider a BEC at $T=0$ trapped inside a single mode Fabry-P\'erot cavity of length $L$ and cavity frequency $\omega_{c}$, with the driving laser frequency $\omega_{L}$ and wave number
$k$. If $\omega_{L}$ is far detuned from the atomic transition frequency of the excited electronic state of the atom, such excited state can be adiabatically eliminated and the atomic cloud act dispersively with the cavity field. In the dipole and rotating wave approximation, the Hamiltonian that will be describing such cold atom-photon system
minimally will be given by $H= H_{atom} + H_{field}$ with
\bea H_{atom}  & = & \int dx \hat{\Psi}^{\dagger} (x) [ \frac{\hat{p}_{x}^{2}}{2M} + \hbar U_{0} cos^{2} (kx) \hat{a}^{+} \hat{a} ] \hat{\Psi}(x) \nn \\
           H_{field}  & = &  \hbar \omega_{c} \hat{a}^{\dagger} \hat{a} \label {hbec} \eea

Here the atoms interact with the light field in the cavity through the  familiar off-resonant coupling
\beq U_{0} = \frac{g_{0}^{2}}{\omega_{L} - \omega_{a}} \nn \eeq
with $g$ is single-photon Rabi frequency. As always the case for a real cavity there will be terms in the full
atom-cavity Hamiltonian which will take into account the external driving of the cavity field, dissipation and collisions.

The hamiltonian defined in Equation \eqref{hbec} suggests two important aspects of the ensuing physics that will occur when an ultracold atomic system will be placed inside such cavity. The cavity atom interaction term
$ \hbar U_{0} cos^{2} (kx) \hat{a}^{+} \hat{a} $ will provide a quantum optical lattice potential \cite{Maschler} which can realize novel quantum phases for the ultracold atoms. Also the cavity atom-interaction can create interesting collective excitation over the many-body ground state of ultracold atoms whose properties can be studied from the cavity transmission spectrum, that is from the photon emitted from the cavity. We shall describe in somewhat detail
these aspects in later section particularly when the ultracold atomic ensemble is experiencing  certain type of synthetic gauge field. To that purpose in the next section, Section \ref{AbnonAb}.
\mbox{we shall} review briefly the properties of ultracold atoms in synthetic abelian and non-abelian gauge field.

\section{Ultracold Atoms in Abelian and Non-Abelian Gauge Field}\label{AbnonAb}

The ultracold atomic Bose-Einstein condensate (BEC), whose properties will be discussed in \linebreak this section, consists of interacting bosonic atoms at a temperature close to absolute zero.
Under the typical experimental condition, the system is described very well by the mean field Gross-Pitaevskii \linebreak equation
\cite{Stringari}.  The Gross-Pitaevskii equation for such trapped condensate looks like
\beq i \hbar \frac{\partial \Psi}{\partial t} = ( -\frac{\hbar^{2}}{2m} \nabla^{2} + \frac{1}{2} m (\omega)^{2} r^{2} + g |\Psi |^{2}) \Psi \label{GP}  \eeq

The above equation is a non-linear Schr\"odinger equation with $\Psi$ being the mean field superfluid order parameter of the $N$-boson
condensate. If we set $g=0$ in Equation \eqref{GP}, it  is mathematically same as the usual Schro\"dinger equation even though $\Psi$ here is not the usual quantum mechanical wavefunction.
 We therefore start our discussion by outlining some general features of the gauge invariance of an usual quantum mechanical system described by Schr\"odinger equation. We first discuss the gauge invariance for abelian gauge field and then extends the discussion for the non-abliean cases.

\subsection{Abelian Gauge Field}
\label{Ab}

According to classical electrodynamics
upto a gauge transformations the vector and scalar potentials are arbitrary.
This non uniqueness of the vector potential for a given magnetic field, however, does not create a problem
in describing the motion of charged particle in classical mechanics under Newton's Laws since the
the Lorenz force is given by the gauge invariant magnetic field through
\beq F = q \bs{v} \times \bs{B} \nonumber \, .\eeq

In quantum mechanics our description of a physical system is through Schr\"odinger equation
\beq  i \hbar \frac{\partial \Psi}{\partial t} = H \psi  \Rightarrow  \Psi (\bs{x}, t) = \mathcal{U}(t) \Psi (\bs{x},0) \nonumber \eeq
where the time evolution operator $\mathcal{U}(t) = \exp (-i \frac{\hat{H} t}{\hbar})$.
Now the Hamiltonian in presence of  magnetic field is given as
\beq H = \frac{1}{2m} (\bs{p} - \frac{e}{c}\bs{A})^{2}. \nonumber \eeq

In the above expression, the canonical momentum $\bs{p}$  is related to the mechanical momentum \linebreak $\bs{\Pi} = \bs{p} - \frac{e}{c} \bs{A}$.
However, the previously proposed idea about gauge invariance now needs a careful scrutiny since the vector potential now appears directly in the Hamiltonian.
For example, consider the case of a uniform magnetic field, namely $\bs{B}=B\hat{z}$.
It can be checked that the commutators of  different components of the mechanical momentum does not vanish,
\beq
[ \Pi_{i} , \Pi_{j} ] = \frac{i \hbar e}{c}\varepsilon_{ijk}B_{k} \nonumber \, .
\eeq

However, this commutator is indeed gauge invariant.
Using the above commutator and the fact that
\beq H = \frac{\bs{\Pi}^{2}}{2m} \nonumber \eeq

It can be straightforwardly derived that the spectrum is given by  so called one dimensional harmonic oscillator like Landau levels
\beq E_{n} = ( n+ \frac{1}{2} ) \hbar \omega_{c}, \quad  \omega_{c} = \frac{eB}{mc} \nonumber \, . \eeq

The energy spectrum is thus a Gauge invariant quantity. However, the gauge invariant form of the energy
and the basic commutation relation not necessarily ensures that relevant physical quantities in quantum mechanics, such as the transition matrix elements two different states under the action of a~given operator are necessarily gauge invariant.

To establish the connection to gauge invariance in classical system one uses Ehrenfest theorem which states that expectation values
of the observables in quantum mechanics behave in the same way like the classical quantities. Therefore
we can expect them to gauge transform in the same way like the classical quantities.
As one can see this is not trivially satisfied, since what appears in the dynamical variable like Hamiltonian
is $\bs{A}$ and not $\bs{B}$. This tells us that under a gauge transformation the operators
indeed gets affected.
To see how the gauge invariance of expectation values can be ensured, let us define a state ket $|\alpha \rangle$
in presence of vector potential $\bs{A}$ and the corresponding state ket $|\alpha'\rangle$ for the same magnetic
field with a different vector potential $\bs{A'} = \bs{A} + \bs{\nabla} \Lambda$.
Our basic requirement for gauge invariance is
\bea  \langle \alpha |\bs{x} | \alpha \rangle & = & \langle \alpha' | \bs{x} | \alpha' \rangle \nonumber \\
\langle \alpha |\bs{p} - \frac{e}{c}\bs{A}  | \alpha \rangle & = & \langle \alpha' | \bs{p} - \frac{e}{c}
\bs{A}'| \alpha' \rangle \eea
apart from the normality of each ket. Now since both kets are normalized  there must be a unitary operator
such that
\beq | \alpha' \rangle = \mathcal{G} |\alpha \rangle \label{U1WF} \eeq

The invariance of position and momentum expectation values then demands
\bea  \mathcal{G}^{\dagger} \bs{x} \mathcal{G} = \bs{x} \nonumber \\
\mathcal{G}^{\dagger} ( \bs{p} - \frac{e}{c} \bs{A}  - \frac{e}{c} \bs{\nabla} \Lambda ) \mathcal{G} & = &
\bs{p} - \frac{e}{c} \bs{A} \label{ginv} \eea

One can immediately see that the unitary operator that does that job is
\beq \mathcal{G} = \exp [\frac{ie}{hc} \Lambda (\bs{r}) ] \label{U1} \eeq

This is actually the generator of $U(1)$ gauge transformation and is same as a phase transformation.
This is also the simplest gauge transformation. What all these tell is that in quantum mechanics
to keep dynamical variables $U(1)$ gauge invariant, the wavefunction needs to acquire an additional phase under a gauge transformation.

\subsection{ Neutral Cold Atoms in Synthetic Abelian Gauge Field: Rotating Ultracold Condensate }\label{RotBose}

Abelian gauge transformation discussed in previous Section \ref{Ab} applies to one of the fundamental interactions in nature, namely the electro-magnetic interaction which occurs
only between charged particles such as electrons.
However, simply going by the behavior of quantum mechanical wave-function under such gauge transformations, one can conclude that
the abelian gauge transformations is mathematically equivalent  to phase transformations.  This naturally raises the question
if such a phase transformation can be induced in a wavefunction by other means even for a charge neutral object such as an ultracold bosonic or fermionic atom
whether that can create artificially a gauge field for the corresponding quantum system even in the absence true electromagnetic
interaction. This question was answered in a profound way by M. V. Berry
in his seminal work \cite{Berry, Shapere} based on a number of other works which already indicated the existence of such different type of gauge fields in a number of physical phenomena, that spans optics \cite{Pancharatnam}, Chemistry \cite{Mead}, Atomic and Molecular Physics \cite{Jackiw} \emph{etc}. Before discussing such general gauge transformations following Berry's argument we  discuss one of the simplest example of realizing abelian gauge field synthetically for ultracold atomic BEC through rotation. The theoretical scheme to be described here, was experimentally realized by a number of experimental group to create vortices in ultracold atoms \cite{rotbec1, rotbec2} such as ENS Group \cite{Dalibard}, MIT group~\cite{Ketterle}
and JILA  group \cite{Cornell}. \textcolor{red}{figure 2 should be mentioned in the main text}
\begin{figure}[H]
\centering
\includegraphics[width=0.65\columnwidth , height=
0.45\columnwidth]{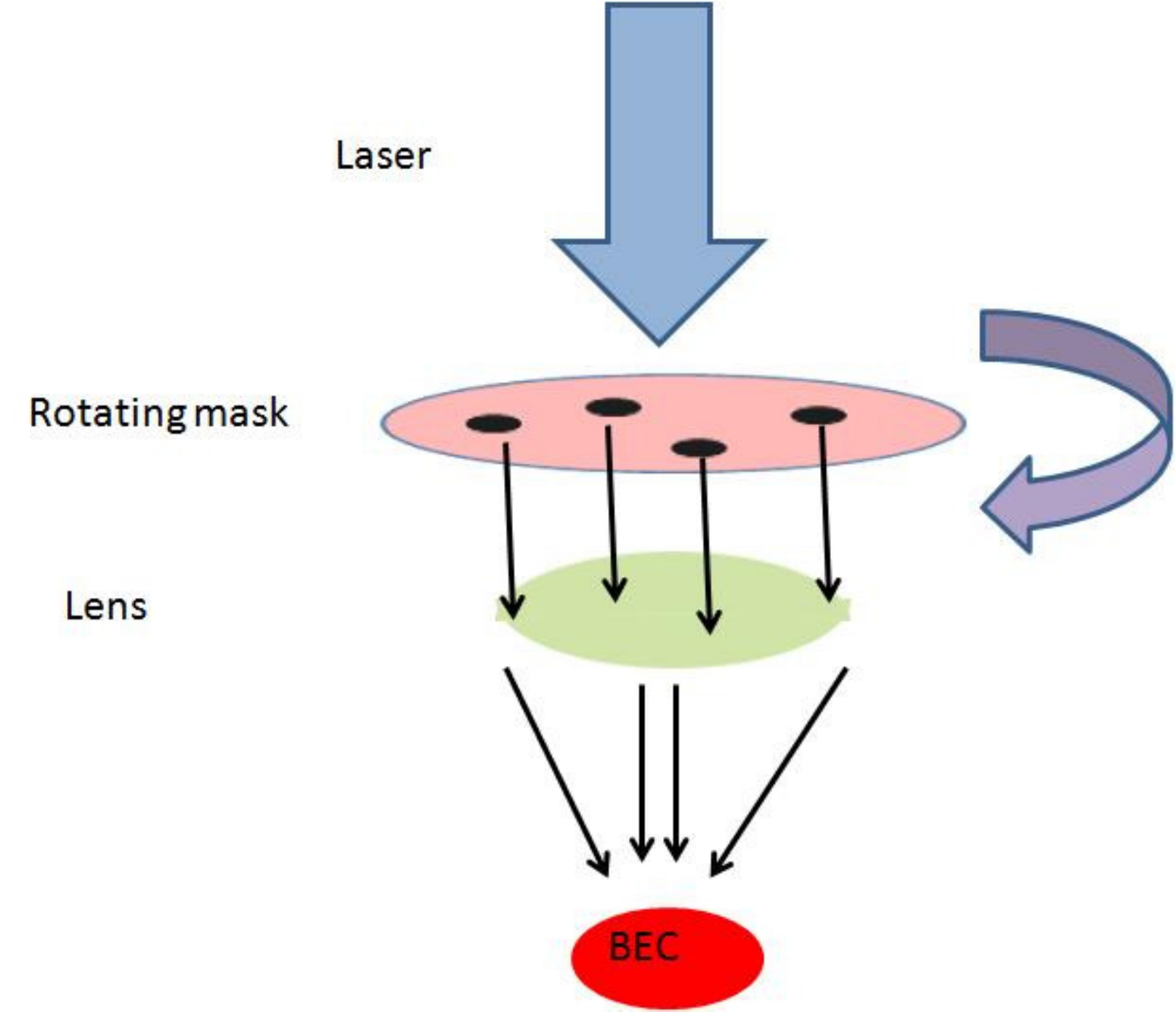}
\caption{A typical configuration for rotated trap which realizes an abelian gauge potential for neutral ultracold atomic condensate. Here the laser induced optical potential imprinted on Bose-Einstein condensate (BEC) is rotated
with the help of a rotating mask. The figure is taken from reference \cite{SGRS}.}
\label{rotation}
\end{figure}
\vspace{-12pt}
To understand how a gauge transformation is similar to the one defined in the previous section, implemented through rotation. Let us consider the
 plane of this rotation is $x$-$y$ plane and the symmetry axis is $\hat{z}$. Now the effect  of such rotation on spatial
co-ordinate is given by
\beq R_{z}(\phi) \begin{bmatrix} x \\ y \end{bmatrix} R_{z}(\phi)^{\dagger} =
\begin{bmatrix} \cos \phi & \sin \phi \\
-\sin \phi & \cos \phi \end{bmatrix}  \begin{bmatrix} x \\ y \end{bmatrix} \nn \eeq

Here $R_{z}( \phi) = exp(-i \frac{\phi \hat{L}_{z}}{\hbar})$ is the rotation operator about $\hat{z}$ axis.
If the rotation is executed at an~uniform angular velocity $\Omega$, then $ \phi = \Omega t $. We can immediately see the connection between $R_{z}(\phi)$  and the $U(1)$ gauge transformation defined in Equation (\ref{U1}). The reason can also be traced to the fact that there is an  equivalence between the Coriolis force in a rotating frame and Lorenz force acting on an electron  in a uniform magnetic field \cite{Fast}. Such a method can be easily implemented experimentally by rotating the trap in which an ultracold condensate is created through a moving laser beam.
The time dependent Hamiltonian that describes a trapped boson in a rotating frame is given
by
\bea H(t) & = & R_{z}(\Omega t) [ \frac{\bs{p}^{2}}{2m} + \frac{1}{2}m (\omega_{x}^{2} x^{2} +
\omega_{y}^{2} y^{2} )] R_{z}^{\dagger} (\Omega t) \nn \\
 & = & \frac{\bs{p}^{2}}{2m} + \frac{m}{2}[(\omega_{x}^{2}(x \cos \Omega t + y \sin \Omega t)^{2} \nn \\
 & + & \omega_{y}^{2} ( -x \sin \Omega t + y \cos \Omega t)^{2} ]  \label{rotham} \eea

The wavefunction for the above Hamiltonian can be obtained by solving the time dependent Schr\"odinger equation (TDSE),
\beq i \hbar \frac{\partial \psi}{\partial t} = H(t) \psi. \label{TDSE} \eeq

To do equilibrium thermodynamics of a systems of such bosons one needs to go to the co-rotating frame from
where the Hamiltonian does not change with time and the rotated frame wavefunction can be obtained from the laboratory frame, with the help of a unitary transformation on the wave function,
\beq \psi ' = R_{z}^{\dagger} (\Omega t) \psi \nn \eeq

As can be straightforwardly seen that this is again equivalent to the gauge transformation given in Equation (\ref{U1WF})
where the $\psi '$ is the wave function in the co-rotating frame.
The transformed  TDSE for $\psi '$ looks like
\beq i \hbar \frac{\partial \psi'}{\partial t} = \left[ \frac{\bs{p}^{2}}{2m} + \frac{m}{2} (\omega_{x}^{2} x^{2} + \omega_{y}^{2} y^{2} )  - \Omega L_{z} \right] \psi'  \label{TDSE1} \eeq

The time independent Hamiltonian on the right hand side can be written
\beq H = \frac{ (\bs{p} - m \bs{A})^{2}}{2m}  + \frac{1}{2}m \left[ \omega_{x}^{2} x^{2} + \omega_{y}^{2} y^{2} - \Omega^{2} r^{2} \right]
\eeq
with the gauge (vector) potential and the gauge field obtained in this way is of the form
\beq \bs{A}  =   - \Omega y \hat{x} + \Omega x \hat{y} ,~\bs{B}  =  2 \Omega \hat{z} \nonumber \eeq

This synthetically created gauge field is similar to the uniform magnetic field and the corresponding synthetic gauge potential is equivalent to symmetric gauge potential for
such  uniform magnetic field.
The transformation also induced a scalar potential
\beq  V_{R} (\bs{r}) = -\frac{1}{2} m \Omega^{2} r^{2} \label{decon} \eeq

Thus, the effective trap potential in the rotating frame gets reduced.
Therefore, in the mean field approximation  the ultracold atomic BEC consists of $N$ interacting bosons near absolute zero temperature in a rotating trap
is described by the gauged (synthetically)  version of time dependent Gross-Pitaevski Equation \eqref{GP} \cite{Stringari} which is

\beq i \hbar \frac{\partial \psi'}{\partial t} = \left( \frac{ (\bs{p} - m \bs{A})^{2}}{2m}  + \frac{1}{2}m [ \omega_{x}^{2} x^{2} + \omega_{y}^{2} y^{2} - \Omega^{2} r^{2}] + g|\psi'|^{2} \right)   \psi' .
\label{RTDSE1}
\eeq

It can be readily verified that the non-linear term is invariant under the action of the unitary operator  $R_{z}(\Omega t)$.
Thus the entire previous discussion on the artificial gauge transformation of single boson Schr\"odinger equation can be applied here for the Gross-Pitaevskii equation also. In the
early days of BEC this was the technique through which vortices and vortex lattice was created in ultracold condensate. \linebreak For detail review on this aspect we refer to
\cite{rotbec1, rotbec2}. An interesting regime is where the rotational frequency $\Omega$
is almost equal to the trap frequency in the transverse plane $\omega_{\perp}$. This means the the trap potential
almost becomes negligible.
Because of the entry of the large number of vortices in the ultracold condensate under this condensation, a number of interesting phases of large
number of vortices appear in this regime. This is the regime of rapidly rotating ultracold gas and have been reviewed  extensively in \cite{rotbec3, rotbec4}.  In the subsequent Section \ref{sec:Berry} we shall now discuss the general theoretical frame work of the occurrence of artificial gauge field of abelian and non-abelian type in diverse quantum mechanical system in a geometrical way.

\subsection{Geometric Phase in Quantum Mechanics and the Related Gauge Fields} \label{sec:Berry}

After considering the case of generating a synthetic abelian gauge field (electro-magnetic field) for ultracold
through rotation, in this subsection we shall briefly describe the theoretical framework following
Berry's argument how such gauge field can be generated geometrically
for a quantum mechanical system.  The details can be found out in many standard book on Quantum Mechanics \linebreak (e.g., \cite{Shankar}).
Let us consider the phase change in a quantum mechanical wavefunction under an
adiabatic change. The adiabatic theorem in quantum mechanics tells us that if the particle Hamiltonian is given by
$H(R(t))$ where $R$ is some external co-ordinate which changes sufficiently slowly (slower than the natural time
scale set by the typical energy spacing in the unperturbed system)
and appears para metrically in $H$,  then the particle will sit in the $n$-th instantaneous eigenstate of $H (R(t))$ at the time $t$ if it started out in the $n$-th eigenstate of $H(R(0))$.
The solution of the time dependent Schr\"odinger equation for this case is
\beq |\psi(t) \rangle = c(t) \exp \left( -\frac{i}{\hbar} \int_{0}^{t} E_{n}(t') dt') |n(t) \rangle \right) \label{Berry1} \eeq
which upon substitution in the time dependent Schr\"odinger equation yields
\beq  \frac{dc(t) }{dt} = -c(t) \langle n(t) | \frac{d}{dt} | n(t) \rangle \nonumber \eeq
with the solution
\beq c(t) = c(0) e^{i \gamma(t)},~~ \gamma(t) = i \int_{0}^{t} \langle n(t') | \frac{d}{dt'} | n(t') \rangle dt'. \nn \eeq

  The important thing here to notice that this phase is arising because the basis state $|n(t) \rangle$ is constantly
changing with time. The instantaneous adiabatic state can therefore be written
as
\beq |n (R) (t) \rangle_{a} = e^{ i \gamma (t)} | n(R(t))  \rangle \nn \eeq
where the subscript  ``$a$''  is used to denote the difference with a time evolved state in the absence
of such phase factor.  The extra phase factor can be rewritten as
\beq \exp \left(- \int_{0}^{t} \langle n(t') | \frac{d}{dt'} | n(t') \rangle dt' \right) = \exp \left( \frac{i}{\hbar}\int_{0}^{t} A^{n}(R) \frac{dR}{dt'}dt' \right) \label{Berry} \eeq
where \beq A^{n}(R) = i \hbar \langle n(R) | \frac{d}{dR} | n(R) \rangle \nn \eeq
is known as the ``Berry curvature'' and plays the role of a vector potential. It can be readily checked that if
the state goes through a phase transformation like
\bea
|n(R) \rangle  &\rightarrow&  \exp ( i \Phi(R) ) | n(R) \rangle = |n'(R) \rangle \nn \\
\text{{Berry Curvature transforms as}}~ A^{n}(R) &\Rightarrow&  A^{n}(R) - \hbar \frac{d \Phi(R)}{dR} \label{Berry2}
\eea

This is the same as the gauge invariance condition that was imposed on the vector potential
in Equation~(\ref{ginv}) in the previous Section \ref{Ab}.  The transformation of the wavefunction defined in Equation~(\ref{Berry2})
is same as the one defined in Equation (\ref{U1WF}) for real electromagnetic field.
Now if the adiabatic parameter $R$ comes back to the same value after a time period $T$ we have
$R(T)=R(0)$ and $H(T)=H(0)$. Under that case the singlevaluedness of the wave function in the parameter (R)
space demands that the line integral of the Berry curvature around the closed loop in the parameter space  must be invariant under such gauge or phase transformation.

Thus the Berry curvature plays the same role as the vector potential due to a real
magnetic field under gauge transformation and its effect on the wave function, namely the integral of the
vector potential around a close loop in the co-ordinate space is gauge invariant as demanded by the single-valuedness of the wavefunction. Also in the full analogy with electromagnetic theory which is a  relativistically invariant theory
and there will be a time like component in the form of a scalar potential  such adiabatic transformation also generates a corresponding scalar potential which has the form
\beq V(R) = \frac{\hbar^{2}}{2m} [ | \frac{d}{dR}| n(R) \rangle|^{2} - \langle \frac{d}{dR} n(R) | n(R) \rangle \langle
n(R) | \frac{d}{dR}| n(R) \rangle ]. \nonumber \eeq

The creation of such ``artificial'' gauge field through rotation discussed in the Section \ref{RotBose}
can also be explained by using the concept of Berry Curvature discussed in section \cite{Berry}. One can recognize here
that the adiabatic parameter is the time dependent rotation angle $\Omega t$ and the Hamiltonian is a function
of this parameter $R=\Omega t$. If the rotational frequency is ramped up adiabatically, then the adiabatic theorem ensures that the system will always stay in the ground state of the rotated Hamiltonian provided the initial system is in the ground state.

The above geometrical way of generalization of Gauge transformation in a quantum mechanical system
can be straight-forwardly extended from the abelian cases to the non-ableian cases if the adiabatic parameter
$\bs{R}$ is a vector having a certain number of components.  This will be discussed in later sections.
The most significant impact of the concept of ``Berry curvature'' or ``Geometric Vector Potential'' is that it  opens the possibility of identifying gauge potential and fields in a wide variety of quantum systems including the system of ultracold atoms.

The general method of creating geometrically induced abelian gauge field was discussed in a number of review articles \cite{dalibardrev, SGRS}.
We shall here briefly refer to the scheme adopted in NIST by the group of I. B. Spielman group which is relevant for the discussion in the  Section \ref{COUCF} where we describe some recent work on cavity optomechanics of ultracold fermionic atoms in such synthetically created abelian gauge field.
In  the scheme developed in NIST \cite{Spielman},
a Landau gauge type artificial vector potential was generated by coupling different hyperfine states of the atoms through Raman lasers, which transfers momentum only along the $\hat{x}$ direction. The coupling can vary spatially and leads to  the effective single atom Hamiltonian
\bea H=H_{1}(k_{x})+[\hbar^{2}(k_{y}^{2}+k_{z}^{2})/2m+V(r)], \nonumber\eea
where \bea H_{1}(k_{x})=\frac{\hbar^{2}\left(k_{x}-\frac{q^{\ast}A_{x}^{\ast}}{\hbar}\right)^{2}}{2m^{\ast}}. \nonumber\eea

Here $A_{x}^{\ast}$ is the engineered vector potential and $q^{*}$ is the fictitious charge.
In this case if the system size is sufficiently large and one considers the bulk of the system,  the Hamiltonian resembles  that
for the charged particle in magnetic field, but with the vector potential in Landau gauge. This Hamiltonian is same as the atomic Hamiltonian we used
in the main paper.

\subsection{Ultracold Atoms in Non Abelian Gauge Field}

\begin{figure}[H]
\centering
\includegraphics[width=0.85\columnwidth , height=
0.65\columnwidth]{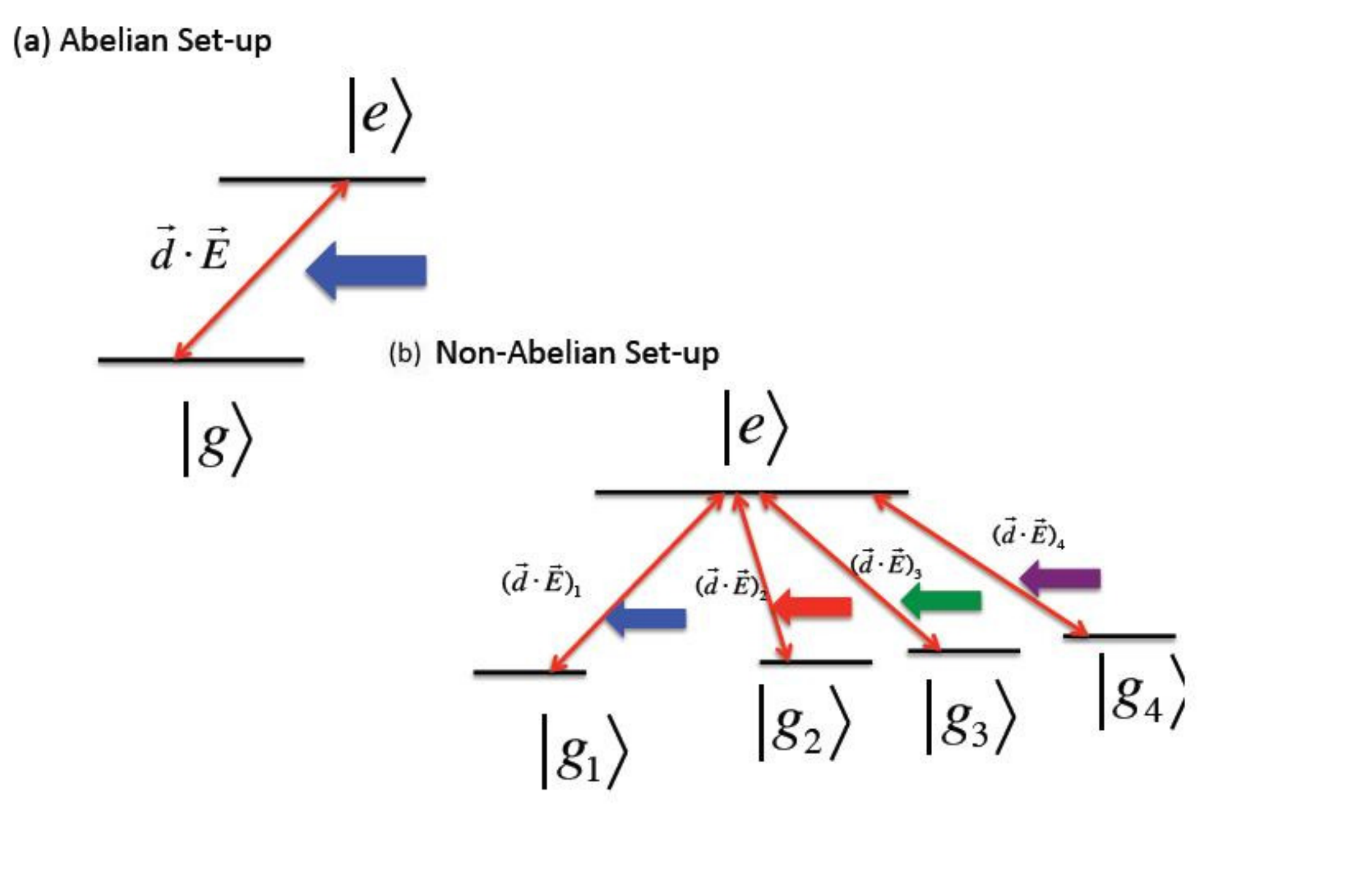}
\caption{Typical atom-laser configuration to create geometrical (\textbf{a}) Abelian and (\textbf{b}) Non Abelian gauge field.}
\label{AtomLaser}
\end{figure}

To understand how an non-abelian gauge field can be introduced in an elementary atomic system
even in the absence of any fundamental non-abelian gauge interaction,
we start by considering the following case.
Consider  a general model of a two level atom with $|g \rangle$ and $|e \rangle$ states
(see Figure \ref{AtomLaser}a)
being respectively its ground state and excited state coupled by a
spatially dependent external field \cite{JCM}.

The general Hamiltonian of  such coupled system can be written as
\beq H_{I} = H_{gg}(\bs{r}) |g \rangle \langle g| + H_{ee}(\bs{r}) | e \rangle \langle e | + H_{ge} |g \rangle \langle e | + H_{eg} |e \rangle \langle g | \eeq
which can be mapped in the spin Hamiltonian
\beq H_{I} = \frac{\hbar \Omega}{2} \bs{n}  \cdot \bs{\sigma}  \nn \eeq
where $\bs{n}$ is a three dimensional unit vector parametrized in terms of polar angle $\theta(\bs{r})$ and azimuthal angle $\phi(\bs{r})$. As one can see the spatial dependence comes from the fact that the coupling between the states is assumed to be spatially dependent since it will depend on electric field of the laser and the atomic wavefunction.  A general state
in this Hilbert space at any point of time can be written as
\beq |\Psi (\bs{r}, t) \rangle = \psi_{\uparrow} (\bs{r}, t) |n_{\uparrow} (\bs{r}) \rangle + \psi_{\downarrow} (\bs{r}, t) |n_{\downarrow} (\bs{r}) \rangle \label{wf}  \eeq
where the basis states  are the local eigenstates of $H_{I}$ at   the spatial point $\bs{r}$
\bea | n_{\uparrow} (\bs{r}) \rangle & = & \begin{bmatrix} \cos \frac{\theta(\bs{r})}{2} \\
\sin \frac{\theta(\bs{r})}{2} e^{i \phi(\bs{r})} \end{bmatrix} \nn \\
|n_{\downarrow} (\bs{r}) \rangle & = & \begin{bmatrix}
-\sin \frac{\theta(\bs{r})}{2} e^{-i \phi(\bs{r})} \\  \cos \frac{\theta(\bs{r})}{2}
\end{bmatrix} ,
\eea
are  referred to as the ``dressed states'' in a quantum optics language or the ``adiabatic states'' in a cold atom language. If the system evolves adiabatically through this space then this means that this local basis of the
Hilbert space is also changing at every point in space. This according to the discussion in Section~\ref{sec:Berry}
will generate Berry curvature as
 \beq \bs{\nabla} (\psi_{i}(\bs{r}) |n_{i}(\bs{r}) \rangle) = \bs{\nabla}\psi_{i}(\bs{r}))|n_{i}(\bs{r}) + \psi_{i} (\bs{r})
|\bs{\nabla} n_{i}(\bs{r})\rangle, i= \uparrow, \downarrow \nn \eeq

\vspace{-6pt}

Now suppose that an initial state the particle is in the state $ | n_{\downarrow} \rangle$ and the motional state is such that it stays in this state all the time ( the transition amplitude to the up-state is negligible). Under this condition we assume $\psi_{\uparrow}=0$ and project the Schr\"odinger equation in the dressed state $|n(\bs{r})_{\downarrow}) \rangle$. This gives us the following ``gauged'' Schrodinger equation for $\psi_{\downarrow}$ \cite{dalibardrev}.
\beq i \hbar \frac{\partial \psi_{1}}{\partial t} = [ \frac{(\bs{P} - \bs{A})^{2}}{2m} + \frac{\hbar \Omega}{2} + V]\psi_{1}
\label{SchAbelian} \eeq

with
\bea \bs{A} (\bs{r}) & = & i \hbar \langle n_{\downarrow} (\bs{r}) |\bs{\nabla} n_{\downarrow} (\bs{r}) \rangle
=-\frac{\hbar}{2} ( \cos \theta -1) \bs{\nabla} \phi \nn \\
\bs{B} (\bs{r}) & = & -\frac{\hbar}{2} \bs{\nabla} \cos \theta  \times \bs{\nabla}\phi \nn \\
 V(\bs{r}) & = & \frac{\hbar^{2}}{2m} |\langle n_{\uparrow}(\bs{r}) | \bs{\nabla} n_{\downarrow} \rangle |^{2}
= \frac{\hbar^{2}}{8m} [ (\bs{\nabla} \theta)^{2} + \sin^2 \theta (\bs{\nabla} \phi)^{2} \nn \eea

\subsubsection{Geometrically Created Non-Abelian Gauge Field}

The above method of inducing gauge field geometrically can be easily generalized for the non-abelian case
\cite{dalibardrev}. Consider now  a $N+1$ state atomic system with $N \ge 3$ with  suitable configuration of laser beams
that induce coupling between these atomic states
A prototype
configuration is displayed in Figure~\ref{AtomLaser}b.
The structure of the the coupling matrix $U(\bs{r})$ will be
\beq
U(\bs{r}) = \begin{bmatrix} \langle 1 | U(\bs{r} ) | 1 \rangle & \langle 1 | U(\bs{r}))  2 \rangle & \cdots &  \langle 1 | U(\bs{r}) | N+1 \rangle \\
\langle 2 |U(\bs{r}) | 1 \rangle & \cdots & \cdots & \langle 2 | U(\bs{r})) | N+1 \rangle  \\

\vdots & \vdots & \vdots & \vdots \\
\langle N+1 | U(\bs{r}) | 1 \rangle & \cdots & \cdots & \langle N+1 | U (\bs{r}) | N+1 \rangle \end{bmatrix} \eeq

For a fixed position $\bs{r}$ the above matrix can be diagonalized to give $N+1$ dressed states $|n_{i}(\bs{r}) \rangle$ with energy eigenvalues $E_{i} (\bs{r})$ where $i$ goes from $1$ to $N+1$. Under certain circumstances
it happens that a~subset $Q$ out of this $N+1$ states are either degenerate or quasi-degenerate  and are well separated from he rest of states energetically. It is under this condition it is possible to realize adiabatic motion
in this low lying degenerate subspace $\mathcal{H}_{Q}$ of dimension $Q$. Assuming that the motional states
are such that there is almost no scattering from this low energy subspace $\mathcal{H}_{Q}$ to $(N+1)$ $-$ $Q$ higher energy state.

Again we can write the full wave function of the
\beq |\Psi \rangle  = \sum_{i=1}^{N+1} \psi_{i} (\bs{r}) | n_{i}(\bs{r}) \rangle \nn \eeq

And then we can project this Schr\"oedinger equation to the reduce Hilbert space $\mathcal{H}_{Q}$ to get an equation for the reduced spinorial wavefunction $\Psi_{Q}= (\psi_{1}, \cdots, \psi_{Q})^{T}$.
We can straightforwardly extent the gauged Schr\"odinger   equation given in  Equation (\ref{SchAbelian}) to its spinorial counterpart, namely
\beq i \hbar \frac{\partial \Psi_{Q}}{\partial t} =
[\frac{(\bs{P} - \bs{A})^{2}}{2m} + \epsilon + V]\Psi_{Q} \eeq
with the important differences that $\bs{A}$ and $V$ are now matrices with their matrix elements given by.
\bea \bs{A}_{i,j}  & = & i\hbar \langle n_{i}(\bs{r}) | \bs{\nabla} n_{j} (\bs{r}) \rangle \nn \\
V_{i,j} & = & \frac{1}{2m} \sum_{l=Q+1}^{N+1} \bs{A} _{il} \cdot \bs{A}_{l,j} \eea

Since different component ( $x,y,z$) these effective vector potentials being matrices will not generally commute
with each other and are therefore called non ableian vector potential.Here $\epsilon$ corresponds to the energy of the unperturbed atomic systems.

The above described atom-light interaction induced  synthetic abelian or non-abelian gauge potential
has been successfully implemented by I. B. Spielman's group \cite{lin1,lin2} in NIST by coupling atomic
states with Raman lasers.
They created synthetic magnetic field, electric field as well as SO coupling in ultracold atomic systems.

%

\subsection{Synthetic Spin-Orbit Coupling for Ultracold Atomic Gases: Case of Non Abelian Gauge Field} \label{sec:SO}

The generation motivation behind creating synthetic spin-orbit coupling for ultracold atoms primarily comes from
the fact that spin orbit coupling plays a very important role in spinotronics \cite{Sinova} and Topological Insulators
\cite{TI} either of which have interesting practical applications. However, more relevant for the current topic for discussion is that
spin-orbit coupling also forms an interesting example for Non abelian gauge potential which we shall describe in the following discussion.
Therefore, before discussing the behavior of such spin-orbit coupled ultracold atoms in a optical cavity we shall discuss how spin-orbit coupling realizes a
particular form non-abelian gauge potential.

In our familiar notation a Non abelian vector potential can be written as
\beq \bs{A} = A_{x} \hat{x} + A_{y} \hat{y} + A_{z} \hat{z} \nn \eeq
where $A_{x}$, $A_{y}$ and $A_{z}$ are now matrices. Field strength for such
Non abelian vector potential given by the expression \cite{NA} can be written as
\beq \bs{B} = \bs{\nabla} \times \bs{A} - \frac{i}{\hbar} \bs{A} \times \bs{A}  \label{NAF} \eeq

The first part of the Expression (\ref{NAF}), is a
generalization of the relation between vector potential and magnetic field for the abelian case, the second
part is only non zero if the gauge potential is Non abelian. For abelian cases, the second part is identically
zero.

In contrast to abelian gauge field, the two non-equivalent Non-abelian gauge potential can lead to the same Non-abelian magnetic field. Following \cite{NA} we shall illustrate this case for the non abelian magnetic field
\beq \bs{B} = 2 \sigma_{z}  \hat{z} = 2 \begin{bmatrix} \hat{z} & 0 \\ 0 & -\hat{z} \end{bmatrix}  \eeq

The above magnetic field is uniform but its direction is opposite for spin-up and spin-down component
of the wavefunction of the particle on which it is applied.

One type of vector potential for such uniform field is a generalization of symmetric gauge vector potential for uniform magnetic field $\bs{B}=B\hat{z}$
\beq \bs{A} = \frac{1}{2} \bs{B} \times \bs{r} = y\sigma_{z}\hat{x} -x \sigma_{z} \hat{y} \eeq

Here the vector potential contributes to the
magnetic field only through the first term (on R.~H.~S.) of the  Expression (\ref{NAF}) for Non abelian
field strength. All component of the vector potential are abelian~matrices.

Another type of vector potential that can also generate the same magnetic field is given by
\beq \bs{A} = -\sigma_{y} \hat{x} + \sigma_{x} \hat{y}. \label{SONA}
\eeq

This is a uniform  non-commuting vector potential (does not depend on local co-ordinate) and hence Non-abelian. The contribution to the field purely comes from
the second term in the Expression \eqref{NAF}. This non abelian gauge potential is also equivalent to spin-orbit coupling.

To see this  let us recall the well known spin-orbit coupling (Thomas term)
which arises due to relativistic correction to the motion of a spin-1/2 electron obeying Schr\"odinger Equation,
namely
\beq H_{SO} = - \frac{e \hbar}{4m_{e}^{2}c^{2}} \bs{\sigma} \cdot (\bs{E} \times \bs{p}) =\frac{e \hbar}{4m_{e}^{2}c^{2}} \bs{p} \cdot ( \bs{E} \times \bs{\sigma})
\label{SO} \eeq

For a uniform electric field along $z$-axis, $\bs{E} \times \bs{\sigma}$ is just the Non abelian uniform vector
potential defined in Equation (\ref{SONA}). Identifying this we can rewrite $H_{SO}  \propto (\bs{p} \cdot \bs{A})$
where $\bs{A}$ corresponds to the vector potential defined in Equation (\ref{SONA}).
The corresponding kinetic energy term of the Hamiltonian with such non-abelian gauge field will take the form,
\beq H_{k} =  \frac{1}{2m}(\bs{p} -m  \bs{A})^{2} \nn \eeq
that describes a free particle in the  presence of Non abelian gauge field Equation (\ref{SONA}). Thus the simulation of such synthetic Non abelian gauge field for ultracold atoms is equivalent to create synthetic spin-orbit (SO) coupling for such systems.
SO coupling plays a crucial role in Spinotronics \cite{Sinova} and Topological Insulator \cite{TI}.
With this background we shall now briefly discuss how such SO coupling is created experimentally for
ultracold BEC.

\subsection{Principle of Spin Orbit Coupling in Ultracold Bosonic Systems: NIST Method}

In the NIST method \cite{lin1} $^{87}$Rb atoms whose ground state electronic structure is $^{2}$S$_{1/2}$ giving  electron spin is $S=1/2$ and nuclear spin is $I=3/2$. Therefore, the total spin $F$
can take value  $F=1$ and $F=2$ due to hyperfine coupling. The low energy manifold therefore consists of
three $F=1$ states characterized by by state vectors $|F,m_{F} \rangle$, and are respectively given as $|1,1 \rangle$, $|1, 0 \rangle$ and $|1, -1 \rangle$. In~presence of Zeeman field these states
have different energy. The resultant system is exposed to \linebreak two counter propagating Raman laser beams
(see {Figure} \ref{fig:schematic})
along the $\hat{x}$ direction. The atom which is moving with velocity $\frac{\hbar k_{x}}{m}$ along
$\hat{x}$ direction will absorb a photon coming from the opposite direction  of Laser I and will have momentum $\hbar( k_{x}-k_{L})$. From this excited state it will emit a a photon in the direction of laser II, making the momentum along $x$-direction will be $\hbar(k_{x} - 2k_{L})$ .
In terms of the hyperfine quantum number $m_{F}$ and the momentum the resultant state can be written as  $|-1, k_{x} - 2k_{L} \rangle$.
Similarly the atoms absorbing photon from laser II and emitting a photon in the direction of laser
I will be finally in the state $|1, k_{x} + 2k_{L} \rangle$.
The final outcome is to have the following three states
\bea |1 \rangle & = & | 1, k_{x} + 2k_{L} \rangle \nn \\
        |0 \rangle  & = & |0, k_{x} \rangle \nn \\
        |-1 \rangle & = &  |-1, k_{x} -2k_{L} \rangle \eea

It is possible to write down the effective Hamiltonian now in $3 \times 3$ matrix form that includes contribution from
atom, field ( laser) and atom-laser interaction. However, tuning the Zeeman energy and the laser frequency  it is possible to restrict
(for details see \cite{zhai}) the system in a two dimensional substance  described by
 the $2 \times 2$ effective Hamiltonian becomes,

\beq H =  \begin{bmatrix}
    \frac{k_{x}^{2}}{2m}+\frac{\delta}{2} & \frac{\Omega}{2}e^{2ik_{L}x}\\
    \frac{\Omega}{2}e^{-2ik_{L}x} &  \frac{k_{x}^{2}}{2m}-\frac{\delta}{2}
  \end{bmatrix} \eeq

Here $2 k_{L}$ is the momentum transfer due to the relative motion between the laser and the hyperfine state of
the atom and $\delta$ is the detuning between the Raman resonance and the energy difference between the spin up and spin-down level. We have also absorbed an overall $\hbar$ factor in various terms. Through a unitary transformation on the two component wave function
defined as $\psi'=U\psi$ with $ U=  \begin{bmatrix}
    e^{-ik_{L}x} & 0\\
    0 & e^{ik_{L}x}\\
  \end{bmatrix} $
the transformed Hamiltonian $ UHU^{\dag}$  now can be written as
\beq  H_{SO} =  \begin{bmatrix}
    \frac{(k_{x}+k_{L})^{2}}{2m}+\frac{\delta}{2} & \frac{\Omega}{2}\\
    \frac{\Omega}{2} &  \frac{(k_{x}-k_{L})^{2}}{2m}-\frac{\delta}{2}\\
\end{bmatrix}  \label{SOH}
\eeq

The above Hamiltonian can be written as
\beq
H_{SO}= \frac{(k_{x}\mathbb{I} + k_{L} \sigma_{z})^{2}}{2m} + \frac{\Omega}{2}\sigma_{x} + \frac{\delta}{2}\sigma_{z}, \label{HSO1}
\eeq
and is the one  realized in NIST Experiment \cite{lin1}.
Even though here the vector potential has only one~component $A_{x}$, since that does not
commute with the scalar potential $\frac{\Omega}{2}\sigma_{x} + \frac{\delta}{2} \sigma_{z}$, this is one of
the simplest realizations of uniform non abelian gauge potential. With a suitable spin rotation it can also be shown
that the first term actually represent and linear combination of equal weight Rashba and Dresselhaus SO coupling.

\section{Cavity Optomechanics of Ultracold Fermions in a Synthetic Gauge Field}\label{COUCF}

In the previous section we reviewed briefly cavity optomechanics with ultracold atoms and ultracold atoms in synthetic gauge field.
As emphasized at the end of  Section \ref{cavopto}, cavity optomechanics with ultracold atoms provides us a
different way of probing the quantum many body states of such ultracold atoms by studying the cavity transmission spectrum. The main issue in this review article is what type of new physics emerges when such
probing technique is employed to study ultracold atoms in presence of a synthetic gauge field. As the following discussion will show the study of cavity transmission spectrum in this case has the potential to detect the cold atom analogue of phenomena like Shubnikov de Haas oscillation which occurs due to the formation of Landau levels of such ultracold atoms in a synthetic magnetic field.

\subsection{Formalism}

Here we briefly discuss how we set up the Hamiltonian that describes the (two-dimensional) system we are interested in. The key ideas from the discussions in the previous sections will be largely employed here, albeit in a many-body setting. The system we consider in this section consists of $N$ ultracold neutral fermionic two-level atoms, each of mass $M$ subjected to a synthetic magnetic field of strength $2\Omega$ and pointing along $\hat{z}$. The atoms are trapped inside a Fabry-P\'erot cavity (on a plane), with mirror area $\mathcal{A}$. The cavity is driven by a pump laser of frequency ${\omega}_p$ and wave vector $\bs{K} = (K_x, K_y)$.  The atoms have transition frequency $\omega_a$, and interact strongly with a single standing wave, the frequency of which is $\omega_c$ when the cavity is empty.

We can study the system Hamiltonian in three components, $ \hh = \hh_A + \hh_C + \hh_I .$ The $\hh_A$ describes the dynamics of the atoms only and the starting point is Jaynes-Cummings Hamiltonian \cite{JCH}. $\hh_C$ describes the dynamics of the cavity photons and $\hh_I$ captures the interaction of the atoms with the photons. Here~we assume the atoms interact with the laser field with a dipole-like interaction. It must be noted that we completely ignore the inter-atomic interaction. This can be achieved through suitable laser configurations (Feshbach resonance method \cite{Feshbach}). Also the interaction between the pump lase and atoms can be ignored if we choose a mode along the z-axis, allowing us to exclude its dynamics on \emph{x}-\emph{y} plane. With all these, starting from a single-particle Hamiltonian we now arrive at a many-body Hamiltonian~\cite{Mekhov1, Mekhov2, Meystre} with the help of the following approximations.

Firstly we want to remove the time-dependence of the Hamiltonian, for this we perform a unitary transformation with $\hat{U}(t) = \exp [i \omega_p t \big ( \ketbra{e}{e} + \ha^\dag \ha \big) ]$. The purpose of this unitary transformation is to describe the system in terms of slowly changing variables by entering into a frame rotating at frequency $\omega_p$. However, since the time scale associated with atomic dynamics is much faster in comparison to the time scale associated with the pump laser frequency all the results obtained in the rotated frame and lab frame stay the same. Since we are working at a very low temperature there is only weak atomic excitations. Also if we set a large detuning (between the pump field and the atom field $\sim$100 GHz), as compared to the life-time linewidth of the atom ($\sim$1 MHz)  then we effectively end up working at \linebreak a time-scale too smaller than that of the transition time-scale. Hence we do not need to include the dynamics of the excited state as all the excited states are already {{ adiabatically eliminated}}. It must be noted that this elimination is performed at the level of the Heisenberg equation of evolution. This~provides us an effective Hamiltonian that described the dynamics of our system but with the inclusion of the above stated approximations.
\bea
\hat{H}_{eff} & = & \hat{H}_A + \hat{H}_I + \hat{H}_C , \label{hfull} \\
\hat{H}_A&= & {\i}{\Pds}\Big{[}{\bf \hat{\Pi}}^2/2M \Big{]}{\Ps},  \label{HL} \\
\hat{H}_I&=  & {\i}{\Pds}\Big{[}\hbar U_0cos^2({\bf K}.{\bf r})\hat{a}^\dagger \hat{a}\Big{]}{\Ps}, \label{HI} \\
\hat{H}_C&=  & {\hbar{\Delta_c}}\hat{a}^{\dagger}\hat{a}-{\imath}{\hbar}{\eta}(\hat{a}-\hat{a}^\dagger). \label{HC}
\eea

Here $U_0 =g_0^2/\Delta_a$ is the effective light-matter coupling constant with $g_0$ is single photon Rabi frequency, $\Delta_{a} = \omega_p-\omega_a$. Here $\hat{H}_A$ is the atomic Hamiltonian in the synthetic magnetic field, $\hat{H}_C$ captures the dynamics of the cavity photons with $\Delta_{c}=\omega_{c} - \omega_{p}  = \Delta \omega_L \approx \kappa$, with $2\kappa$ being the cavity decay line-width. $\eta$ is the coupling between the pump and the cavity.  Here ${\bf \Pi}={\bf p} - M{\bf A}$ is the kinetic momentum, with the effective vector potential $ {\bf A} = {\bf \Omega \times r}$ in symmetric gauge. The eigenstates of this operator are Landau levels (LLs) with effective cyclotron frequency $\omega_0 = 2\Omega$, and eigen-energies $E_{n,m}  = 2\hbar\Omega(n + 1/2)$. The effective magnetic length in this problem is $l_0 = \sqrt{\hbar/2M\Omega}$. Since we are interested in probing the LLs with the help of the cavity photons, we expand the atomic field operator $\Ps$ in the LL basis (with symmetric gauge choice),
\bea
\Ps & = &\sum\limits_{m,n}\c \langle{\bf r}|n,m\rangle
= \sum\limits_{m,n}\frac{e^{-|z|^2/4l_0^2}}{\sqrt{2\pi l_0^2}}G_{m+n,n}(iz/l_0) \c   \label{basis} \, .
\eea
with $z=x+\imath y$. $|n,m\rangle$ is the Landau-eigenket. $\langle{\bf r}|n,m\rangle$ is the symmetric gauge wavefunction. The special function $G_{n+m,m}(z) = (\frac{m!}{(n+m)!})^{1/2}(- \frac{\imath z}{\sqrt {2}})^nL_m^n (\frac{|z|^2}{2}),$ where $L_m^{n}$ is the generalized Laguerre polynomials of degree $m$, detailed properties of this function are discussed in the supplemental information of \cite{ourPRL} . $\cd$ is the fermionic creation operator that creates the state $|n,m \rangle$, namely
a fermion in the $n^{th}$ Landau level (LL), with the guiding center $m$ obeying, $\{\cd,\hat{c}^\dag_{n',m'}\} = \{\c,\hat{c}_{n',m'}\} = 0  \, ,\{\hat{c}^\dag_{n,m},\hat{c}_{n',m'}\} = \delta_{n,n'}\delta_{m,m'} $ ($n=0,1,2...,\nu-1$ and $m = 0,1,2,..., N_\phi -1$).  $\nu= N/N_\phi$ is called the filling factor where $N_\phi=\mathcal{A}/(2\pi l_0^2)$ is the degeneracy of each Landau level.

We want to detect the LLs through the photons leaking out of the cavity (called the cavity transmission spectrum). For this we approach more like a photo-emission experiment, by scattering the cavity photons from the atoms at different LLs. The photons energize the atom to jump to a higher LL (leaving behind an atom-hole), but after a certain time this electron gets de-excited by emitting a photon. These photons carry a signature of the atom they interacted with before. In the absence of atom-photon interaction the ground state of our system will be a direct product state of photonic vacuum and excitonic vacuum, where the later is given by completely filling the first $\nu$ Landau levels of each guiding center, namely, $|GS\rangle = \prod \limits_{m=0}^{N_\phi-1} \prod \limits_{n=0}^{\nu-1} c_{n,m}^{\dagger} |0\rangle$. Since these inter Landau level excitations only involves the change in the LL index, they can be studied using the language of bosonization \cite{Castro} by introducing bosonic operator,
\beq
\hat{b}^\dag_p({\bf k})=\frac{1}{\sqrt{pN_\phi J_p^2(k R_\nu)}}
e^{-(l_0|k|)^2/2} \sum\limits_{n=0}^{\infty} \sum\limits_{m,m'=0}^{\infty} \cdp \c \Big (G_{n+p,n}(l_0 k^*)G_{m',m}(l_0k)\Big ).
\label{boperator}
\eeq

This operator creates a bosonic particle-hole excitation by shifting an atom from $n$-th LL to the $n+p$-th LL, where  $J_p$ is the Bessel function of first kind, $R_\nu=\sqrt {2\nu}l_0$. This being a bosonic operator it obeys the usual bosonic commutation relations. With the help of this operator the bosonized effective Hamiltonian~becomes,
\bea
\hat{H}_{eff} = \hbar \sum\limits_{p=1}^{\infty} \Big \{\sum\limits_{\bf k} p\omega_0 \hat{b}_p^\dagger({\bf k})\hat{b}_p({\bf k}) +  \delta^\nu_p \sqrt{p}\Big (\hat{b}_p^\dagger(2{\bf K}) +\hat{b}_p(2{\bf K})\Big ) \hat{a}^\dagger \hat{a} \Big \} + {\hbar{\Delta}}\hat{a}^{\dagger}\hat{a} -{\imath}{\hbar}{\eta}(\hat{a}-\hat{a}^\dagger) \, , \label{hboson}
 \eea
where the operator $\hat{N}$ is replaced with its steady-state expectation value, subsequently the term $\frac{\hat{N}\hbar U_0}{2}\hat{a}^\dagger a$ is  incorporated into ${\hbar{\Delta_c}}a^{\dagger}a$ of $H_C$ to get the effective cavity detuning $\Delta = \omega_c - \omega_p + \frac{NU_0}{2}$. Notice~the atoms interact with the photons (second last term in the above Hamiltonian) with an~exchange of momentum $\pm 2 |\bs K|$, this is because of two photons moving in the opposite directions. $ \delta^\nu_p = \frac{U_0}{4} \sqrt{N_\phi J_p^2(2K R_\nu)}$ is the atom-photon coupling constant that couples the excited levels with the photon field, where $J_p(x)$ is the Bessel function of first kind. This coupling linearly depends on  atom-photon coupling constant $U_0$ and is enhanced by the Landau level degeneracy ${\sqrt{N_\phi}}$, which is different as compared to the case of ordinary fermions \cite{Meystre} and  akin to the scaling of the atom-photon coupling constant by $\sqrt{N}$ for a N-boson condensate \cite{Eslng}. In Figure \ref{delp_B}, with increasing $\Omega$, the coupling constant oscillates along with jump discontinuities. This is the usual Shubnikov de Hass (SdH) effect occurring for synthetic magnetic field  when the Fermi level makes a jump to the previous level at some increased value of the field.
\begin{figure}[H]
\centering
\includegraphics[width = 10.5 cm, height = 9cm]{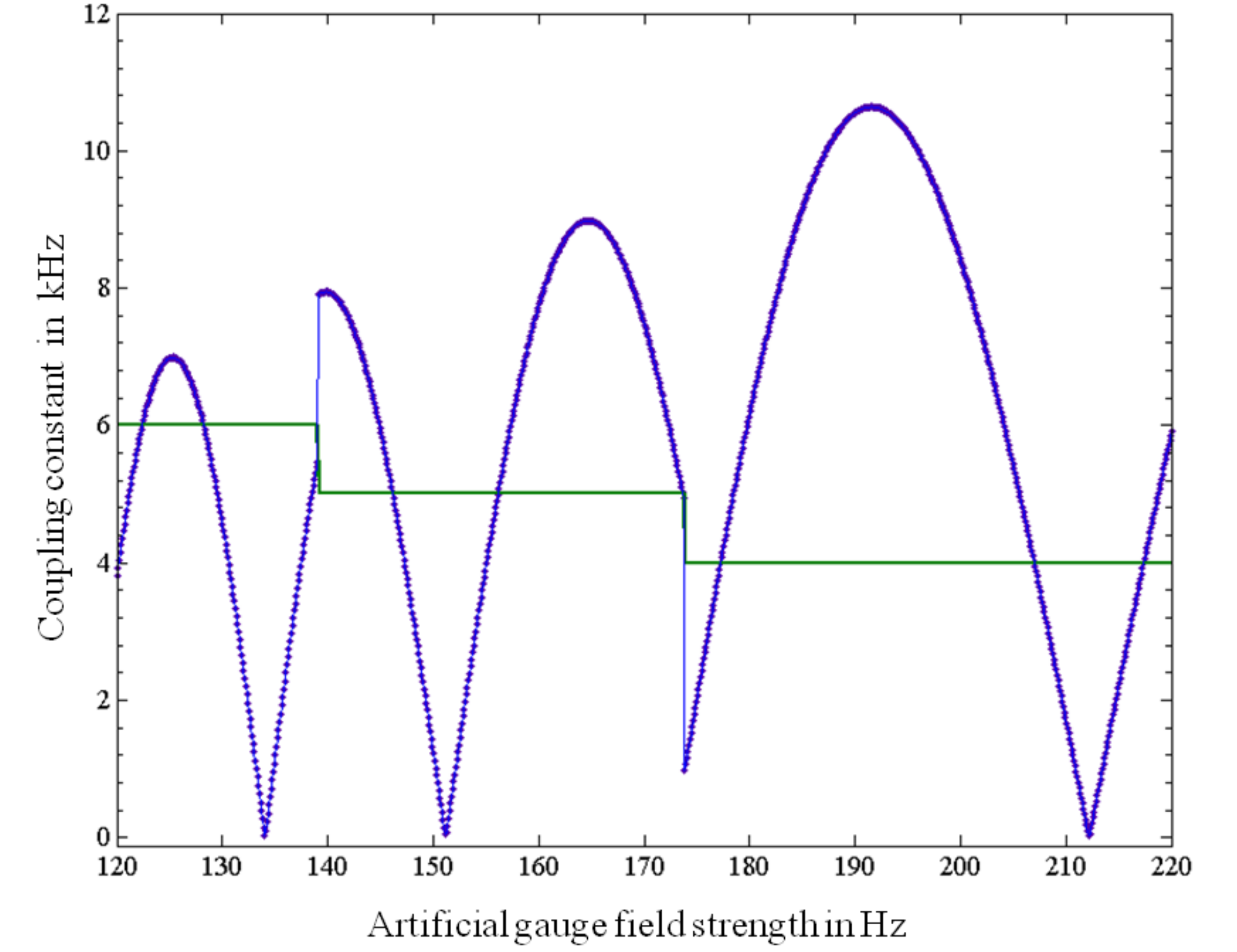}\\
(\textbf{a})\\
\includegraphics[width=10.5 cm, height= 9cm]{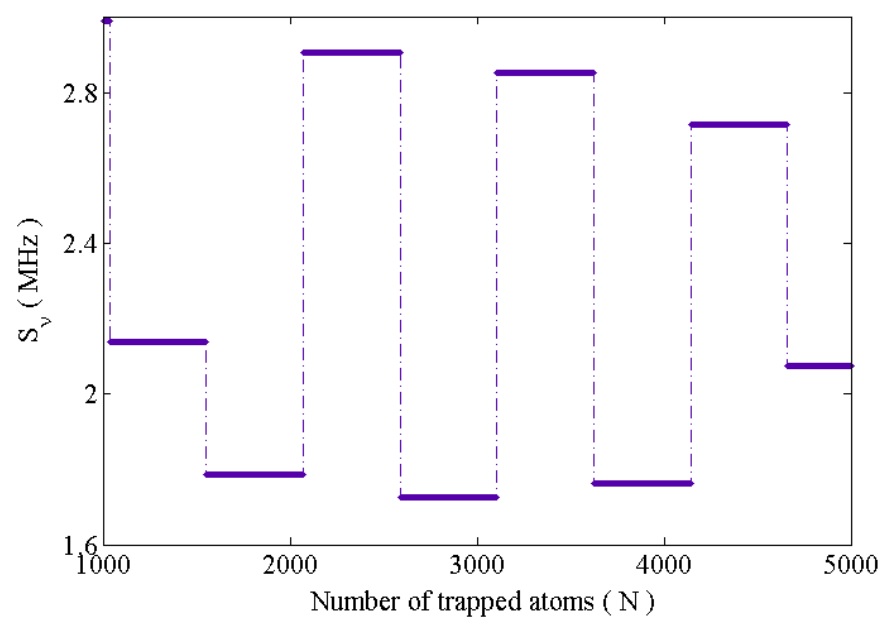}\\
(\textbf{b})

\caption{\small (\textbf{a}) Variation of coupling constant (for \emph{p} = 1) with synthetic field strength. The~green colored steps in the background correspond to the corresponding first empty LL, $\nu$ = {6,5,4}; (\textbf{b})~variation of $S_\nu$ with number of trapped atoms. Figures taken from \cite{ourPRL}. (Reprinted figures with permission from Padhi, B.; Ghosh, S. Phys. Rev. Lett. 2013, 111, 043603. Copyright (2013) by the American Physical Society. Source: http://dx.doi.org/10.1103/PhysRevLett.111.043603)}
\label{delp_B}
\end{figure} 		

\newpage

\subsection{The Shubnikov de Hass Oscillation }

The oscillatory behavior of $\delta^{\nu}_p$ is an early indication of the fact that this cold atomic system is successfully mimicking the physics of LLs and  there is indeed SdH present in this system. A closer look at this reveals this oscillation is simply due to the length scales associated with the current problem.

In presence of synthetic gauge field the cyclotron radius of the ultracold atoms ($l_0 \sim$ 200--800 nm) is comparable with the wavelength of the probing photon ($\lambda \sim$ 600 nm). With increase in field strength the cyclotron radius decreases and the number of wavelengths that fits within this radius also changes,
leading to the oscillatory behavior of the atom-photon coupling strength as a function of the field strength. In comparison, in the corresponding electronic problem the electron cyclotron radius is much smaller ($l_0 \sim$ 20 nm ) so the incident photon can not actually {{see}}  the individual cyclotron orbit, making such oscillation hard to be observed in electronic LL spectroscopy \cite{Spec1}.

Next we proceed to see how the SdH oscillation manifest itself through the cavity spectrum. For that we look at the time-evolution of the cavity photon operators. By setting up the Heisenberg equation one~obtains the steady solution $ \hat{a}^{\dag(s)} ( \hat{a}^{(s)} )$
\beq
\hat{n}_{ph} = \hat{a}^{\dag(s)}\hat{a}^{(s)}=
\frac{\eta^2}{\kappa^2 + (\Delta - S_\nu \hat{n}_{ph})^2}, \quad
S_\nu = \frac{2}{\omega_0}\sum\limits_{p=1}^{\infty} (\delta^\nu_p)^2 = \frac{U_0^2\mathcal{A}M}{32\pi\hbar} \left(1-J_0^2(2KR_\nu) \right)  \, .
\label{Snu}
\eeq

The cavity spectrum is obtained by taking the expectation value of the above equation and it turns out to be, $S_\nu^2n_{ph}^3 - 2 S_\nu \Delta n_{ph}^2 + (\kappa^2+\Delta^2)n_{ph} = \eta^2$. Such non-linear cubic equation is characteristic of optical multistability \citep{Eslng}, which is also apparent from Figure \ref{eta}.

The distance between the two turning points of the bistability curve in Figure \ref{eta}b is calculated to be $ h(S_\nu) = \frac{4(\Delta^2 - 3\kappa^2)^{3/2}}{27 S_\nu}$. This is the quantity which can be measured experimentally and as shown in Figure~\ref{delp_B}b this also carries the signature of the SdH oscillation. The  experimentally measurable cavity spectrum \cite{Eslng, DSK2} can be used to extract the corresponding $h(S_\nu)$, and hence $S_\nu$ which can be compared with the theoretical value  obtained from Equation (\ref{Snu}), provides one the information about the  LL inside the cavity.

\begin{figure}[H]
\centering
\includegraphics[width=8cm]{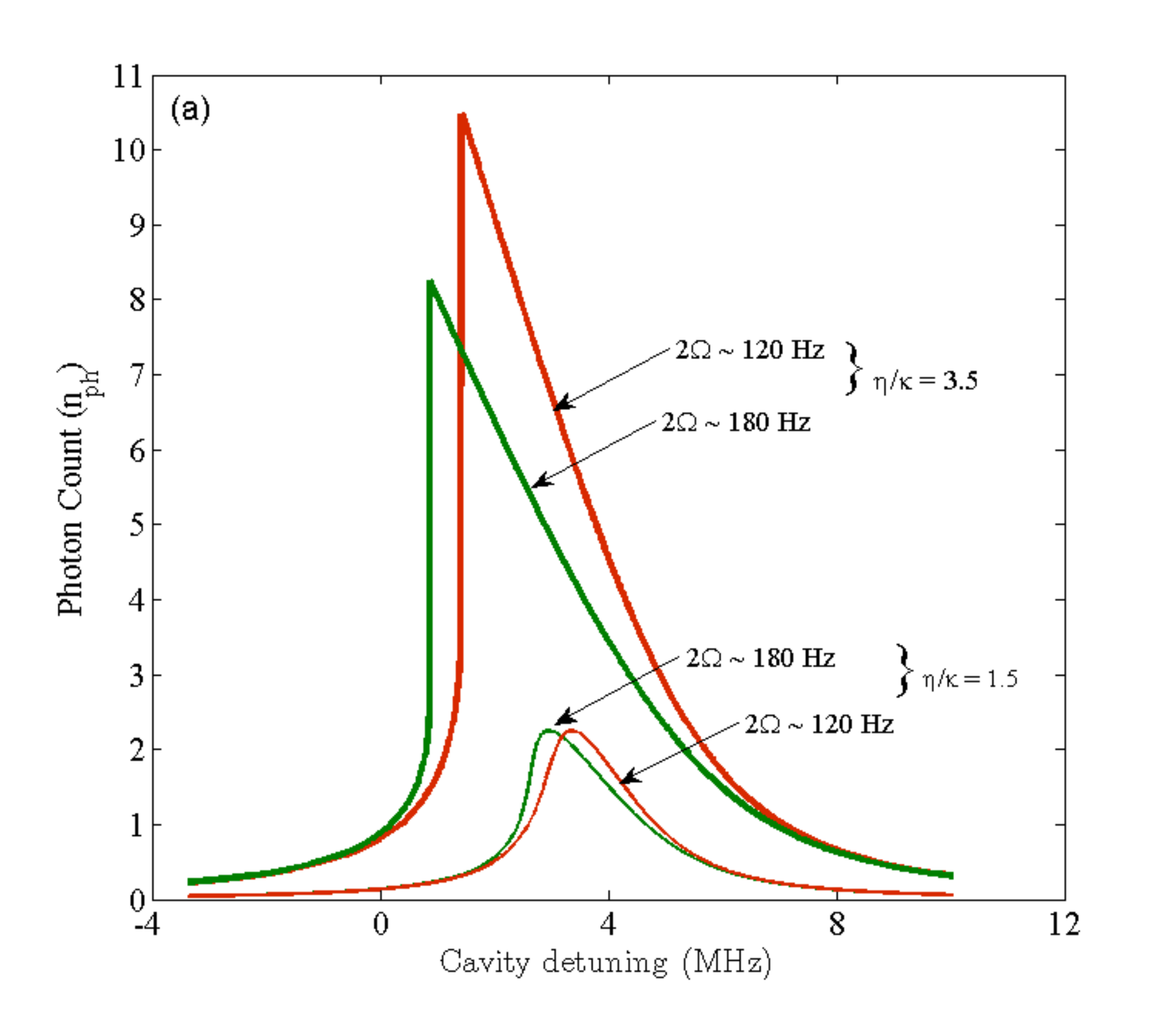}\\
(\textbf{a})
\caption{\textit{Cont.}}
\label{eta}
\end{figure}

\begin{figure}[H]\ContinuedFloat
\centering	
\setcounter{subfigure}{1}
\includegraphics[width=8cm]{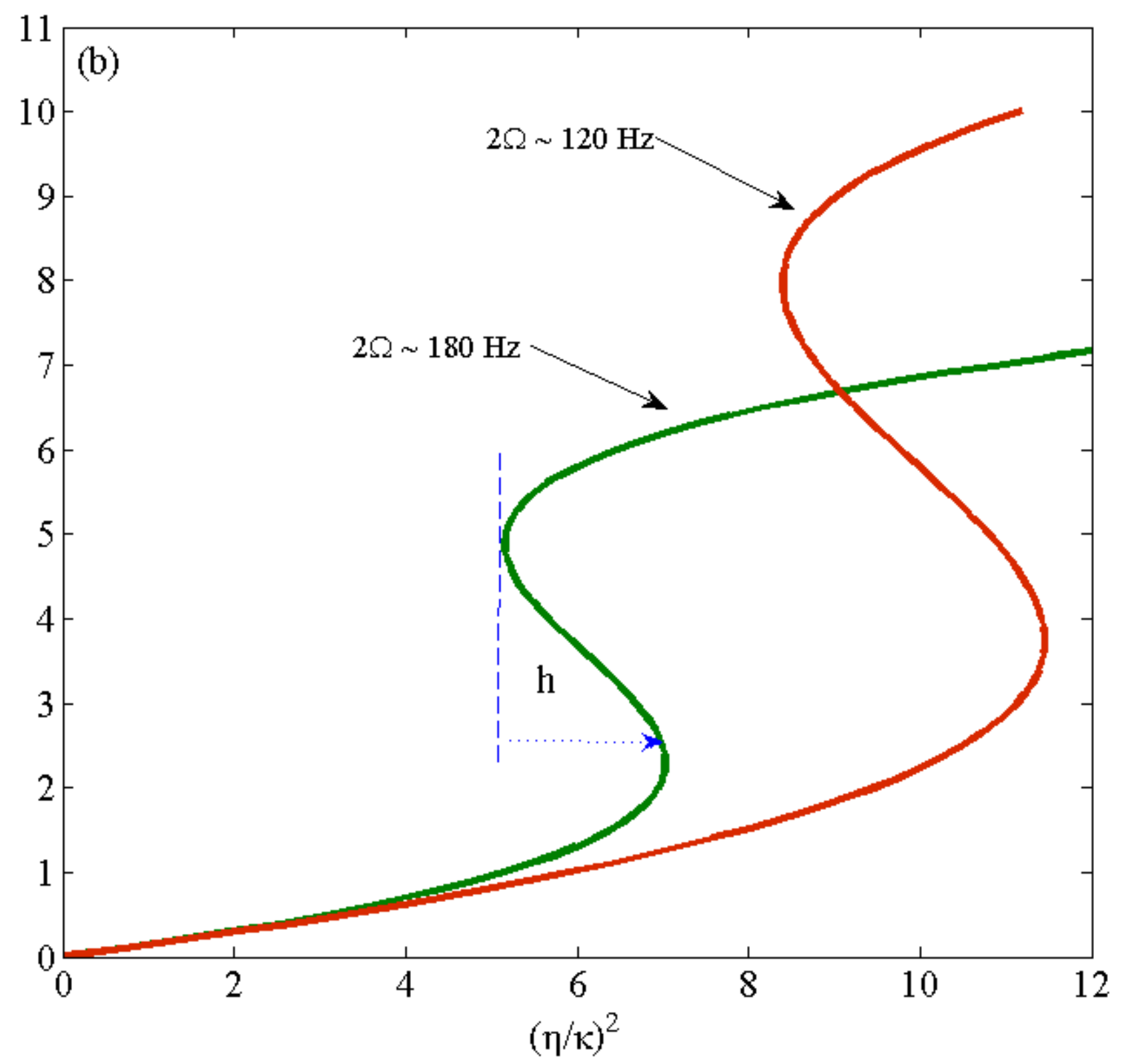}\\
(\textbf{b})
\caption{ Steady-state interactivity photon number as a function of (\textbf{a}) pump cavity detuning for a set of synthetic field and $\eta/\kappa$; (\textbf{b}) pump rate for the same set of synthetic fields and cavity detuning $\Delta = 2\pi \times$ 2.5 MHz. Figure taken from \cite{ourPRL}. (Reprinted figures with permission from Padhi, B.; Ghosh, S. Phys. Rev. Lett. 2013, 111, 043603. Copyright (2013) by the American Physical Society. Source: http://dx.doi.org/10.1103/PhysRevLett.111.043603)}
\label{eta}
\end{figure}

To summarize this section here, we saw that cavity optomechanics could be very useful tool to explore ultracold atomic systems in a synthetic gauge field, providing us a direct access to the atomic Landau levels. We will explore more possibilities in the next section, in a more general situation. As part of further studies on the system discussed here one may try to study the physics of the lowest LLs through the cavity spectrum. One can also include inter-atomic interaction and explore how they affect the~spectrum.

\section{Dynamically Created Spin-Orbit Coupling inside a Cavity}\label{DSOC}

In the previous Section \ref{COUCF} we have discussed the interesting physics that comes to picture when ultra cold fermionic atom in a laser induced synthetic gauge field is coupled to a cavity
field. An alternative way is to dynamically induce a synthetic gauge field for such  ultra cold atomic ensemble using the tricks of cavity quantum electrodynamics. A number of schemes \cite{ring, dong, Deng} is proposed in this direction.

\subsection{Synthetic SO Coupling in Ring-Cavity}

In the scheme proposed by Mivehvar and Feder \cite{ring} a three level atom is coupled with the help of \linebreak two counter-propagating laser modes which form a ring cavity in the well known $\Lambda$ scheme. The resulting system can be described by the atom-photon hamiltonian
\vspace{-6pt}
\bea H^{r}  & = & \frac{\hbar^{2} q_{z}^{2}}{2m} I_{3 \times 3} + E_{a} \sigma_{aa} + E_{b} \sigma_{bb} + E_{e} \sigma_{ee} \nonumber \\
& & \mbox{+} \hbar ( \omega_{1} a_{1}^{\dagger} a_{1} + \omega_{2} a_{2}^{\dagger} a_{2} ) \nonumber \\
& & \hbar(g_{ae}(z) a_{1} \sigma_{ea} + g_{be}(z) a_{2} \sigma_{eb} + H.c.) \nonumber \eea

In the above expression $E_{e} > E_{b} > E_{a}$ are the energy of the three internal states of the atom, with the
condition that $(E_{ea}, E_{eb}) \gg E_{ba}, ( E_{ij}= E_{i}-E_{j}$). $\sigma_{ij}= | i \rangle \langle j |$, and $g_{ae}(z)$ is the coupling strength between the levels $|a\rangle$ and the excited state $|e \rangle$ through the laser $a_{1} \exp(ik_{1}z)$ and $g_{be}(z)$ is the coupling strength between the levels $|b\rangle$ and the excited state $|e \rangle$ through the laser $a_{2} \exp(-ik_{2}z)$. $\hbar q_{z}$ is the center of mass momentum of the atom, multiplied by the $3 \times 3$ identity matrix in the internal space of this atom.

Now under the condition that frequency detuning $\hbar \Delta_{1} = \hbar \omega_{1} - E_{ea}$ and $\hbar \Delta_{2} = \hbar \omega_{2} - E_{eb}$ are very large in comparison to the energy gap $E_{ba}$, the exited state can be eliminated adiabatically (for details see~\cite{ring}). Now after making a unitary transformation to the resulting hamiltonian so that it moves with the resulting momentum transferred to it through the atoms photon interaction, one finally gets an~effective Hamiltonian
\bea H_{eff}^{r} & = & \frac{\hbar^{2}}{2m}[q_{z} - ( k_{1} \sigma_{aa} - k_{2}\sigma_{bb})]^{2} \nonumber \\
                         &  & \mbox{} \frac{1}{2} \hbar \tilde{\omega}_{0} \sigma_{z} +\hbar(\omega_{1}^{\dagger} a{1} +
\omega_{2} a_{2}^{\dagger} a_{2} ) \nonumber \\
& & \mbox{} \hbar \Omega_{R}(a_{2}^{\dagger} a_{1} \sigma_{ba} + \text{h.c.} ) \, . \nonumber
\eea

Here the Rabi frequency $\Omega_{R}= g_{ae}g_{be}(\frac{\Delta_{1} + Delta_{2}}{\Delta_{1} \Delta_{2}})$
and $\hbar\tilde{\omega}_{0}=\tilde{E}_{b} - \tilde{E}_{a}$, where $\tilde{E}$ is the Stark-shifted atomic energy $E$.
The interesting point to note in the above effective hamiltonian is that the last term is similar to the Jaynes-Cummings hamiltonian described in the section, but with $a$ replaced by $a_{2}^{\dagger} a_{1}$ which is a two photon operator.
One can now define the following spin operators for the atomic system
\bea \sigma_{+} & = & \sigma_{ba} = \frac{1}{2}(\sigma_{x} + i \sigma_{y})=\frac{1}{\hbar}s_{+} \nonumber \\
        \sigma_{-} & = & \sigma_{ab}=\frac{1}{2}(\sigma_{x} - i\sigma_{y})=\frac{1}{\hbar}s_{-} \label{atomspin} \eea

Similarly we can define Schwinger angular momentum operators for the photon field
       \bea j_{x} & = & \frac{\hbar}{2}(a_{1}^{\dagger}a_{2}+a_{2}^{\dagger}a_{1})  \nonumber \\
        j_{y} & = & \frac{\hbar}{2i}(a_{1}^{\dagger}a_{2} -a_{2}^{\dagger}a_{1}) \nonumber \\
        j_{z} & = &  \frac{\hbar}{2}(a_{1}^{\dagger}a_{1} -a_{2}^{\dagger}a_{2})  \label{photonam} \eea
\bea \tilde{H}_{eff} & = & \frac{\hbar^{2}}{2m} \{q_{z} I_{2 \times 2} - [\frac{\Delta k}{2}I_{2 \times 2} -k \sigma_{z}]\}^{2} \nonumber \\
& + & \tilde{\omega}_{0} s_{z} + \frac{\hbar}{2} ( \omega_{1} + \omega_{2})N + (\omega_{1} - \omega_{2})j_{z}  \nonumber \\
& + & \frac{\Omega_{R}}{\hbar}( j_{-}s_{+} + j_{+}s_{-}) \label{MFSO}  \eea

Here $k = \frac{(k_{1} + k_{2})}{2}$ and $\Delta k = k_{1} - k_{2}$. The first term in the above hamiltonian Equation
(\ref{MFSO}) represents spin-orbit coupling with equal contribution from the Dresselhaus and Rashba term.
The rest of the terms that represents atom-photon interaction is a generalized Jaynes-Cummings hamiltonian.
 One can define total angular momentum $\bs{J}= \bs{j} + \bs{s}$ combining the photonic angular momentum and atomic pseudo-spin and analyze the spectrum of such hamiltonian. In the work \cite{ring} Mivehver and Feder studied the eigenvalue problem corresponding to the above hamiltonian and particularly discussed in detail superposition states of atomic and photonic excitations known as cavity polaritons. Particularly~diagonalizing the full hamiltonian in the dressed state basis they showed under what condition two photon process can lead to the typical double well like dispersion structure which is a hallmark of the spin-orbit coupling. They have also considered the strong and weak atom-photon coupling regime and described how they can be achieved.

\section{Cavity Mediated Spin-Orbit Coupling}\label{CMSOC}

In an alternative suggestion by Deng {\it et al.} \cite{Deng} synthetic spin-orbit coupling is generated for $\Lambda$-type atoms placed inside a cavity. The electronic ground state of such system can exist in $|\uparrow\rangle$ and $|\downarrow \rangle$ state are respectively denoted as $| g \uparrow \rangle$ and $|g  \rangle \downarrow$ where as the excited state is denoted as $| e \downarrow \rangle$ and the transition frequency between the ground state is given as $\omega_{a}$.  A  Zeeman splitting of magnitude $\hbar \omega_{z}$ is created between $|\uparrow \rangle$ and $| \downarrow \rangle$ state.  The atomic transition $|\downarrow \rangle \rightarrow |e \rangle$ is driven by the pump laser by illuminating the atoms along the $y$-axis by a standing-wave pump laser with frequency $\omega_{L} + \Delta \omega_{L}$ and the atomic transition
$|\uparrow \rangle \rightarrow |e \rangle$ is driven by a plane wave probe laser with frequency $\omega_{L}$.
These lasers are respectively polarized along $x$ and $z$ axis.

In the large atom-pump detuning limit, $|\frac{\Omega_{1,2}}{\Delta}| \ll 1$ and $|\frac{g_{0}}{\Delta}| \ll 1$, the excited state of the atom can be eliminated and the single-particle part of the many body hamiltonian is given by
\cite{Deng}
\beq \hat{\bs{h}} = \frac{\bs{p}^{2}}{2m} \hat{\bs{I}} + \hbar \begin{bmatrix} -\frac{\delta}{2} & \hat{M}_{-}(x,y) \\
\hat{M}^{\dagger}_{-}(x,y) & \frac{\delta}{2} + \hat{M}_{z}(x,y) \end{bmatrix} \eeq

Here
\beq \delta  =    \omega_{z} + \Delta \omega_{L} + \frac{\Omega_{2}^{2}}{\Delta} \nonumber \eeq
is the effective two-photon detuning.
$ \hat{M}_{-} = (-\Omega \cos ky +  \Omega_{c} \hat{a} \cos kx)e^{-iky} $ with $\hat{a}$ being the annihilation operator of the cavity photon. $\Omega =-\frac{\Omega_{1} \Omega_{2}}{\Delta}$ and $\Omega_{c}=\frac{g_{0} \Omega_{2}}{\Delta}$ are the effective Raman coupling strength of the classical and cavity field and thus realizes Spin-Orbit coupling through Raman transitions. The other term
$M_{z} = U_{1} \cos ^{2} ky + \eta(\hat{a} + \hat{a}^{\dagger}) \cos kx \cos ky + U_{0}\hat{a}^{\dagger}\hat{a}\cos{kx}$ realizes a quantum optical lattice with $\eta =-\frac{g_{0} \Omega_{1}}{\Delta}$, $U_{0}=-\frac{g_{0}^{2}}{\Delta}$ and $U_{1}=-\frac{\Omega_{1}^{2}}{\Delta}$.The effective second quantized hamiltonian of such spin-orbit coupled pseudo-spin -$\frac{1}{2}$ system is
\bea \hat{H} & = &\sum_{\sigma, \sigma'} \int d \bs{r} \hat{\psi}^{\dagger}_{\sigma}[\hat{h}_{\sigma, \sigma'} + V_{ext}(\bs{r}) \delta_{\sigma, \sigma'}]\hat{\psi}_{\sigma'}(\bs{r}) \nonumber \\
& & \mbox{} + \frac{1}{2} \sum_{\sigma, \sigma'} \frac{4 \pi \hbar^{2} a _{\sigma \sigma'}}{m} \int
d \bs{r} \hat{\psi}_{\sigma}^{\dagger}(\bs{r})\hat{\psi}_{\sigma'}^{\dagger}(\bs{r}) \hat{\psi}_{\sigma'}(\bs{r})
\hat{\psi}_{\sigma}(\bs{r}) \eea

Here $m$ is the mass of the atom, $\hat{\psi}_{\sigma=\uparrow, \downarrow}$ is the field operator for spin -$\sigma$ atom, $V_{ext}(\bs{r})$ is the spin-independent trapping potential, and $a_{\sigma, \sigma'}$ is the
$s$-wave scattering length between spin $\sigma$ and $\sigma'$~atoms.

For large decay rate $\kappa \gg |\eta|, |\Omega_{c}|$, the cavity field reaches a steady state on a much faster time scale than the time scale of atomic motion in this case one can set $\frac{\partial \hat{a}}{\partial t}=0$
for the Heisenberg equation of motion for the photon field operator
substitute the resultant expression in the Heisenberg equation of motion for the atomic field operator.
The atomic field operator is then replaced by their expectation value $\psi_{\sigma}(\bs{r})= \langle
\hat{\psi}(\bs{r}) \rangle$. This finally yield the spinorial Gross-Pitaevskii like equation
\beq i \hbar \frac{\partial \psi_{\sigma}}{\partial t} =
\sum_{\sigma'} ( \hat{h}_{\sigma \sigma'} + V_{ext} \delta_{\sigma \sigma '} + \frac{4 \pi \hbar^{2} a_{\sigma \sigma'}}{m} \psi_{\sigma'}^{*} \psi_{\sigma})\psi_{\sigma'} \eeq

This equation was solved to explore the quantum phases of the spin-orbit coupled pseudo-spin-$\frac{1}{2}$ bosonic atoms in the $\eta, \Omega$ parameter plane and a number of interesting phases such as checkerboard phase, striped phase was obtained with exotic spin-ordering. Using somewhat similar mechanism in the next
Section \ref{SOC} we shall explain in detail the phase diagram of SO coupled bosonic atoms realized in NIST type experiments placed in side a high finesse optical cavity.

\section{Spin-Orbit Coupled Ultracold Bosons in a Cavity}\label{SOC}

In this section we study a two component BEC interacting with a cavity, in presence of spin-orbit coupling (SOC). We explore how the SOC combined with cavity dynamics give rise to various magnetic phases and how cavity spectrum can be used to detect them.

\subsection{Formalism}

We borrow much of the formalism from the previous section, however, present here with more detail. The assumptions discussed in the previous section for constructing the effective Hamiltonian still hold here. However, there are few significant differences. Due the presence of SOC, along with the $U(1)$ gauge field we have a $SU(2)$ gauge field appearing \cite{Kubasiak} in the canonical momentum, $\hat{\bs \Pi}^2/ 2m =(\bs p + m \bs A)^2/2m$ , where $\bs A = \bs A_{U(1)}+ \bs A_{SU(2)}$ and we choose $\bs A_{SU(2)} = (\alpha \sigma_y, \beta \sigma_x, 0) $ which is a combination of Rashba and Dresselhouse \cite{RD-SOC, RD-SOC2} type spin-orbit coupling. Such a scenario has already been realized in~\cite{lin1}. When $\beta = -\alpha$ the spin orbit coupling is purely of Rashba type. Here $\alpha , \beta$ actually denote the strength  of SOC in the unit of ${\hbar K}/{\pi m}$, $\hat{\sigma}_{x, y,z}$ are $2 \times 2$ spin\-1/2 representation of Pauli matrices.

Another difference in this case is, we have a three level system. For simplicity we assume both the transitions $\ket{2} \leftrightarrow \ket{1}$ and $\ket{3} \leftrightarrow \ket{1}$ have the same coupling with the cavity. In this case also we can perform an adiabatic elimination of the excited state provided the following conditions are respected. \linebreak The excited state vary with a time scale of $1/\gamma$ (atomic line-width) and the ground state and cavity photons evolve with a time scale of $1/\Delta_a$, in this case $\Delta_a$ stands for $ \Delta_{12}^a + \Delta_{13}^a$.  Hence by choosing a large atom-pump detuning, $\Delta^a_{ij} \gg \gamma$ we can adiabatically eliminate the excited states from the dynamics of our system. In other words, the two laser-dressed hyperfine states $\ket{F=1, m_F= 0}$ and $\ket{F=1, m_F= 1}$ of the \textsuperscript{87}Rb atoms are now mapped to a synthetic spin-1/2 system (hence pseudo-spin), with states labeled as $\ket{\upa}$ and $\ket{\dna}$. It must be noted that there exists no real spin-1/2 bosonic systems in nature due to spin-statistics theorem, but with the help of lasers we could realize such a system in ultracold atomic condensate  \cite{lin1}.  With all these machinery we arrive at the effective Hamiltonian of the system
\bea
\mH^{(1)}_{eff} = \int d \bs x \bs \hp^\dag(\bs x) \Big ( \frac{\hat{\bs
\Pi}^2}{2m} + U_{lat}  \Big ) \bs \hp(\bs x) + \hat{H}_c
\nn \\
+ \frac{1}{2} \int d \bs x \sum_{s,s'} U_{s,s'} \dsi{s} \dsi{s'} \si{s'} \si{s} ,
\label{eq:Heff}
\eea


Here $s,s' \in \{ \upa, \dna\}$. For simplification of notations we have defined a column vector $\bs \hp = (\hp_{\upa}, \hp_{\dna})^T$. The atom-atom interaction strength is denoted as $U_{\upa, \upa} = U_{\upa, \upa} = U$ and $U_{\upa, \dna} = U_{\dna , \upa} = \lambda U $. \linebreak Here $U = 4\pi a_s^2  \hbar^2/m$ and $a_s$ is the s-wave scattering amplitude, the parameter $\lambda$ is decided by the laser configuration. One can note the atom-cavity coupling has lead to the formation of an optical lattice~\cite{Mekhov1}, which is $U_{lat} = V_0 [\cos^2(K x) + \cos^2 (K y)]$. Here $V_0$ is the depth of the well, $V_0 = \hbar U_0 \ha^\dag \ha $ and $U_0 \sim 1/ \Delta_a$ is the effective atom-photon coupling strength. Now since the lattice depth has become a (photon number) operator, it is no longer a classical lattice but a quantum lattice. In our calculations we have considered a Nd:Yag (green) laser source of $\lambda = $1064 nm (hence the lattice constant is $a_0 = \lambda /2 =$ 532 nm). The kinetic energy of an atom carrying one unit of photon momentum, $|\bs p| = \hbar K$ describes the characteristic frequency of the center of mass motion of the cloud. Thus the relevant energy scale is $E_r = \hbar^2 K^2/2m$ (recoil energy), in the units of which we measure all other energies involved in the problem. For our case the lattice recoil frequency is $\omega_r = E_r/ \hbar = 12.26$ kHz.

To investigate various interesting phases of this system through the cavity spectrum,  first we establish an equivalence of the effective Hamiltonian obtained in Equation \eqref{eq:Heff} in a cavity induced quantum optical lattice
with a prototype Bose-Hubbard model
in a classical optical lattice. Using  tight binding approximation this is done as follows. By constructing maximally localized eigenfunctions at each site of the lattice we expand each component of the atomic field operator $\hp_s$ in the basis of Wannier functions $ \hp_s (\bs r) = \sum_{i} \hat{b}_{s i} w(\bs r - \bs r_i ),$ where $\hat{b}^\dag_{s i}$ is a bosonic operator that creates an atom in pseudo-spin state $| s \rangle$ ($s = \{ \upa, \dna \}$) at site $i$ of the optical lattice.  However, in presence of a gauge potential the Wannier functions pick up a gauge dependent phase and should be modified as
\beq
w(\bs r - \bs r_i) \rightarrow W(\bs r - \bs r_i) = e^{-i \frac{ m}{\hbar}\int_{\bs r_i}^{\bs r} \bs A(\bs r') \cdot d\bs l}w(\bs r - \bs r_i).
\label{Wann}
\eeq

Under nearest neighbor approximation (\emph{i.e.}, hopping is permitted in between two adjacent sites only), the gauge transformed Wannier function in Equation \eqref{Wann} forms a valid basis for the Hilbert space hence we expand the effective Hamiltonian in Equation \eqref{eq:Heff} in this basis.  Note, however, the following approximation is done in order to write the Wannier function in the above form \cite{Larson}: The Wannier function depends on the atomic density, which in turn depends on the Wannier function as well. So by assuming the thermodynamic limit (by letting the number of atoms and cavity volume go to infinity), yet keeping finite number of atoms on each site one decouple the above mentioned feedback mechanism and write the function in this way.

Unlike the case of the BH model in
a classical optical lattice \cite{Jaksch}, for a lattice generated by quantum
light one should treat the matrix elements of the potential and kinetic energy
separately. It is because of the presence of the term $\ha^\dag \ha$ in the potential term. So the modified BH Hamiltonian becomes
\bea
\mH^{(2)}_{eff} =& E_0 \hN + E_1 \hB + \hbar U_0 \ha^\dag \ha (J_0 \hN +J_1 \hB)
- \hbar \Delta_c \ha^\dag \ha
\nn \\
& - i \hbar \eta (\ha - \ha^\dag) + \frac{1}{2} \sum_{i, s,s'} U_{s,s'}
b^\dag_{is} b^\dag_{is'} b^{\pdag}_{is'} b^{\pdag}_{is} ,
\label{eq:Hubbard}
\eea

{Here,} $E_0$ ($E_1$) and $J_0$ ($J_1$) are the on-site (off-site) elements of $E_{ij}$ and $J_{ij}$, respectively and these are:
\bse
\begin{align}
E_{ij} &= \frac{\hbar^2}{2m} \int d^2r w^*_i(\bs r ) \bs \nabla^2 w_j(\bs r ), \\
J_{ij} &= \int d^2r w^*_i(\bs r )
[\cos^2(Kx) + \cos^2(Ky)] w_j(\bs r ) .
\end{align}
\ese

$\hN = \sum_{s , i} \hat{b}^\dag_{s i} \hat{b}_{s i}$ is the total atom number operator and $\hB = \sum_s \sum_{<i,j>} \hat{b}^\dag_{s i}  e^{-i \phi_{ij}}\hat{b}_{s j} $ is the nearest neighbor hopping operator. Here $\phi_{ij}$ is the phase acquired by an atom while hopping from lattice site $i$ to $j$:
Here is a $2 \times 2$ unit matrix. Because of the dynamical nature of the lattice ( the coefficient
term for the lattice potential involves operators) $E_{ij}$ and $J_{ij}$ are treated separately, otherwise the hopping amplitude would be identified with $t = E_1 + J_1$ and the chemical potential with $\mu = E_0 + J_0$.

Now we do the final job of eliminating the cavity degrees of freedom to arrive at our working Hamiltonian. The interplay of energy scales associated with the spin orbit coupling, motion of atoms in a dynamical lattice and atom-atom interactions brings out a richer and more complex dynamics, as compared to the usual BH model \cite{Jaksch, Mekhov1}, which we try to capture through the light coming out of the cavity.  To facilitate  further discussion on dynamics governed by  Equation \eqref{eq:Hubbard}  we shall do certain simplifications based on the typical
experimental systems. Following typical experimental situation~\cite{Esslinger, CavityExp1, DSK1, CavityExp3} we work under so called "bad cavity limit", where we assume the cavity field reaches its stationary state very quickly than the time scale involved with atomic dynamics. Hence it is reasonable (at least for $t > 1/\kappa $) to replace the light field operators with their steady state values, and thus adiabatically eliminate the cavity degrees of freedom from the Hamiltonian Equation \eqref{eq:Hubbard} so that it depends only on the atomic variables. It will be useful to remember this process is distinct from the adiabatic elimination of the excited state $\ket{1}$. By constructing the time evolution of the photon field operator we arrive at its steady state solution
\beq
\ha^{(s)} = \frac{\eta}{ \kappa + i [ U_0 (J_0 \hN + J_1 \hB) - \Delta_c ]} \approx \frac{\eta}{\kappa - i \Delta_c'} \Big [ 1 - \frac{i U_0
J_1}{\kappa - i \Delta_c'} \hB - \frac{U_0^2 J_1^2}{(\kappa - i \Delta_c')^2}
\hB^2 + ... \Big ]  \,
\label{eq:Photon}
\eeq
where $\Delta_c' = \Delta_c - U_0J_0 N_0$ and $N_0 = \langle \hN \rangle
$ is the total number of atoms. Substituting this in the previous Hamiltonian Equation \eqref{eq:Hubbard} we obtain the effective Hamiltonian, expressed \textit{only} in terms of atomic variables :
\bea
\mH^{(3)}_{eff} = - \mj \hB + \mf \hB^2 + ... + \frac{1}{2} \sum_{i, s,s'}
U_{s,s'} \hb^\dag_{is} \hb^\dag_{is'} \hb^{\pdag}_{is'} \hb^{\pdag}_{is}.
\label{eq:cavity-Hubbard}
\eea
\bse
\begin{align}
\mj / J_1 &= U_0 \eta^2 \frac{\kappa^2 - \Delta_c'^2}{(\kappa^2 + \Delta_c'^2)^2}
- E/ J_1, \\
\mf/ J_1^2 &= 3 U_0^2 \eta^2 \Delta_c' \frac{3\kappa^2 - \Delta_c'^2}{(\kappa^2 + \Delta_c'^2)^4} .
\end{align}
\ese

{The} parameter $\mj$ is the rescaled hopping amplitude, where the scaling
factor is introduced by the cavity parameters and that of atom-photon
interaction strength. Note $\tilde{J}_0$ can be made to vanish by setting $\Delta_c' = \kappa$, and similarly $\tilde{J}_1$ vanishes when $\Delta_c' = \sqrt{3}\kappa
$.

 It is clear from Equation (\ref{eq:cavity-Hubbard}) that cavity-atom coupling induces higher order hoppings feasible through terms like $\hB^{(n)}$.  Also the amplitude of there terms are well controllable through cavity parameters allowing to study higher order atom-atom correlations in these systems. Through suitable choice of cavity parameters, we suppress all higher order terms starting from $\hB^2$. This renders $\mH^{(3)}_{eff}$ to a tight-binding Hamiltonian
\cite{Kittel},  which has incorporated in itself the effects of cavity, abelian and non-abelian gauge field altogether :
\beq
\mH^{(4)}_{eff} = - \mj \hB + \frac{1}{2} \sum_{i, s,s'}
U_{s,s'} \hb^\dag_{is} \hb^\dag_{is'} \hb^{\pdag}_{is'} \hb^{\pdag}_{is}.
\label{eq:TB}
\eeq

 This is our effective Bose Hubbard  Hamiltonian, on which rest of the work is built on. The hopping amplitude is $\mj$. The hopping operator $\hB$ now contains all the information about spin orbit coupling. However, it may be pointed out that
apart from modifying bare hopping amplitude $J_0$ to the rescaled $\mj$, the
cavity also triggers long-range correlations via higher order terms in $\hB$
which we ignored. We'll~discuss this issue later on. In the following subsection we analyze the complete energy spectrum in the non-interacting limit of the effective Hamiltonian in Equation \eqref{eq:TB}.


\subsection{Non-Interacting Limit}

\begin{figure}[H]
\centering
{\includegraphics[width=0.5 \columnwidth,height= 0.35 \columnwidth]{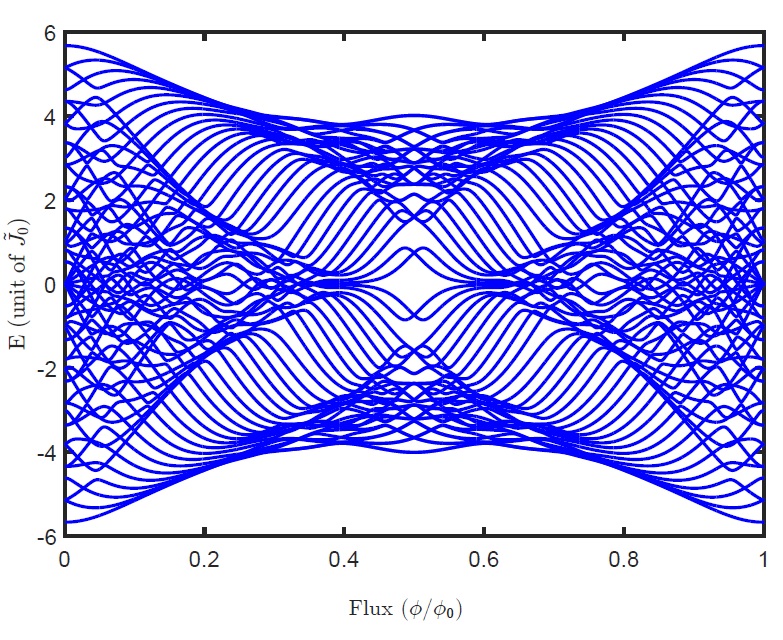}\\
(\textbf{a}) }\\
{\includegraphics[width=0.5 \columnwidth,height = 0.35
\columnwidth]{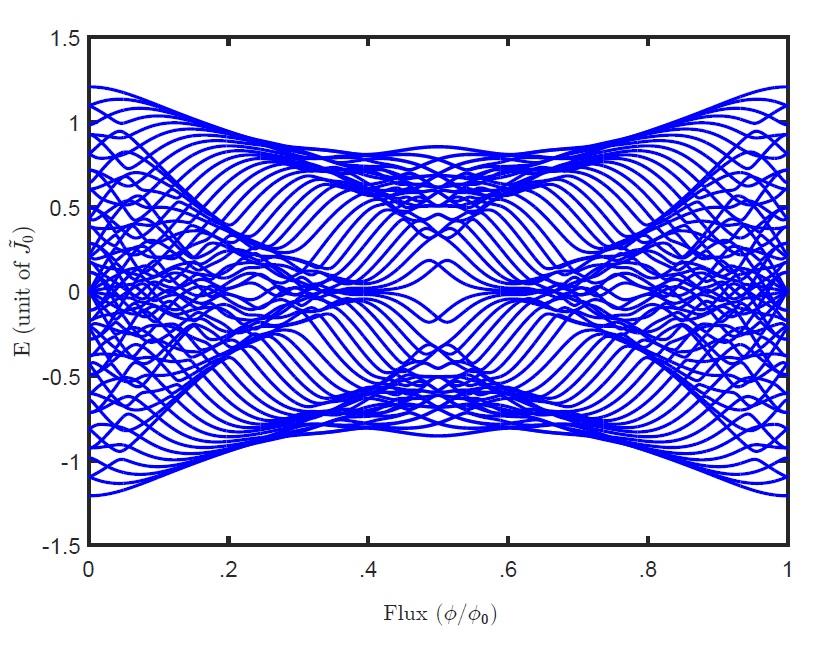}\\
(\textbf{b}) }
\caption{ The spectra obtained in presence of a purely abelian gauge field, for different cavity detunings: $\Delta = -45$ MHz; (b) $\Delta = 45$ MHz. Here we have taken a 8 $\times$ 8 lattice with a filling factor equal to one. }
\label{fig:butterfly}
\end{figure}

\begin{figure}[H]
\centering
{\includegraphics[width=0.45\columnwidth,height=0.35
\columnwidth]{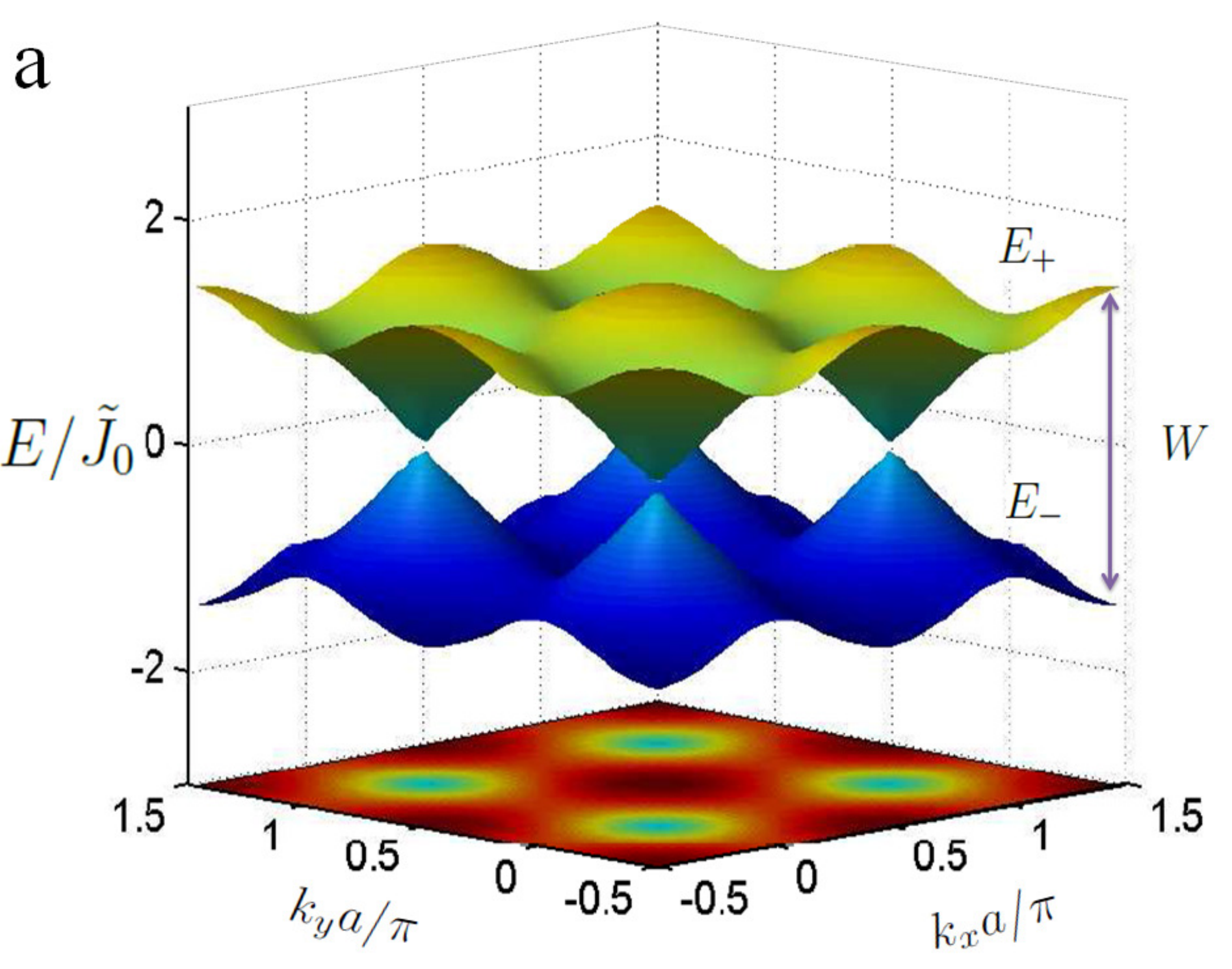} }
{\includegraphics[width=0.45\columnwidth,height=0.35
\columnwidth]{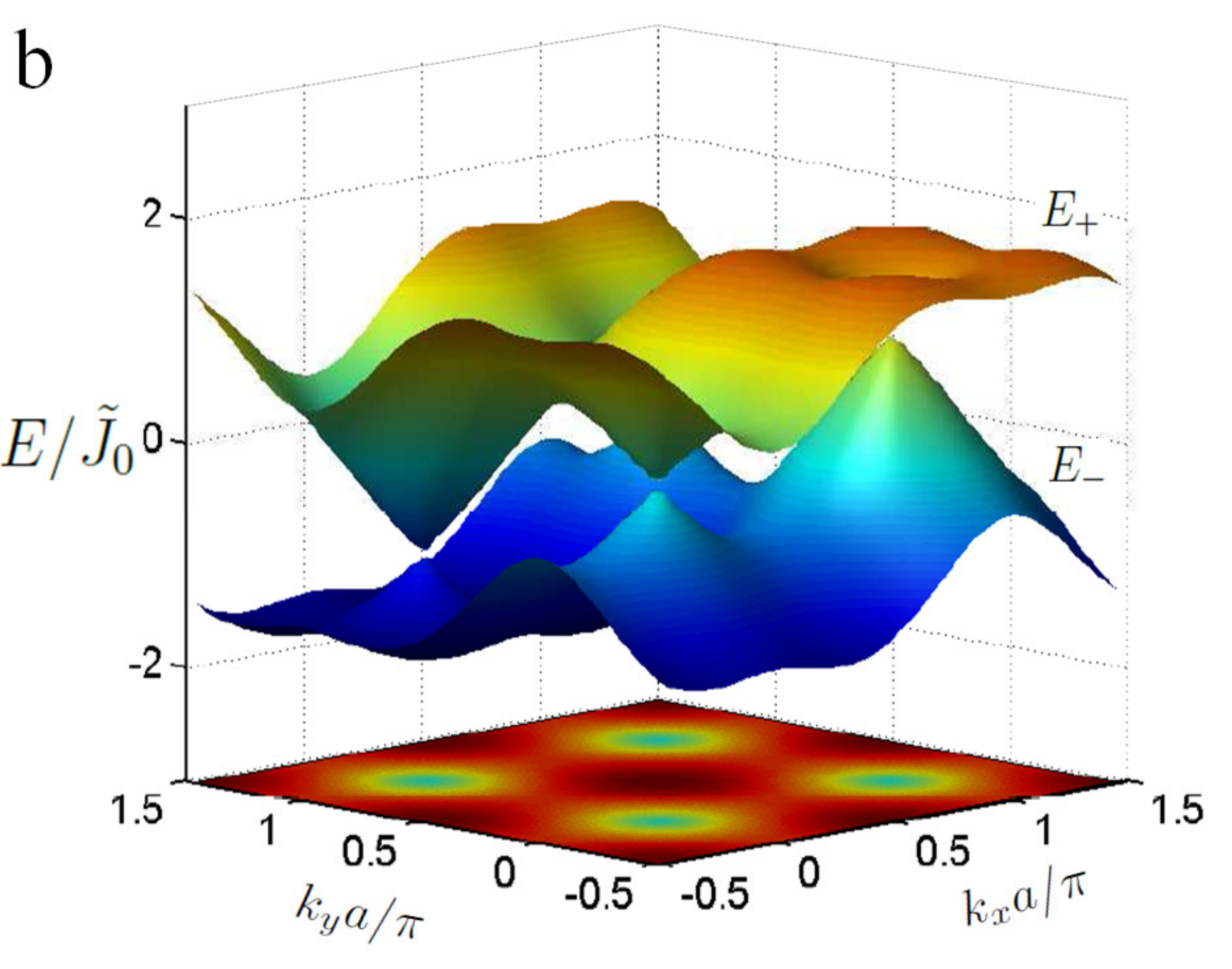} } \\
{\includegraphics[width=0.45\columnwidth,height=0.35
\columnwidth]{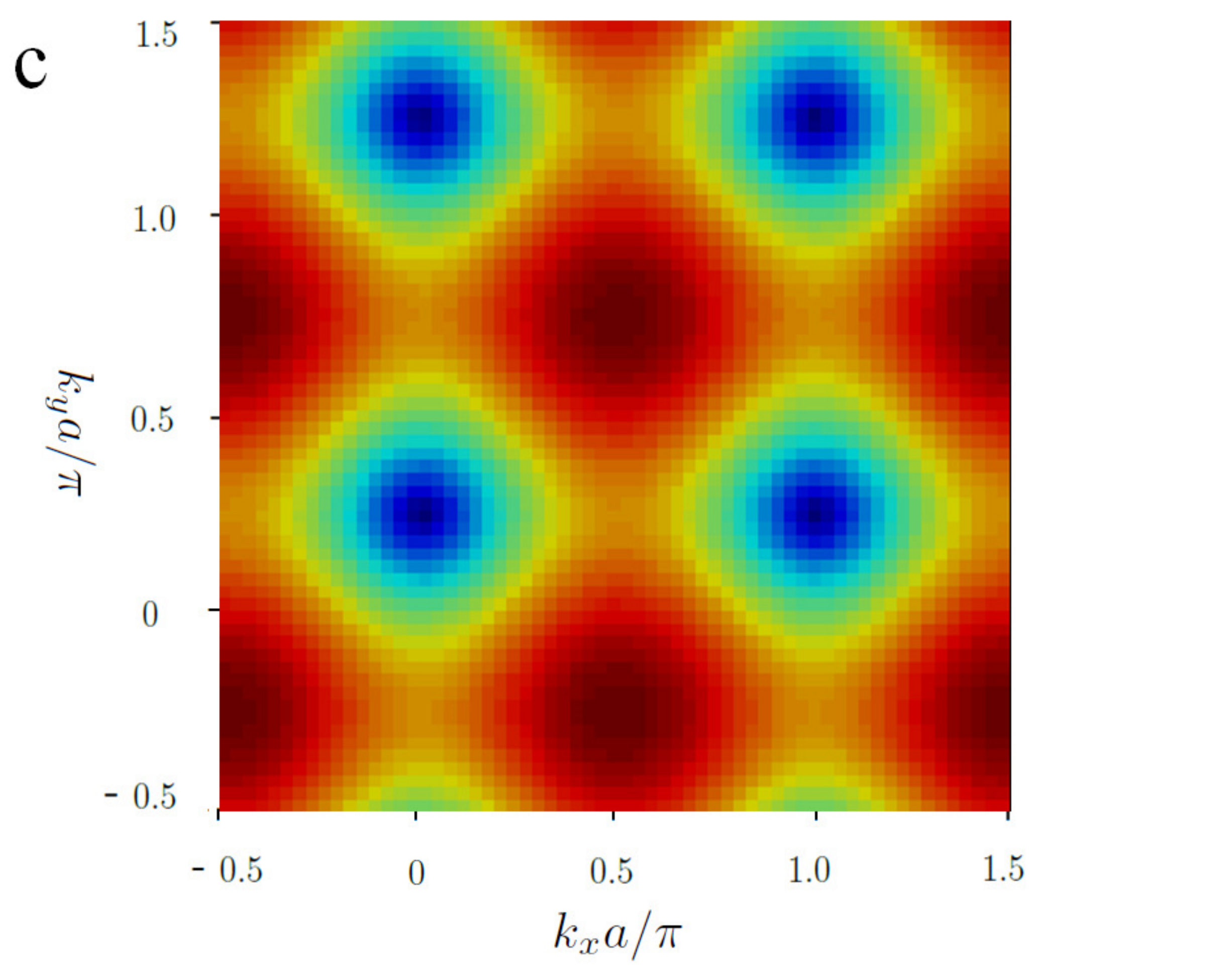}
\label{fig:diracpoints-c} }
{\includegraphics[width=0.45\columnwidth,height=0.35
\columnwidth]{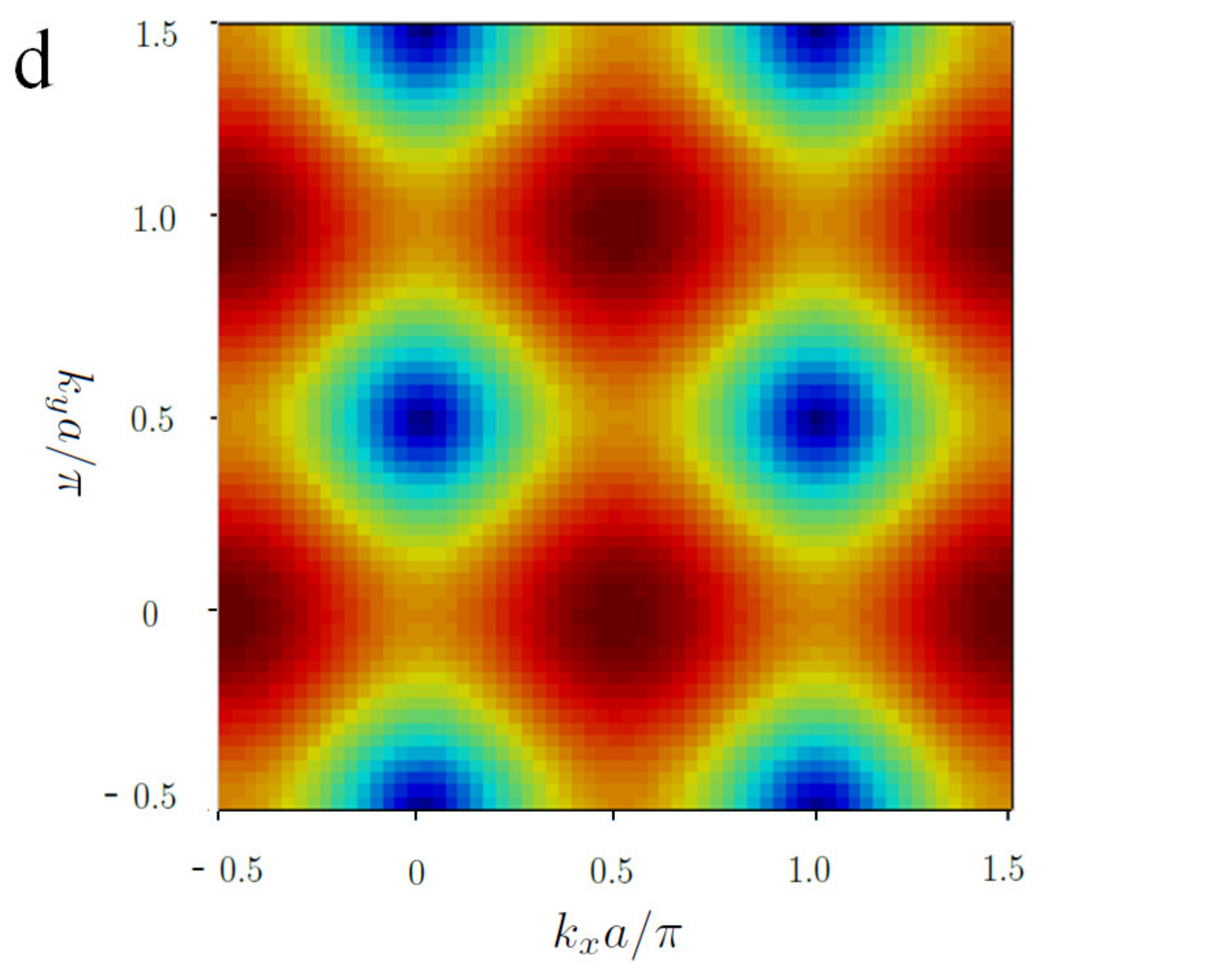}
\label{fig:diracpoints-d} }
\caption{ A three dimensional view of the energy spectrum plotted for a~purely ($\Phi = 0$ non-abelian gauge field. The strength of SOC is (\textbf{a}) $\alpha = \pi/2 = \beta$; (\textbf{b})~$\alpha = \pi/2 + 0.25, \beta = \pi/2 - 0.25$. The
surface plot is an intensity map of the energy difference between $E_+$ and $E_-$. The four green spots on the surface correspond to the four~(Bosonic)~Dirac points (at the zone centers) where the energy gap between
the two~bands~vanishes. The red band and the blue band correspond to $E_+$ and
$E_-$, respectively. $W$ is the maximum band-gap, that occurs at the zone boundaries; In (\textbf{c},\textbf{d}) the location of the Dirac points on the momentum space are shown for $2m\pi \Phi=$ 0.75 and 1.5, respectively. With increasing $\Phi$ the Dirac points move along +ve $k_y$ axis.  From \cite{ourPRA}. (Reprinted figures with permission from Padhi, B.; Ghosh, S.  Phys. Rev. A 2014, 90, 023627. Copyright (2014) by the American Physical Society. Source: http://dx.doi.org/10.1103/PhysRevA.90.023627)}
\label{fig:diracpoints}
\end{figure}

The rescaling of the hopping amplitude by cavity parameters allows a number of
physical properties to be controlled through such parameters. We study
the spectrum of this tight-binding Hamiltonian obtained in Equation \eqref{eq:TB}. We shall show that the resulting system yields
 two interesting spectra namely, the Hofstadter butterfly spectrum, see Figure \ref{fig:butterfly} and the Dirac spectrum, see Figure \ref{fig:diracpoints}.

  The emergence of Hofstadter spectrum is natural as the considered non interacting bosonic system mimics the motion of a Bloch particle (a quantum mechanical particle in a periodic lattice potential) in presence of a uniform $U(1)$ gauge field. The Hofstadter spectrum is revealed when the energy values of the Bloch particle is plotted against the abelian Flux inserted. Such is the case in the absence of Spin Orbit coupling  ($\alpha = 0$) where the Hamiltonian in Equation \eqref{eq:TB} becomes identical with a~Harper Hamiltonian, which can be obtained through Peierl's substitution in the usual tight-binding Hamiltonian~\cite{Hofstadter}. Recently, two groups at the M.I.T and in Munich have experimentally realized such butterfly spectrum in cold atomic systems \cite{Bloch-Ketterle, BK1}. However, compared to those systems, in the present case  one can control (through suitable choice of $\mj$) the energy scale of the butterfly structure just by suitably tuning the cavity parameters. The effects of non-abelian gauge field  on such butterfly structure, was also studied \cite{Kubasiak}.
  
  Next we show how the Dirac spectrum emerges. For this the Hamiltonian in Equation \eqref{eq:TB} is diagonalized and the spectrum obtained is:
\bea
{E_\pm}/{\mj}= 2 \cos \alpha \cos k_x + 2 \cos \beta \cos (k_y - 2m\pi \Phi) \nn \\ \pm \sqrt{\sin^2 \alpha \sin^2k_x + \sin^2 \beta \sin^2 (k_y - 2m\pi \Phi)} ,
\label{diracspec}
\eea
where $(m,n)$ is a lattice point, see Figure \ref{fig:schematic}b. The energy values are plotted against particle momentum and a Dirac like spectrum is obtained in Figure \ref{fig:diracpoints}.

The band-splitting in the spectrum becomes evident as
soon as the effects of SOC is incorporated, showing a band gap ($E_g$) of
$ E_g/ \mj = 4\sin \alpha \sqrt{\sin^2k_x + \sin^2(k_y -2m\pi\Phi)} $, where the gap can be tuned by the cavity as well (through $\mj$). Also in the first Brillouin zone the band gap is maximum when $(k_x,k_y) \in \{ (\pm \pi/2 , \pm \pi/2) \}$ and $E_g^{max} /\mj= 4\sqrt{2} \sin \alpha \equiv W$. It is possible to carry out a
bandgap measurement in such systems through Bragg spectroscopy \cite{DiracExpt, DE2},
through which one can measure the non-abelian flux inserted in the system. However, the gap vanishes when both $\sin k_y = \sin k_x = 0$. In the first Brillouin Zone (by setting $\Phi = 0$) this can happen for $(k_x , k_y ) \in \{ (0,0) , (\pm \pi , 0) , (0 , \pm \pi) , (\pm \pi , \pm \pi) \} \equiv \bs k_D$. In the vicinity of these points the effective low energy behavior can be described by a Dirac like Hamiltonian,
\beq
\hat{H}_{eff} = - \sum_{\bs p} \hp_{\bs p}^\dag \hat{H}_D \hp_{\bs p}  ,
\hat{H}_D = c_x \gamma_x p_x + c_y \gamma_y p_y.
\label{eq:Dirac}
\eeq

Here, $\hat{H}_D $ is a Dirac Hamiltonian, $\bs p = \bs k - \bs k_D$, but the field operators $\hp_{\bs p}$ are
bosonic annihilation operators. The
gamma matrices $\gamma_0 = \bs 1, \gamma_1 = \gamma_x = \sigma_y, \gamma_2 = \gamma_y = \sigma_x$ are the $2+1$ dimension representation of Clifford algebra, $\{ \gamma_i, \gamma_j\} = 2\delta_{ij}$. The speeds of light $c_x = 2\sin \alpha, c_y=2\sin \beta$ are now anisotropic. As shown in the Figure \ref{fig:diracpoints}, through this anisotropy the SOC strength can be used as a handle to controlling the shape of the Dirac cones. We refer the ``Dirac-like'' points $\bs k_D$ in our bosonic system also as Dirac points. Near $\bs k_D$ the excitation quasi particles are mass-less bosons having a dispersion relation linear in $\bs k$, the slope of which is controlled by adjusting the spin-orbit coupling strength.

\begin{figure}[H]
\centering
{\includegraphics[width=0.45\columnwidth,height=0.3
\columnwidth]{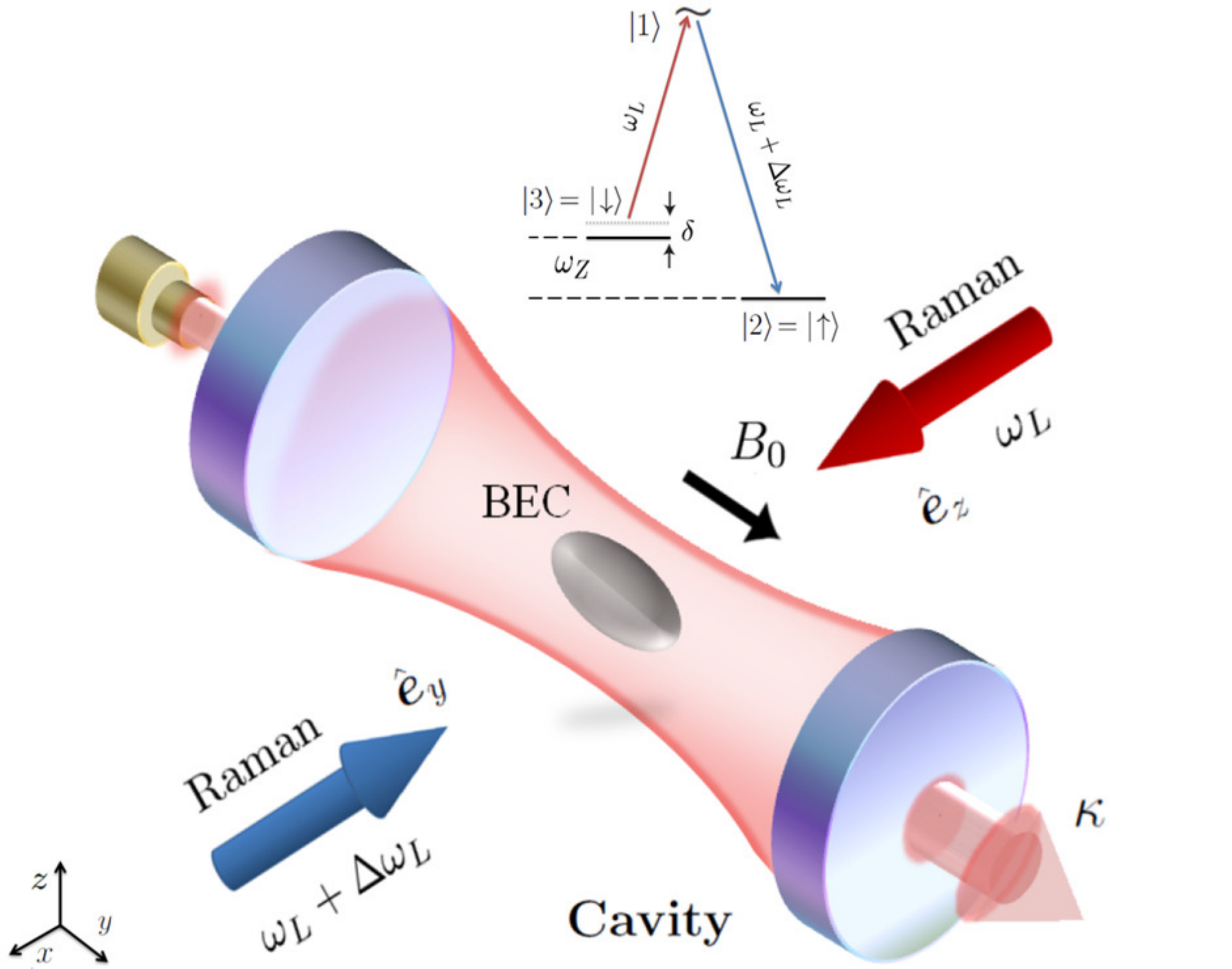}  \label{fig:schematic}\\
(\textbf{a})}\\
{\includegraphics[width=0.5\columnwidth,height=0.4
\columnwidth]{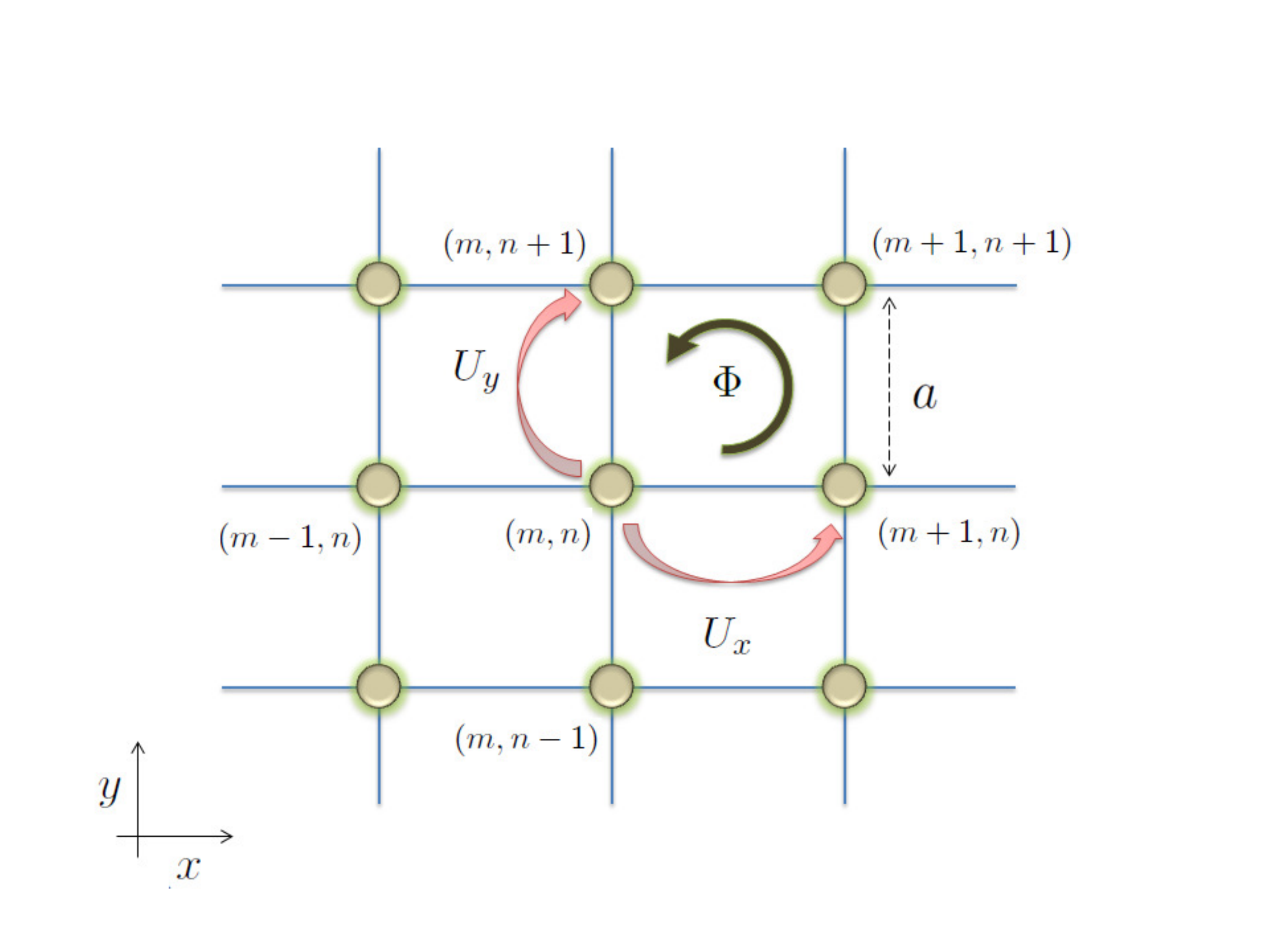} \label{Lattice}\\
(\textbf{b})}
\caption{ (\textbf{a}) \textsuperscript{87}Rb BEC inside an optical cavity: spin-orbit coupling (SOC) is created by
two~counter-propagating Raman lasers with frequencies $\omega_L$ and $\omega_L + \Delta \omega_L$ that are applied along $\hat{x}$. The Raman beams are polarized along $\hat{z}$ and $\hat{y}$ (gravity is along -$\hat{z}$). A bias field $B_0$ is applied along $ \hat{y}$ to generate the Zeeman shift. (Inset) Level diagram of the \textsuperscript{87}Rb atom. Internal states are denoted as $\ket{1}, \ket{2}, \ket{3}$. The coupling of these states is shown schematically; (\textbf{b}) Schematic of an optical lattice. The phase operator $U_x$ determines
the phase acquired by an atom when it hops from site $(m,n)$ to the site
$(m+1,n)$. Similarly,~the operator $U_y$ determines the phase acquired by
hopping along the positive \emph{y}-axis. The~operators $U_x^\dag$ and $U_y^\dag$
determine the phase acquired in hopping along negative \emph{x} and \emph{y} axes,
respectively. Figure taken from \cite{ourPRA}. (Reprinted figures with permission from Padhi, B.; Ghosh, S.  Phys. Rev. A 2014, 90, 023627. Copyright (2014) by the American Physical Society. Source: http://dx.doi.org/10.1103/PhysRevA.90.023627)}
\label{fig:schematic}
\end{figure}


It must be emphasized that such massless bosonic quasiparticles which mimic the massless Dirac fermions
in relevant fermionic systems \cite{DiracExpt, DE2} arise in this system  as a consequence of the spin-1/2 nature of the bosons. Such spin-1/2 bosons have no natural analogue because of Pauli's spin-statistics theorem. However, this constraint can be lifted by synthetic symmetries \cite{Leggett} and synthetic bosonic (pseudo) spin-half system can be realized. After the preliminary proposals on simulation of Dirac fermions in cold atom system \cite{Duan-Dirac} they were soon realized experimentally \cite{DiracExpt, DE2}, using density profile measurement methods or Bragg spectroscopy. Similar techniques may also be exploited to observe the bosonic quasiparticles that follows massless Dirac equation. In the above derivation of Dirac like hamiltonina we have not considered inherent atom-field nonlinearities that can lead to the formation of loop like structures in the energy band \cite{Dell}. An exploration of this issue will be very interesting, but is kept out of further discussion in this article.

As evident from Equation \eqref{diracspec}, the effect of an abelian field would be to move these points on the momentum space (see Figure \ref{fig:diracpoints}c,d. With finite abelian field there also emerges a Hofstadter spectrum as discussed previously. This can be verified by plotting the energy as a function
of the  abelian (magnetic) flux \cite{Kubasiak}.  For the same system here in   Figure \ref{fig:diracpoints}c,d and  we plotted the energy  against the Bloch momentum for a given value
of the abelian flux to show the location of the Dirac points. From~the Equation \eqref{diracspec} it is also suggestive that with the use of a spatially modulated abelian flux one may control the separation between the Dirac points. Motion and merging of Dirac points has also been very interesting as they lead to topological phase transitions \cite{Dirac-Phases, DP2, DP3}. One can also switch on the interaction and study its effects on the spectrum \cite{Interaction-Dirac, Wang1}.

\subsection{Interaction and Magnetic Order}

The primary effect of turning on inter-particle interaction is the onset of various magnetic orders in the ground state of the Hamiltonian. This can be shown by mapping the Hamiltonian in Equation \eqref{eq:TB} to an effective spin Hamiltonian ( one treats the interaction part of Equation \eqref{eq:TB} as the zeroth-order Hamiltonian and then the hopping part ($\mj \hB$) is treated perturbatively to get the effective spin Hamiltonian matrix elements). Readers interested in details may look into \cite{Altman, Kuklov, Kuklov2, Lukin}. Using such analysis the effective spin Hamiltonian have been studied in cold atomic systems, in presence \cite{Trivedi, Sengupta, Radic, Cai} or absence~\cite{Altman, Kuklov, Kuklov2, Lukin} of SOC. We realize that the mathematical structure of our effective mBHM Hamiltonian in Equation \eqref{eq:TB} is the same to the one considered in \cite{Trivedi, Sengupta, Radic, Cai}, provided we switch off the abelian field part. However, since we have considered a cavity induced quantum optical lattice, instead of the hopping amplitude  $t$  in a classical optical lattice, which was the case studied in those works, here we have a~rescaled hopping parameter $\mj$, which essentially captures the information of the quantum light. Thus in the parent Hamiltonian of  references \cite{Trivedi, Sengupta, Radic, Cai}, if we substitute $\mj$ in place of $t$ we arrive at the same conclusion. In fact, since $\mj$ can be controlled by means of the cavity parameters thus one can also maneuver the entire phase diagram by suitably adjusting these parameters.

The effective spin Hamiltonian turns out to be a combination of
two-dimensional Heisenberg exchange interactions, anisotropy
interactions, and Dzyaloshinskii-Moriya interactions \cite{DMInteraction, Moriya}. These~terms collectively stabilize the following orders \cite{Radic}: ising ferromagnets (zFM), antiferromagnets (zAFM), Stripe phase, Spiral phase (commensurate with 3-sites or 4-sites periodicity, respectively denoted as 3-Spiral and 4-Spiral), and the vortex phase (VX). Detailed discussion of these phases can be found in \cite{Radic}; for completeness we provide a brief description of each of these phases below.

\begin{figure}[H]
\centering
{ \includegraphics[width=0.45\textwidth, height= 0.35\textwidth]{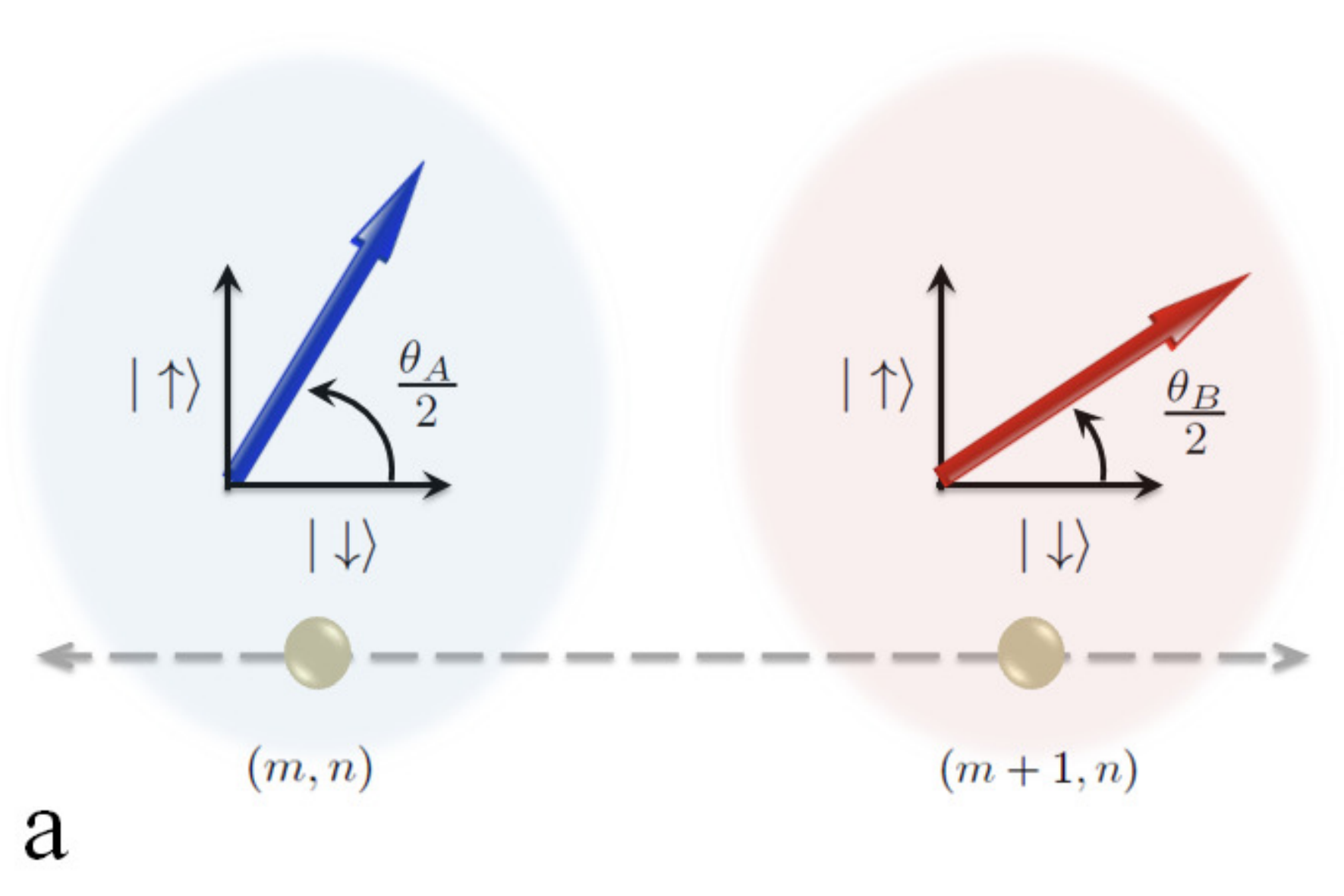}
\label{fig:CavSpecRII-a}}
{ \includegraphics[width=0.45\textwidth,
height=0.35\textwidth]{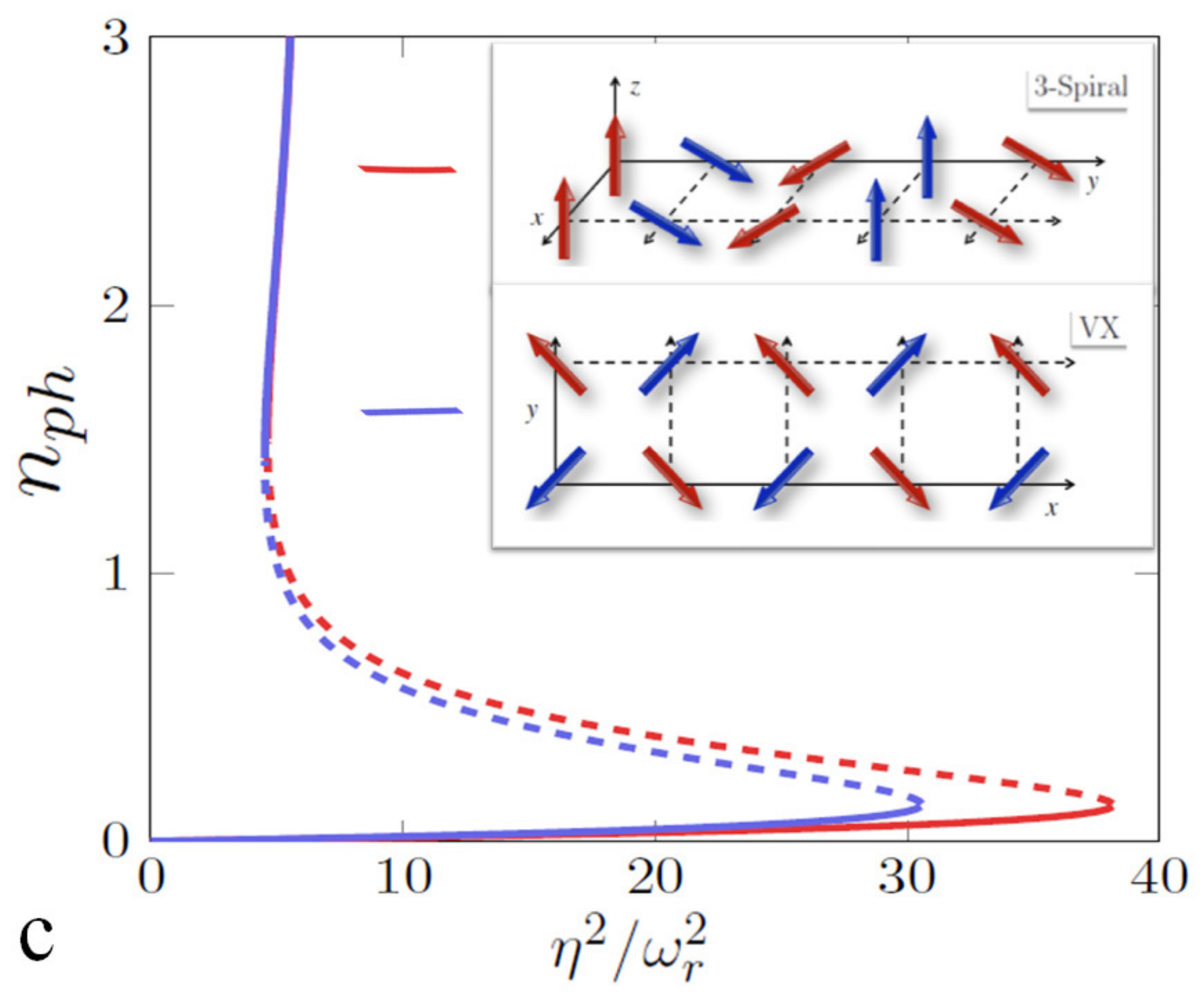}
\label{fig:CavSpecRII-c}}
\\
{\includegraphics[width=0.45\textwidth,
height=0.35\textwidth]{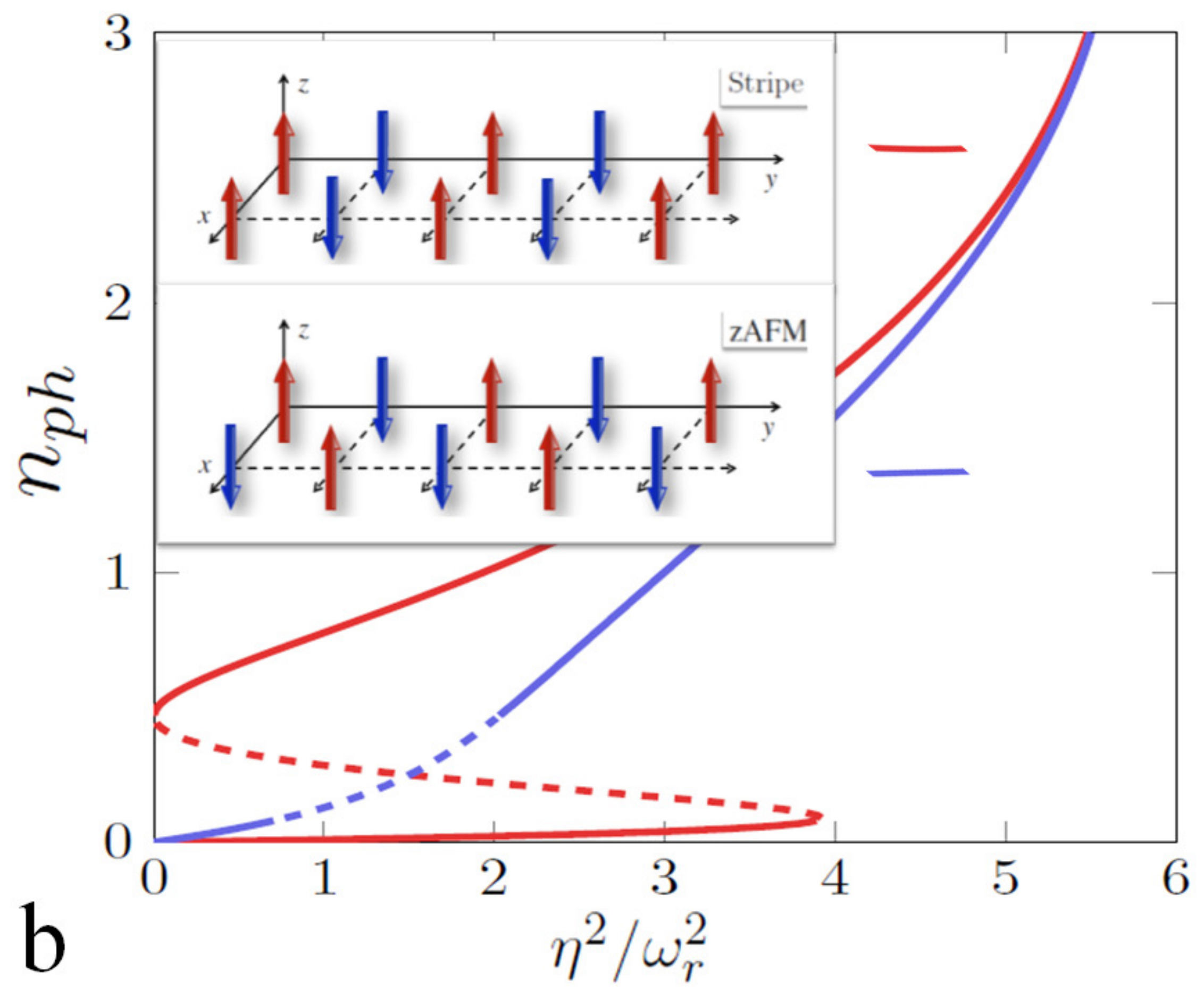}
\label{fig:CavSpecRII-b}} \qquad
{ \includegraphics[width=0.45\textwidth,
height=0.35\textwidth]{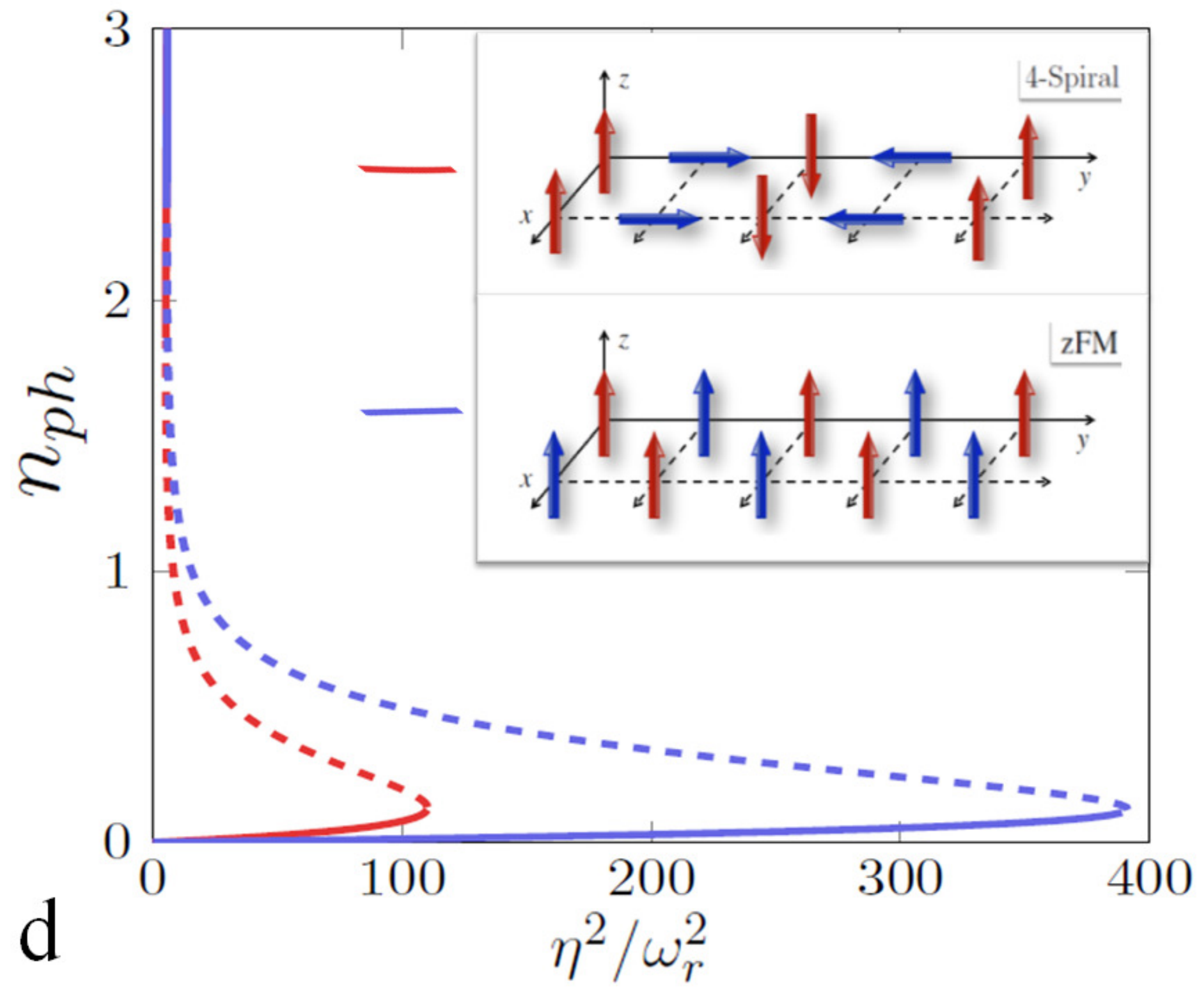}
\label{fig:CavSpecRII-d}}
\caption{ (\textbf{a}) The spin vectors in internal spin spaces of two neighboring sites; (\textbf{b}--\textbf{d}) Cavity spectrum for different phases in the MI region for different
non-abelian flux insertions. The SOC strength for all the phases are $(\alpha,
\beta)/\pi$ = (0.01,0.01) zFM; (0.2,~0.2) 4-Spiral; (0.3, 0.3) 3-Spiral;
(0.5, 0.5) Stripe; (0.34, 0.34) VX. Note the turning points are highly dependent
upon the phases. The dotted part shows the unstable region of the spectrum. The red and blue legends correspond to the magnetic order, shown in boxes.  From~\cite{ourPRA}. (Reprinted figures with permission from Padhi, B.; Ghosh, S.  Phys. Rev. A 2014, 90, 023627. Copyright (2014) by the American Physical Society. Source: http://dx.doi.org/10.1103/PhysRevA.90.023627)}
\label{fig:CavSpecRII}
\end{figure}

A schematic of the spin configurations of these phases are given in the insets of {Figure} \ref{fig:CavSpecRII}. The zFM order is a uniformly ordered phase where all the spins are aligned along the \emph{z}-axis; however, in the zAFM phase the direction of the spin vectors alternate as parallel or anti-parallel to the \emph{z}-axis. There is a subtle difference between the stripe phase and the zAFM: in the stripe phase, along a given axis on the \emph{xy}-plane all spins are up but for the other axis they alternate as up and down. In zAFM they the spins alternate along both the axes. Two types of spiral waves appear for this system. In both the cases, all the spins along one axis on \emph{xy}-plane are parallel; however, along the other axis, the spin vectors make an angle with the \emph{z}-axis which changes (starting from 0) as we move along the axis. However, there exists a period in number of lattice sites after which the angles are repeated like wave. In 4-spiral, 4 sites make one period: the angles progress with site as $\pi, \pi/2, 0, -\pi/2, \pi ...$. In 3-spiral, 3 sites make one period: the angles progress with site as $\pi, \pi/3, -\pi/3, \pi ...$. The vortex phase is one of the \emph{{XY}} phases, in which all the spin vectors lie on the \emph{XY} plane.

Now we turn our attention to the spectrum of the non-interacting SOC bosons in a cavity induced quantum optical lattice potential. The cavity spectrum can be obtained from Equation \eqref{eq:Photon} by setting $\partial_t \hat{a} = 0$ as:
\beq
n_{ph} = \expect{ \ha^{\dag (s)}\ha^{(s)}}_\Psi = \frac{\eta^2}{\kappa^2 +
(\Delta_c' - U_0 J_1 \expect{\hB}_\Psi)^2} .
\label{eq:nph}
\eeq

The many body wavefunction $\Psi$ above has an orbital part and a spinorial part. We focus on the spinorial part of the wavefunction since they characterize the magnetic orders. Detection of various phases in the orbital part of the wavefunction, through the cavity spectrum was carried out in \cite{Mekhov1}. To simplify our discussion we assume that the orbital (optical lattice site) part of the  wavefunction corresponds to a Mott insulator (MI) state with one atom per lattice site (by restricting the lattice depth $\geq20E_r$ \cite{Fisher}).  In our work we propose a method which enables us to probe the spinorial part of the wavefunction (hence the magnetic orders) with the help of the cavity spectrum. The key idea is by forming a quantum lattice with the help of the cavity a feedback mechanism (of cavity light) is triggered, causing the cavity spectrum to non-linearly depend on $n_{ph}$ through this modified Lorentzian \cite{Meystre-Book}. \linebreak In addition, the spectrum is also dependent upon the state $\ket{\Psi}$ through the expectation value of the hopping operator $\expect{\hB}_\Psi$. This dependence is pronounced only when $J_1$ is finite. In further discussions we will show how this dependence can be used to probe the spinorial part of the quantum many-body ground state wavefunction.

Following \cite{Girvin} the wave function for various orders can (in the Mott
phase only) be written as
\beq
\ket{\Psi_{MI}} = \prod_{i \in A , j \in B} \ket{\psi_A}_i \ket{\psi_B}_j ,
\label{eq:MI}
\eeq
with site indexes $i, j$ and $\ket{\psi_{A,B}} = \cos \frac{\theta_{A,B}}{2}
\ket{\upa} + e^{ i \phi_{A,B}} \sin \frac{\theta_{A,B}}{2} \ket{\dna} $. The
entire lattice is divided into two sub-lattices $A, B$ and we assume alternating
sites belong to different sub-lattices. The parameters $\theta , \phi$ are
projection angles in the internal spin space. We assume there are exactly equal
number of lattice sites in sub-lattices $A$ and $B$, hence the total number of
sites is $K^2$ even, also assuming unit filling we set $K^2 = N_0$. Please note
$K$ was earlier used to denote the wave number of the cavity photon and here we use the
same notation for a different thing. We calculate the expectation value of the tunneling
operator, $\expect{\hB}$ for various magnetic orders and summarize in the Table
\ref{tbl}. This will be the basis of further discussions. We can distinguish between different magnetic orders because each order
can now be associated with a corresponding $\expect{\hB}$, hence a
cavity spectrum, { {provided}} there is non-vanishing \emph{z}-axis component
of the spin vector (the reason will be clear later on). Thus one can not distinguish between any of the \emph{XY} phases, such as the vortex phase or the anti-vortex phase \emph{etc}. However, the other various magnetic orders, which can arise in a spin-orbit coupled system through experimental control of the free parameters ($\alpha, \beta $) \cite{Radic} or ($\alpha, \lambda$) \cite{Trivedi, Cai} can be well distinguished.

We further divide the MI regime into two regions separated at a potential depth of $25 E_r$ (see Figure~ \ref{fig:CavSpec}a). In one region of the depth values the $J_1$ vanishes, hence it becomes impossible to probe the spinorial part of ground state through the cavity spectrum. In the other region the $J_1$ is finite, enabling us to probe the ground state. We name these regions as region I: Shallow MI regime.

Lets first consider region II. As evident from Figure \ref{fig:CavSpec}a in this
region $J_0$ vs $V_0$ can be approximated by a linear function ($J_0 = aV_0+b$)
and $J_1$ can be assumed to be zero. The variation of $n_{ph}$ with respect to pump amplitude $\eta^2$ is shown in Figure \ref{fig:CavSpec}b and that with respect to detuning $\Delta_c'$ is shown in Figure~\ref{fig:CavSpec}c. There exists a bi-stable region in the spectrum which is shown by red dashed line. In the strong MI regime the atoms get tightly localized at their site resulting in a negligible hopping amplitude. The atoms can sense the presence of the abelian or non-abelian field only through the hopping term, and now since the hopping amplitude is almost negligible the cavity spectrum is insensitive to the abelian or non-abelian gauge field.

\begin{table}[H]
\centering
\small
\caption{\label{tbl} Expectation of the hopping operator and the steady-state
photon number for different phases in the  Mott insulator (MI) state.}
\begin{tabular}{cc}
\toprule
{\bf Order}  & \quad $\expect{\hB}_\Psi$ \\ \hline
zAFM & $0$ \\
Stripe & $2K(K-1)\cos \beta$ \\
VX & $K(K-1)(\cos \alpha + \cos \beta)$ \\
3-Spiral & $3K(K-1)(\cos \alpha + 4\cos \beta)/8$ \\
4-Spiral & $K(K-1)(\cos \alpha + 3\cos \beta)/2$ \\
zFM & $2K(K-1)(\cos \alpha + \cos \beta)$ \\
\bottomrule

\end{tabular}
\end{table}

As the pumping amplitude $\eta$ decreases the photon number decreases (see Figure \ref{fig:CavSpec}b; however, at a~certain point (point D) the photon number abruptly drops to a very small value (point A), hence the lattice suddenly becomes very shallow. This causes a phase transition from Mott insulator to superfluid phase. Similarly, as $\eta$ increases the photon number also increases, so does the lattice depth as well. At~the point B it suddenly jumps to a large value of $n_{ph}$ (point C) hence a phase transition from super fluid to Mott insulator occurs. This is an instance of bistability driven driven phase transition, which was previously pointed out {in} \cite{Larson,Meystre} in different contexts. Points B or D are often referred to as turning points or critical points. When the photon number gets lowered one might end up at a super fluid phase or one might stay in the shallow MI region. So to determine the phase exactly one needs to obtain the exact phase diagram and locate the appropriate turning points. We do not extend this discussion further.

Now we turn to the case of shallow MI regime (or region I). We separate the following section where we show that in this region it is feasible to probe the ground state of the SOC BEC through the cavity spectrum. When $J_1 \neq 0$, the Lorentzian in Equation \eqref{eq:nph} can sense the presence of the magnetic orders through~$\expect{\hB}$.

Before getting to our results, it is worthwhile to point out that after the realization of spin-orbit coupling for bosonic clouds
\cite{reviewSOC} or condensate \cite{lin1} by Spielman's group the phase
diagram of such a system was theoretically obtained by various groups in
\cite{Trivedi,Sengupta,Radic,Cai}. Experimental verification of these phases
might not be very trivial, most importantly detecting all the emergent phases
using a single experimental setup is a formidable  task. So far, the method
of spin structure factor measurement through Bragg spectroscopy \cite{Bragg} has
been commonly used. Other methods include measurement of spatial noise
correlations \cite{AltmanNoise}, polarization-dependent phase-contrast imaging
\cite{Kurn-Magnetization}, direct imaging of individual lattice sites
\cite{SingleSite} \emph{etc}. However, each of these techniques come with
their own set of  complications.

\begin{figure}[H]
\centering
{\includegraphics[width=0.45\columnwidth]{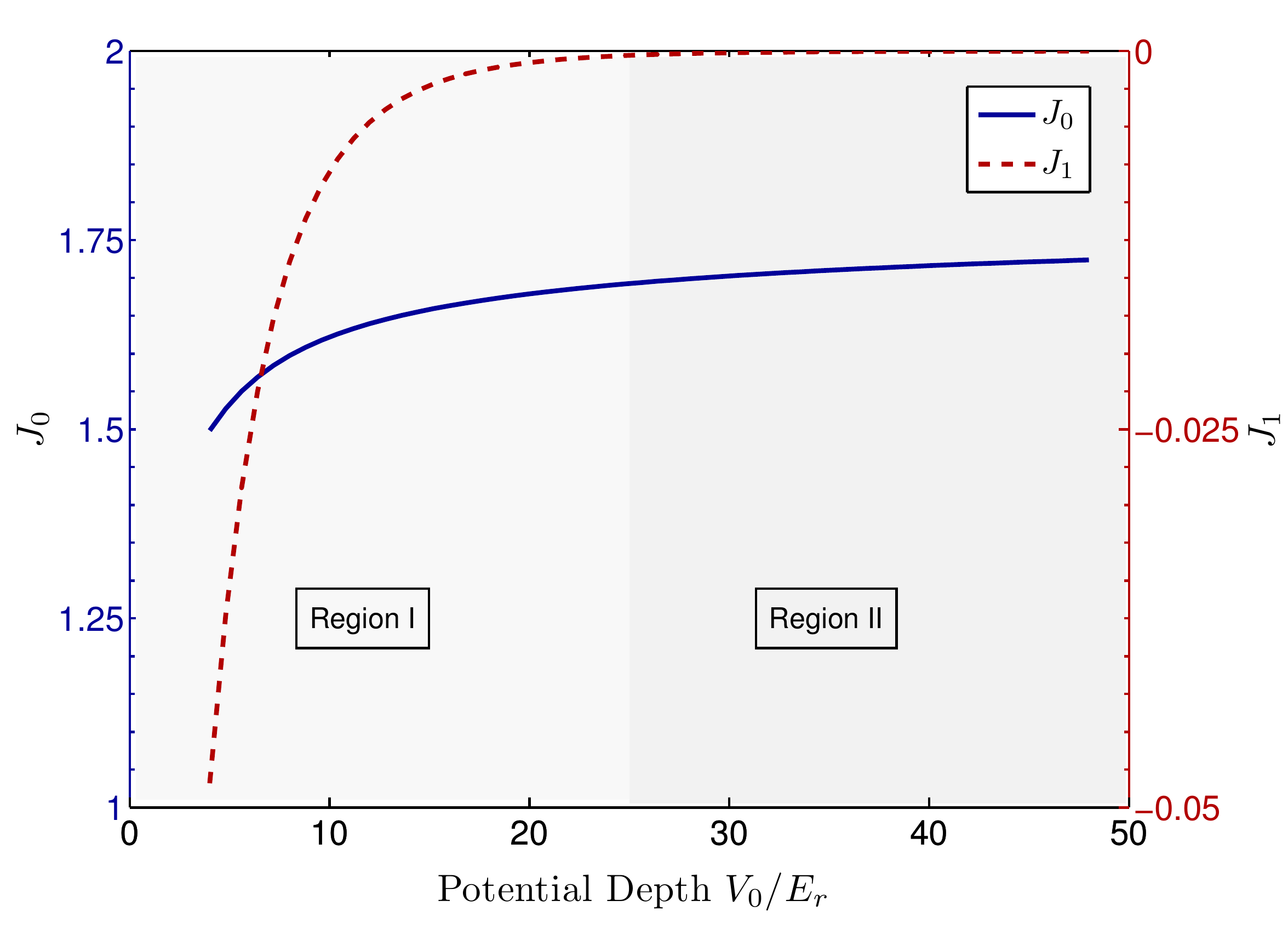}
\label{fig:CavSpec-a}\\
(\textbf{a})}\\
{\includegraphics[width=0.45\columnwidth]{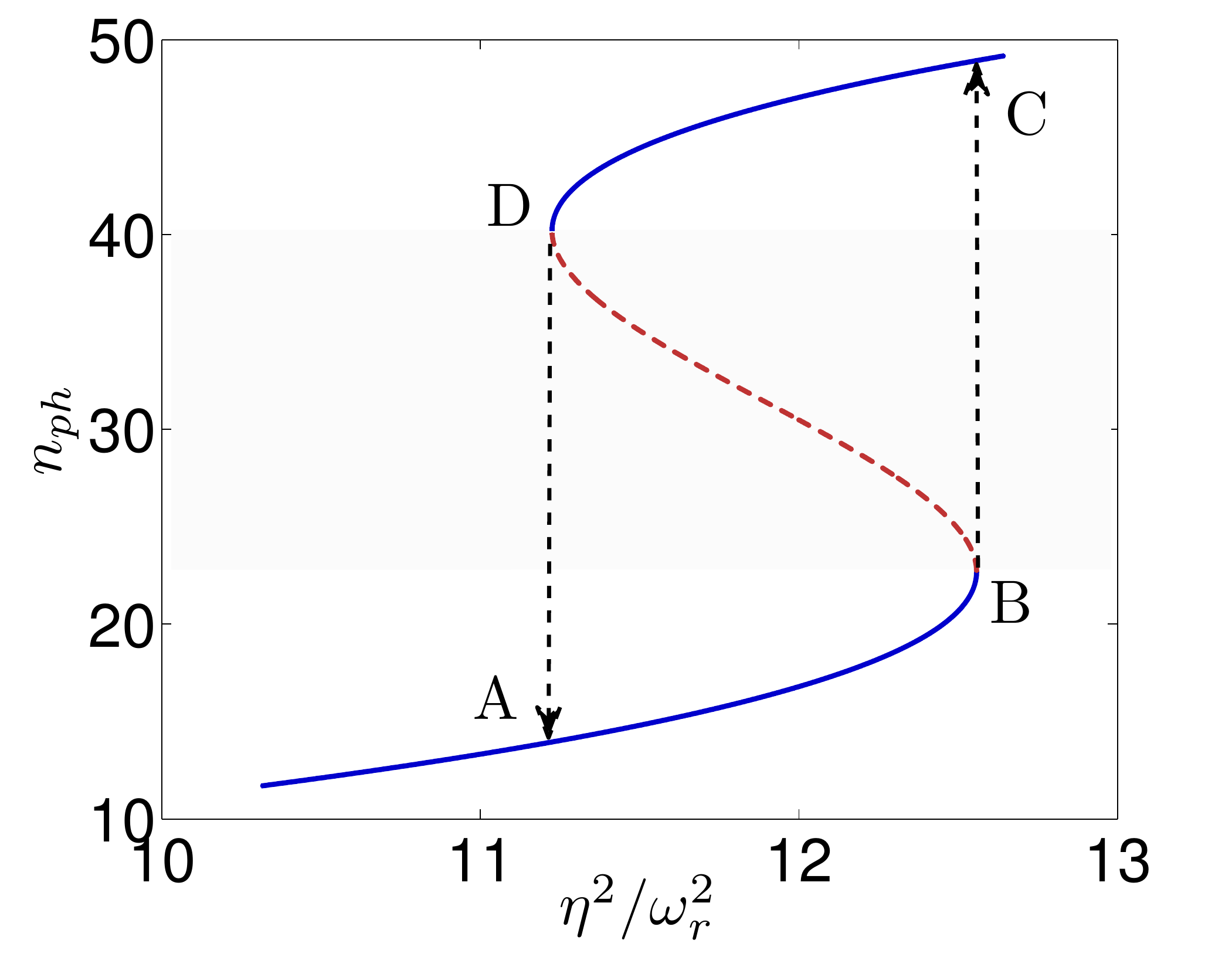}
\label{fig:CavSpec-b}\\
(\textbf{b})}\\
{\includegraphics[width=0.45\columnwidth]{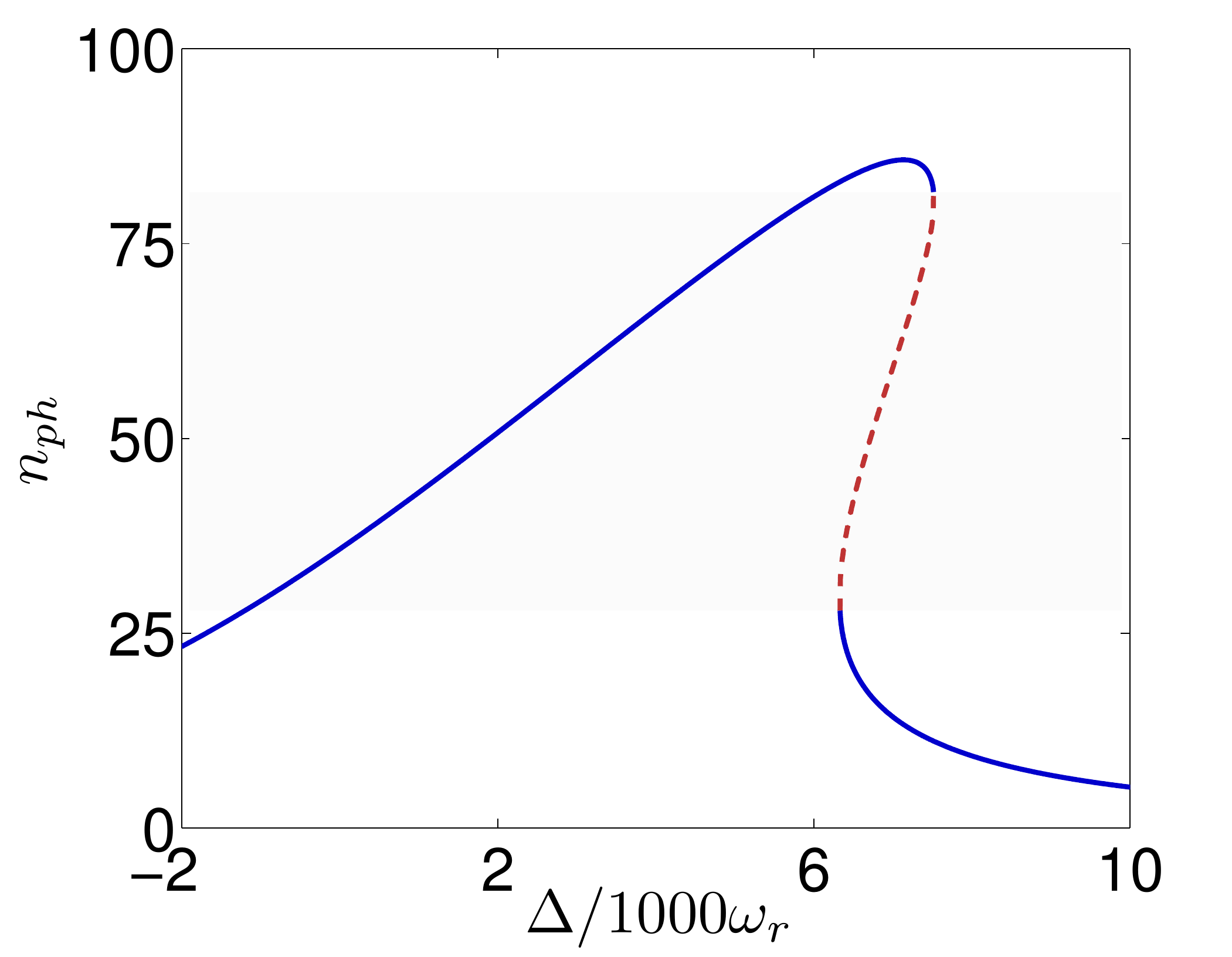}
\label{fig:CavSpec-c}\\
(\textbf{c})}
\caption{ ({\bf a}) The variation of overlap integral elements with potential depth. We study the variation in two regions, which are (arbitrarily) separated at $V = 25 E_r$; we have used a $6 \times 6$ lattice, $\{U_0, \kappa \}= \{ 12, 1 \} \omega_r $; ({\bf b}) with pump amplitude $\eta$ for $\Delta_c = 5000 \omega_r$; ({\bf c})~with detuning $\Delta_c$ for $\eta = 6 \omega_r$. The red dotted lines are the unstable regions of photon count.  From \cite{ourPRA}. (Reprinted figures with permission from Padhi, B.; Ghosh, S.  Phys. Rev. A 2014, 90, 023627. Copyright (2014) by the American Physical Society. Source: http://dx.doi.org/10.1103/PhysRevA.90.023627)}
\label{fig:CavSpec}
\end{figure}

Extending the idea which
was originally espoused for BEC without spin degrees of freedom
\cite{Mekhov1} here we propose a different scheme of experiment where
such magnetic orders can be ascertained without making a direct measurement
on the atomic system. The relation between such approach and
 ``quantum non-demolition measurement'' technique  was also discussed \cite{SelfOrganization2, Supersolid}. The method facilitates the detection all possible phases arising in
the Mott regime of a SOC BEC and this can also be extended to the superfluid
(SF) regime.

The cavity spectra for each of these orders are obtained in Figure \ref{fig:CavSpecRII}. The spin-orbit coupling strength $(\alpha , \beta)$ chosen for
a particular order is such that, that specific order gets stabilized \cite{Radic}. As we gradually increase the pump value the photon number gets increased, but at the turning point ($\eta_c$) it suddenly jumps to a higher value of photon number, since the photon intermediate count corresponds to the unstable region. Clearly, the behavior of the spectra for different orders are different, specifically the value of $\eta_c$ varies widely. The zAFM will not show any such jump, and the stripe phase will have a very small value of $\eta_c$. For zFM phase, $\eta_c$ will always be the largest and for 4-spiral phase it would be quite comparable with the $\eta_c$ of zFM. For the XY phase and 3-spiral, the $\eta_c$ are always between these two extremes.

The above discussion is supported by the following observation. In Figure \ref{fig:CavSpecRII}a the internal spin (by ``spin'' we actually refer to ``pseudo-spin'') spaces of two neighboring sites are shown as red or blue blobs. The basis vectors of the spin spaces are the eignvectors of $\hat{S}_z$. If a spin vector makes an angle $\theta$ with the \emph{z}-axis in the real space, then in the spin space it makes an angle $\theta/2$ with the $\dna$ axis.  A~particular magnetic order is nothing but a specific spatial distribution of these $\theta$ and $\phi$ values. The~value of $\expect{\hB}$ is a measure of the probability of spin-dependent hopping across neighboring sites, which hence captures this variation of $\theta$ values over the configuration space. We proceed in the following way: if a spin vector creates an angle $\theta_A$ with the \emph{z}-axis and the spin vector at the site nearest to it makes an angle $\theta_B$ then in their internal spin spaces they make an angle $\theta_A/2$ and $\theta_B/2$ with $\dna$. \linebreak Hence the projection of the spin vectors on the $\dna$ axis are $\cos \theta_{A,B}/2$ and that on the $\upa$ axis are $\sin \theta_{A,B}/2$. The probability for a hopping of $\upa$ to $\upa$ (or $\dna$ to $\dna$) is the modulus squared product of the projection lengths along $\upa$ ($\dna$) axes. Hence for hopping of $\upa$ to $\upa$ has a probability of $( \sin \frac{\theta_A}{2}\sin \frac{\theta_B}{2})^2$ and for hopping of $\dna$ to $\dna$ it is $(\cos\frac{\theta_A}{2} \cos\frac{\theta_B}{2})^2$. Since $\upa$ and $\dna$ are orthogonal vectors hopping associated with a spin flip is found to have vanishing $\expect{\hB}$.

To illustrate the implication of the above technique consider the case of zAFM. In zAFM on alternative sites spin vectors are oriented parallel or anti-parallel to the \emph{z}-axis, \emph{i.e.}, $\theta_A = 0,$ $\theta_B = \pi$. Hence any reordering of the spin vectors (mediated by the cavity light) which do not alter the magnetic order should consist of hopping from $\upa$ to $\dna$ or visa-versa. However, the matrix element $\expect{\hB}$ for such a hopping is zero. Hence $\expect{\hB}_{zAFM} = 0$ (see the Table). Similarly, in the case of zFM all spin vectors are aligned along the\emph{ z}-axis, \emph{i.e.}, $\theta_A = \pi = \theta_B $. Hence any hopping other than $\upa$ to $\upa$ will have vanishing contribution in $\expect{\hB}_{zFM}$ and $\expect{\hB}_{zFM} \propto (\sin \pi/2 \sin \pi/2)^2$. It must be noted that the value of $\expect{\hB}$ in turn controls the value of $\eta_c$, hence the trend of variation of $\expect{\hB}$ with respect to the phases gets mapped to that in the values of $\eta_c$. The $\cos \alpha$ or $\cos \beta$ are just scaling factors introduced because of SOC. This is the central result of our work. Now we show that other than the phase information the cavity spectrum can also be used to extract the amount of abelian or non-abelian flux inserted in the system.


\begin{figure}[H]\ContinuedFloat
\centering
{ \includegraphics[width=0.45\columnwidth]{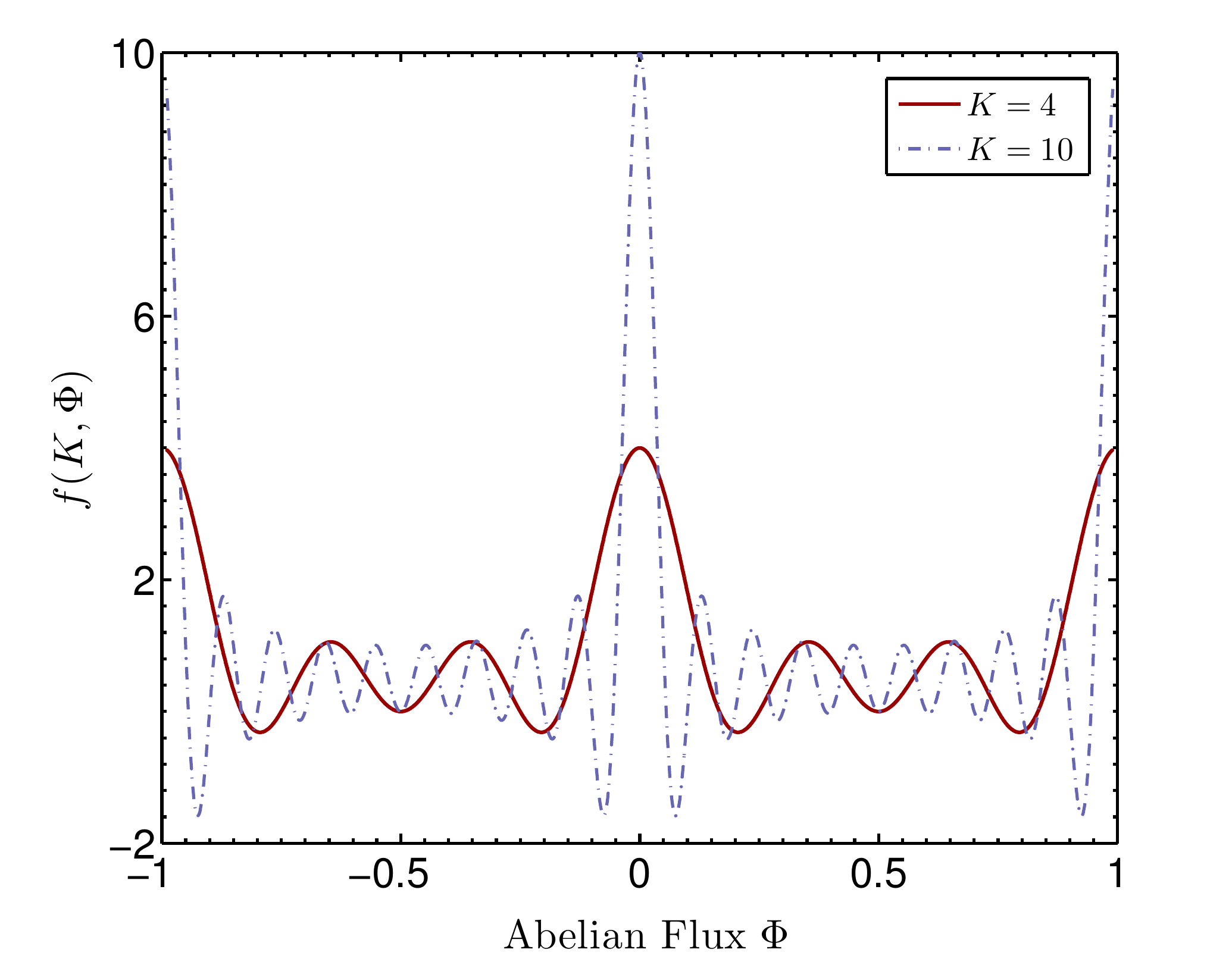}
\label{fig:flux-a}\\
(\textbf{a})}\\
{ \includegraphics[width=0.4\columnwidth]{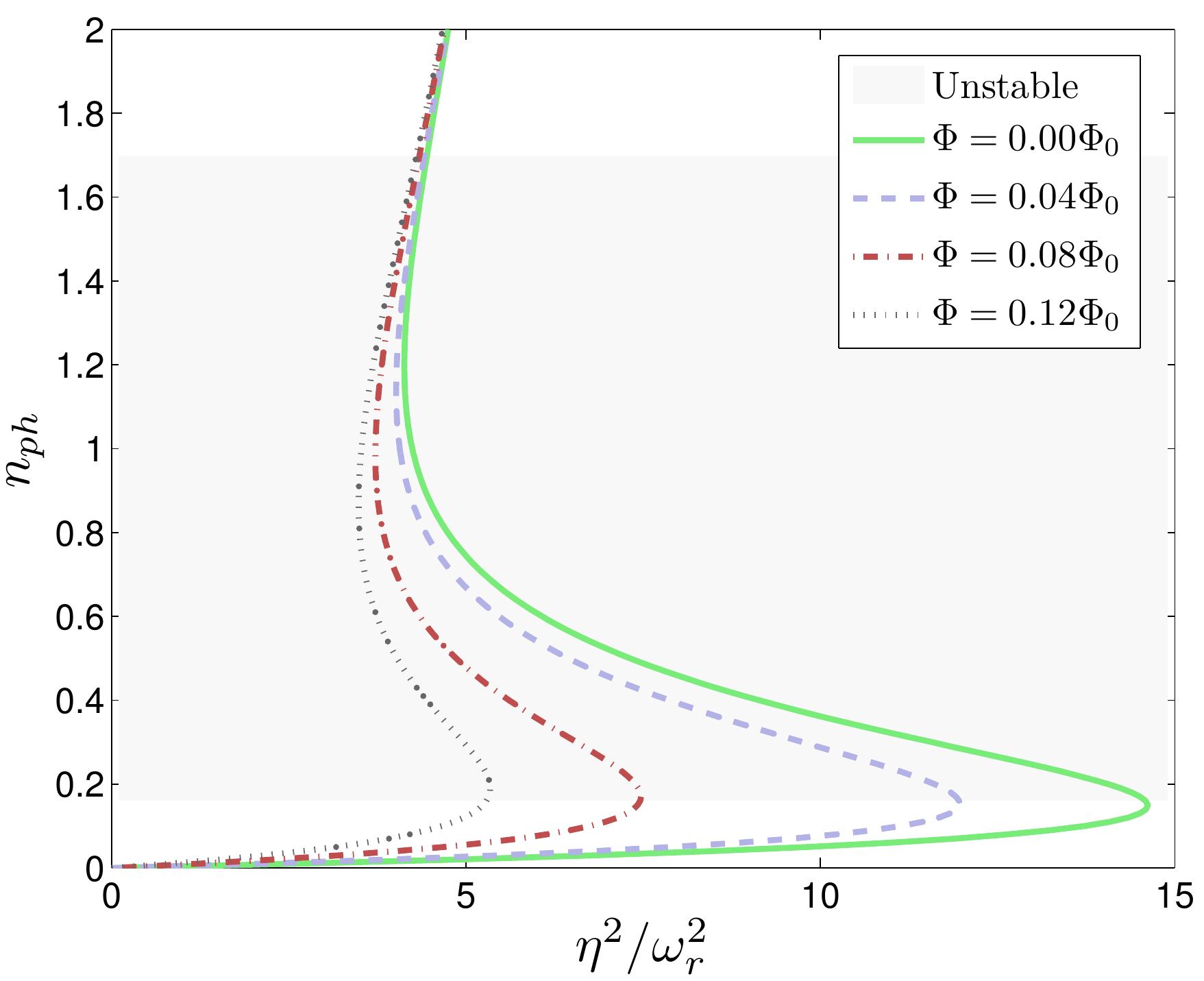}
\label{fig:flux-b}\\
(\textbf{b})}\\
{ \includegraphics[width=0.4\columnwidth]{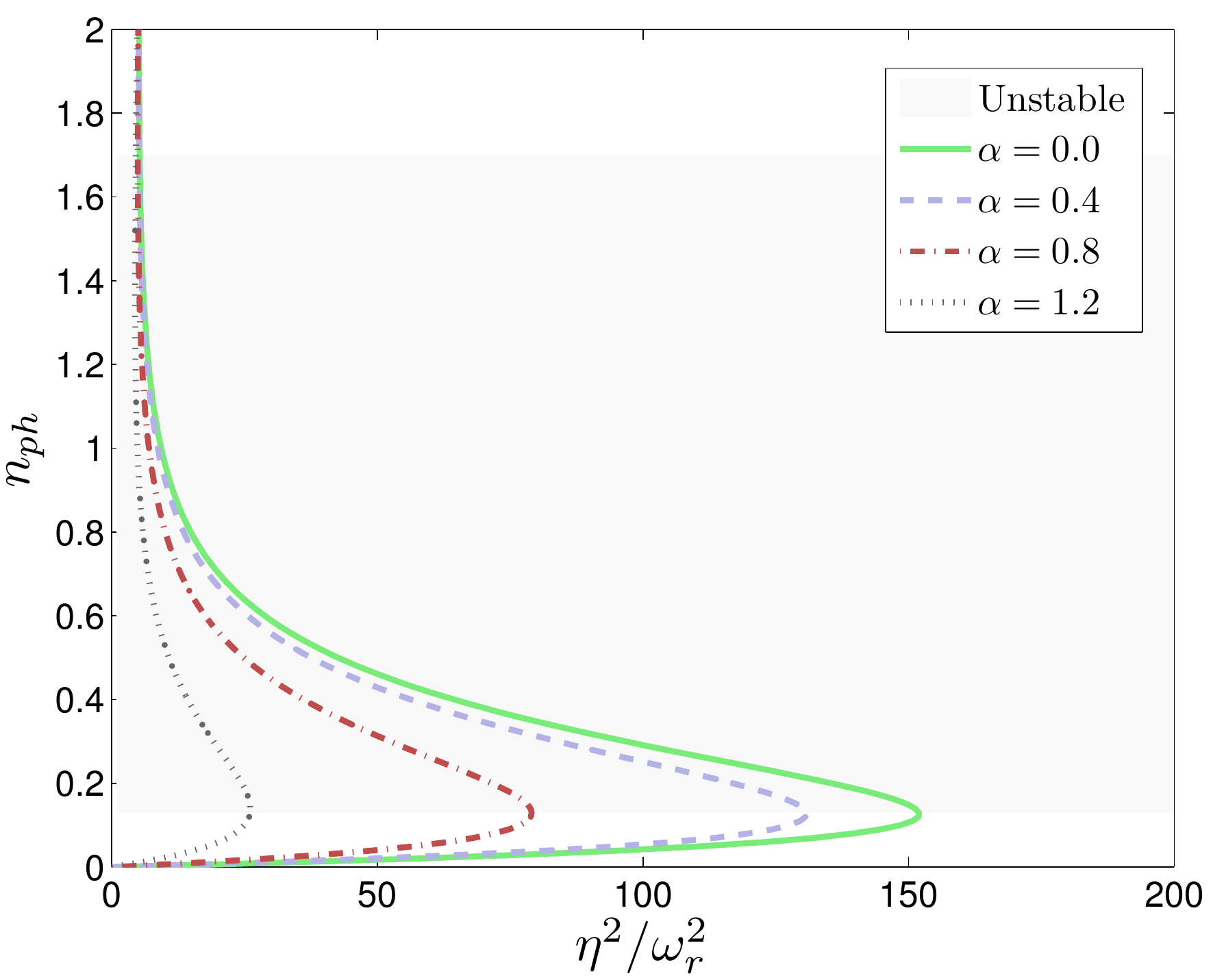}
\label{fig:flux-c}\\
(\textbf{c})}
\caption{ (\textbf{a}) The variation of grating function $f(K,\Phi)$ with the
inserted abelian flux. The graph legends indicate the size of the lattice. In a
large lattice limit the grating function does not sense the variation of $\Phi$;
The cavity spectrum for different fields: (\textbf{b})~abelian fields (with fixed
non-abelian field, $\alpha = -\beta = \pi/2 - 0.15$); (\textbf{c})~non-abelian fields
(with fixed abelian field, $\Phi = 0.08 \Phi_0$). The negative slope region is
the unstable (gray) part of the spectrum.  From \cite{ourPRA}. (Reprinted figures with permission from Padhi, B.; Ghosh, S.  Phys. Rev. A 2014, 90, 023627. Copyright (2014) by the American Physical Society. Source: http://dx.doi.org/10.1103/PhysRevA.90.023627)}
\label{fig:flux}
\end{figure}


In order to show how the cavity spectra can be used for flux detection we consider the zFM phase, which is stabilized in presence of both an abelian and a non-abelian field \cite{Grass}. In presence of an abelian flux, the expectation value of the tunneling operator for zFM order becomes $\expect{\hB}_{FM} = 2 \cos \alpha (K-1) (K + f(K, \Phi))$. The presence of the abelian flux gives additional phases to the hopping thus resulting in a overall phase factor of $f(K, \Phi) = \frac{\sin(K \pi \Phi)}{\sin(\pi \Phi)} \cos[\pi \Phi (K-1)]$. This function is plotted in Figure \ref{fig:flux}a. The similarity of the functional form of $f(K, \Phi)$ with that of an \emph{N}-slit grating function is just because in this case the phases arising due to the presence of this field gets summed over to yield such a function. Evidently the optical lattice acts as a quantum diffraction grating \cite{Adhip, Mekhov1}.

\section{Conclusions }

The review  article analyzes some recent progress in cavity optomechanics with ultracold atomic condensate in synthetic gauge field after providing a general introduction
to the field of cavity optomechanics (with single atom an atomic ensemble) and the physics of cold atoms in artificial gauge field. At this moment the field is nascent and many
interesting problems in this direction can be addressed. However, one of the main issues will be experimental implementation of such scheme which involves ultracold atomic condensate
in a synthetic gauge field inside an optical cavity. This requires further detailed theoretical analysis of the relevant systems which for example will consider the effect of measurement backaction on the quantum phases of the cold atoms \cite{DSK4}.

One more thing is, in the presence of a dynamical lattice, both the atom and
photon operators evolve, in accordance with their corresponding (coupled)
Heisenberg equations \cite{Mekhov1, Mekhov2}. One can solve this pair of equations
simultaneously to study the full self-organization. However, assuming the atoms
fall through the cavity light field sufficiently faster (much before the atoms affect the cavity photon) we ignore the back action of the atoms on the cavity light \cite{Aspelmeyer}. Self-organization of atoms in the lattice~\cite{SelfOrganization, SOrg2, SelfOrganization2} can in itself be a separate direction to pursue, facilitating the study of self-organized checkerboard phase \cite{Deng}, supersolid phase \cite{Supersolid}, or quantum spin-glass phase \cite{Spin-glass}.

In most of the analysis presented in this review, only the average photon number is calculated from the cavity transmission spectrum and was related with the quantum phases of ultracold atoms in the cavity.
Another interesting quantity that can be calculated from the cavity transmission spectrum is the quantum fluctuations around the mean photon number \cite{MekhovQF}
which could be again be used to quantify the back action of the measurement process on the many-body state of the ultracold atoms. This apart it will also be useful to explore if topological characterization of ultracold atomic states \cite{TI}, detection of edge or surface states in such states \cite{Edge} can be carried out for ultracold atomic system inside an optical cavity.

Hopefully, our analysis will  augment the theoretical and experimental study of the behavior synthetically gauged ultracold atoms inside high finesse cavity.


\acknowledgments{Acknowledgments}
One of us (Sankalpa Ghosh) is supported by a DAE SRC Outstanding Investigator Fellowship and Project Grant, given by Department of Atomic Energy, Govt. of India.
\newpage

\authorcontributions{Author Contributions}

Sankalpa Ghosh conceived the idea of the work. With the assistance of Sankalpa Ghosh, Bikash Padhi carried out the theoretical and numerical works. Both authors contributed to writing and revising the manuscript.


\conflictofinterests{Conflicts of Interest}
The authors declare no conflict of interest.

\bibliographystyle{mdpi}
\makeatletter
\renewcommand\@biblabel[1]{#1. }
\makeatother


\end{document}